\documentclass[11pt]{article}
\pdfoutput=1
\usepackage{amsmath,amssymb,epsfig,amsfonts}
\usepackage{graphicx}
\usepackage{subfig}
\usepackage[usenames, dvipsnames]{color}

\usepackage[dvipsnames]{xcolor}
\usepackage{cite}
\usepackage{multirow}
\usepackage{slashed}
\usepackage{verbatim} 
\usepackage{enumitem}
\usepackage{enumerate}
\usepackage{rotating}
\usepackage{xcolor}
\usepackage{setspace}
\usepackage{multirow}
\usepackage{bm}
\usepackage[normalem]{ulem}

\usepackage{booktabs} 
\usepackage{tikz-cd} 

\usepackage[fleqn,tbtags]{mathtools}

\usepackage{ytableau}

\usepackage{tikz}
\usepackage{diagbox}
\usetikzlibrary{positioning}
\usetikzlibrary{calc}
\usetikzlibrary{decorations.pathreplacing,calligraphy}

\usetikzlibrary{arrows}

\usepackage{xstring}
\usetikzlibrary{decorations.pathmorphing} 
\usetikzlibrary{decorations.markings} 
\usetikzlibrary{arrows} 
\usetikzlibrary{shapes} 
\usetikzlibrary{matrix} 
\usetikzlibrary{positioning} 
\usepackage[english]{babel} 
\usepackage[autostyle]{csquotes}

\usepackage{CJK}
\usepackage{tcolorbox}

\usepackage{titlesec}

\usepackage{hyperref}
\usepackage{cleveref}

\titlespacing{\paragraph}{%
  0pt}{
  0.2\baselineskip}{
  0.5em}

\makeatletter

\newcount\hour \newcount\minute
\hour=\time \divide \hour by 60
\minute=\time
\def\now{%
\ifnum \hour<13
  \ifnum \hour=0 \advance \hour by 12 \number\hour:\else \number\hour:\fi%
     \ifnum \minute<10 0\fi%
     \number\minute%
\ A.M.%
\else \advance \hour by -12 \number\hour:%
  \ifnum \minute<10 0\fi%
  \number\minute%
  \ P.M.%
\fi%
}

\makeatother

\begin{document}

\baselineskip=18pt  
\numberwithin{equation}{section}  
\allowdisplaybreaks  


%
%


\thispagestyle{empty}

\vspace*{0.8cm} 
\begin{center}
{{\LARGE \bf  Nontrivial bundles and defect operators in $n$-form gauge theories }}

 \vspace*{1.5cm}
Shan Hu\\

\vspace*{1.0cm} 
{\it   Department of Physics, Hubei University \\
Wuhan 430062, China}\\

\vspace{6mm}

{\small \tt hushan@hubu.edu.cn} \\

\vspace*{0.8cm}
\end{center}
\vspace*{.5cm}

\noindent
In $(d+1)$-dimensional $1$-form nonabelian gauge theories, we classify nontrivial $0$-form bundles in $ \mathbb{R}^{d} $, which yield configurations of $D(d-2j)$-branes wrapping $(d-2j)$-cycles $c_{d-2j}  $ in $Dd$-branes. We construct the related defect operators $ U^{(2j-1)} ( c_{d-2j} )   $, which are disorder operators carrying the $D(d-2j)$ charge. We compute the commutation relations between the defect operators and Chern-Simons operators on odd-dimensional closed manifolds, and derive the generalized Witten effect for $U^{(2j-1)} ( c_{d-2j} ) $. When $c_{d-2j}$ is not exact, $ U^{(2j-1)} ( c_{d-2j} ) $ and $ U^{(2j-1)} (- c_{d-2j} ) $ can also combine into an electric $(2j-1)$-form global symmetry operator, where the $(2j-1)$-form is the Chern-Simons form. The dual magnetic $(d-2j)$-form global symmetry is generated by the $D(d-2j)$ charge. We also study nontrivial $1$-form bundles in $(d+1)$-dimensional $2$-form nonabelian gauge theories, where the defect operators are $\mathcal{U}^{(2j)} ( c_{d-2j-1} )   $. With the field strength of the $1$-form taken as the flat connection of the $2$-form, we classify the topological sectors in $2$-form theories.

\newpage

\tableofcontents


\section{Introduction}

Although the nonabelian gauge theory for the $1$-form is well-established, its higher-form extension is mysterious. In mathematics, the definition of a higher connection requires the generalization from principal bundles to gerbes \cite{0, 00, 000}. On physical side, the construction of the nonabelian higher-form gauge potential is usually obstructed by no-go theorems \cite{01, 01a, 001} and a global higher-form symmetry must be Abelian \cite{G}.

In Abelian gauge theories, a nontrivial $n$-form bundle could produce an $(n+1)$-form field with the non-vanishing field strength, so the lower-form field could at least capture part of features of the higher-form. Here by nontrivial $n$-form bundles, we mean $ n $-forms that do not have a global definition in $\mathbb{R}^{d} $ and must be defined in patches, although $ \mathbb{R}^{d} $ is topologically trivial. The related $(n+1)$-form field strength is globally defined in $ \mathbb{R}^{d}$ but must contain singularities.

The simplest example is in the $ 2d $ space $ \mathbb{R}^{2} $ with the coordinate $(r,\theta)\sim (x^{1},x^{2})$. The $ 1 $-form is $ A $ with the field strength $ F=dA $. The gauge transformation rule is $ A\rightarrow A+d\phi $ for a globally defined $ 0 $-form $ \phi $. The loop operator on a loop $  \mathcal{M}_{1} \subset \mathbb{R}^{2}  $ with $ \partial \mathcal{M}_{2}=\mathcal{M}_{1} $ is
\begin{equation}\label{4.3q}
O(A, \mathcal{M}_{1})=\frac{i}{2\pi}\int_{\mathcal{M}_{1}} A=\frac{i}{2\pi}\int_{\mathcal{M}_{2}} F\;.
\end{equation}
Without the $1$-form $A$, a $0$-form $ \phi $ characterized by $  u=e^{\phi} \in U(1)$ produces a $1$-form $ a=u^{-1}du =-ind\theta$, when $ u $ is taken to be $ u =e^{-in\theta} $ for $ n\in  \mathbb{Z}$. The field strength 
\begin{equation}
 f_{\mu\nu}(x)=(da)_{\mu\nu}(x)=-2 \pi n i\epsilon_{\mu\nu}\delta^{(2)}(x)\;\;\;\;\;\;\; \mu,\nu=1,2 
\end{equation}
is supported at the origin. $ u $ is globally defined but $ \phi = - in\theta$ has a branch cut at $ \theta=2\pi $. In $ \mathbb{R}^{d} $, the generic $ \phi $ bundle with $ \phi \sim \phi +2n\pi i$ is characterized by 
\begin{equation}\label{4.3}
 O(u^{-1}du , \mathcal{M}_{1})=w(u^{-1}du , \mathcal{M}_{1}) \in \mathbb{Z} \cong  \Pi_{1}[U(1)]
\end{equation}
for all loops $\mathcal{M}_{1} \subset  \mathbb{R}^{d}$, where $ w(u^{-1}du , \mathcal{M}_{1})  $ is the winding of $ U(1) $ around $\mathcal{M}_{1}$. $ \phi $ with the non-zero $ w(u^{-1}du , \mathcal{M}_{1}) $ gives a $ 1 $-form $  u^{-1}du$, for which $ f $ is supported on a $ (d-2) $-cycle $ c_{d-2} $ (the Poincar\'e dual of $ f $). Gauging the trivial $ \phi $ imposes the equivalence relation $ a \sim a+d\phi $. The ordinary $A/F $ can be obtained as the linear combinations of $ a/f $. A $0$-form gauge transformation operator with the transformation parameter $ \phi $ is denoted by $ U^{(1)} (c_{d-2})$, which is a defect operator localized at $c_{d-2}$ implementing a $1$-form transformation with $A\rightarrow A+a$.

Similarly, in the $ 3d $ space $ \mathbb{R}^{3} $, the $ 2 $-form is $B$ with the field strength $ H=dB $ and the gauge transformation rule $ B\rightarrow B+dA $ for a globally defined $ A $. The surface operator 
\begin{equation}
E(B, \mathcal{M}_{2})=\frac{i}{2\pi} \int_{\mathcal{M}_{2}} B=\frac{i}{2\pi}  \int_{\mathcal{M}_{3}} H
\end{equation}
can be defined for a closed surface $\mathcal{M}_{2}  \subset \mathbb{R}^{3}  $ with $ \partial \mathcal{M}_{3}=\mathcal{M}_{2} $. Without the $2$-form $B$, we may construct a nontrivial $A$ bundle. Each $ S^{2} $ centered at the origin can be covered by two patches $ P_{N} $ and $ P_{S} $ with the $1$-forms $ A_{N} $ and $ A_{S} $ assigned on them. At $ P_{N} \cap P_{S}\sim S^{1} \times I$, $ A_{N} =A_{S}+u^{-1}du$ for a globally defined $ u $ in $ S^{1} \times I $. The $ A $ bundle is also characterized by the winding $w(u^{-1}du,S^{1})$ in (\ref{4.3}). Such configuration is just the Dirac (Wu-Yang) monopole \cite{1,2,3}. From $ A $, we get a $2$-form $b= F=dA $ with the field strength
\begin{equation}
h_{\mu\nu\lambda}(x)=(db)_{\mu\nu\lambda}(x)=-2\pi n i\epsilon_{\mu\nu\lambda}\delta^{(3)}(x)\;\;\;\;\;\;\; \mu,\nu,\lambda=1,2 ,3
\end{equation}
supported at the origin. Generic $A$ bundles in $ \mathbb{R}^{d} $ are characterized by 
\begin{equation}
 E(F , \mathcal{M}_{2}) \in  \mathbb{Z}\cong \Pi_{1}[U(1)] 
\end{equation}
for the arbitrary $\mathcal{M}_{2} \subset \mathbb{R}^{d}  $ with $  \partial \mathcal{M}_{2}=0  $. An $A$ bundle with the non-zero $ E(F , \mathcal{M}_{2})$ gives a $ 2 $-form $b=F$, whose field strength $h=db$ is supported on a $ (d-3) $-cycle $ c_{d-3} $ (the Poincar\'e dual of $ h$). Gauging the trivial $A$ imposes the equivalence relation $ b \sim b+dA $ for a globally defined $A $. The ordinary $B/H$ is the linear combinations of $b/h$. A $1$-form gauge transformation operator with the transformation parameter $ A $ is denoted by $ \mathcal{U}^{(2)} (c_{d-3})$, which is a defect operator localized at $c_{d-3}$ making $B \rightarrow B+b$.

The generalization to the $ n $-form situation is straightforward. A nontrivial $( n-1) $-form bundle $  \omega^{(n-1)} $ produces an $n $-form $d  \omega^{(n-1)}  $ with the field strength supported on a $ (d-n-1) $-cycle $ c_{d-n-1} $. In an $n$-form gauge theory, a gauge transformation with $\omega^{(n-1)} $ the transformation parameter is a defect operator localized at $c_{d-n-1}$.

In this paper, the discussion is extended to nonabelian theories, which could have the nontrivial $\Pi_{k}(G)$ when $k >1$. For a $ (d+1) $-dimensional $1$-form gauge theory in $ \mathbb{R}^{d,1}$ with the gauge group $ G=SU(N) $ or $U(N)$, the integral of a Chern-Simons $(2 j -1)$-form $ \text{CS}_{2j-1}(A) $ over a closed $ (2j-1) $-manifold $\mathcal{M}_{2j-1} $ gives $ O(A,\mathcal{M}_{2j-1} ) $, a generalization of (\ref{4.3q}). $O(A,\mathcal{M}_{2j-1} ) $ measures the $D(d-2j) $-brane charge enclosed by $ \mathcal{M}_{2j-1}  $ in $Dd$-branes \cite{5}. For a globally defined $u$, 
\begin{equation}
O(u^{-1}du,\mathcal{M}_{2j-1} ) =w(u,\mathcal{M}_{2j-1})   \in \mathbb{Z} \cong  \Pi_{2j-1}(G) 
\end{equation}
is the winding of $u$ around $\mathcal{M}_{2j-1}$. $ u $ with the non-zero $ w(u,\mathcal{M}_{2j-1})  $ produces a $1$-form $a=u^{-1}du$, whose $ j $th Chern character $ P_{j}(f) $ is supported on a $ (d-2j) $-cycle $ c_{d-2j} $ (the Poincar\'e dual of $ P_{j}(f)$). $ a $ is also the configuration of $D(d-2j) $-branes wrapping $ c_{d-2j}$ in $ Dd $-branes \cite{5}.

A gauge transformation operator $ U^{(2j-1)} (c_{d-2j} )$ with the transformation matrix $ u $ is a defect operator localized at $ c_{d-2j}  $ carrying the $D(d-2j) $-brane charge. $ U^{(1)} ( c_{d-2} )$ and $ U^{(3)} ( c_{d-4} )$ are just the codimension-2 't Hooft operators \cite{6,7,7a, th1} and the codimension-4 instanton operators \cite{8,9}. We calculate the commutation relation between the defect operator $ U^{(2j-1)} ( c_{d-2j} ) $ and the Chern-Simons operator $\exp \{ig  O(A,\mathcal{M}_{2k-1} ) \}$ for the generic $d,j$ and $k $. Aside from (\ref{519}), which gives actions of electric $(2j-1)$-form and magnetic $(d-2j)$-form local transformation operators on each other. When $j \neq k$, relations like (\ref{6.100}) and (\ref{j}) are also obtained with a Chern-Simons operator on $\mathcal{M}_{2k-1}\cap c_{d-2j} $ generated. In an even-dimensional $1$-form gauge theory, $   \exp \{igO (A,\mathbb{R}^{d})\} $ is the operator implementing a T-transformation (\ref{ka}) which changes the $ \theta $-angle $tr F^{\frac{d+1}{2}}$. We compute the action of $ \exp \{igO (A,\mathbb{R}^{d})\}  $ on $  U^{(2j-1)} ( c_{d-2j} ) $ for $j=1,2$, and get (\ref{6100}) and (\ref{61000}), which are Witten effects \cite{wi} for defect operators. The interpretation in an odd-dimensional gauge theory is that when the action contains a Chern-Simons term, the defect operator $  U^{(2j-1)} ( c_{d-2j} ) $ will get some electric $(d-2j)$-charge to couple with the Chern-Simons $(d-2j+1)$-form \footnote{There is a difference between the electric $(d-2j)$-charge and the magnetic $D(d-2j)$-charge. $ U^{(2j-1)} ( c_{d-2j} ) $ always carries the $D(d-2j)$-charge and could couple with an external R-R $(d-2j+1)$-form field $C^{(d-2j+1)}$ \cite{5, Li}.}.

The distinction between the defect operator $ U^{(2j-1)} ( \Sigma_{d-2j} )$, a local $ 0 $-form gauge transformation operator and an electric $(2j-1)$-form  global symmetry operator \cite{G} can be explicitly demonstrated when $G=U(1)$ and $j=1$. Multiplied by a suitable local gauge transformation, $ U^{(1)} ( \Sigma_{d-2} ) $ can always be written as 
\begin{equation}\label{sing}
 U^{(1)} ( \Sigma_{d-2} )= \exp \left\lbrace - 2\pi\int_{\Sigma_{d-1}} \;\ast \Pi \right\rbrace\;,
\end{equation}
where $ \Pi $ is the conjugate momentum of $A$ and $\partial \Sigma_{d-1}=\Sigma_{d-2}$. $U^{(1)} ( \Sigma_{d-2} )  $ is a Gukov-Witten operator \cite{gw1, gw2} in $\mathbb{R}^{d}$ making $ A\rightarrow A+a $, where $a$ is the Poincar\'e dual of $\Sigma_{d-1}  $. When $\partial \Sigma_{d-1}=\Sigma_{d-2} \neq 0 $, $ f=da\neq 0 $, (\ref{sing}) is a defect operator implementing a local $1$-form transformation and will become a symmetry operator in $2$-form theories. When $\Sigma_{d-1}$ is compact with $\partial \Sigma_{d-1}=0$, $ f=0 $, (\ref{sing}) is a local $0$-form gauge transformation. When $\Sigma_{d-1}$ is unbounded with $\partial \Sigma_{d-1}=0$, $f=0$, (\ref{sing}) is an electric $1$-form global symmetry transformation acting on Wilson lines. In section \ref{8}, when $ G=SU(N) $ or $U(N)$ and $j=1,2$, we will also study the corresponding electric $1$-form and $3$-form global symmetry operators which act nontrivially only on $\exp \{ig  O(A,\mathcal{M}_{1} ) \}$ and $\exp \{ig  O(A,\mathcal{M}_{3} ) \}$. The dual magnetic $(d-2j)$-form global symmetry arises from the conservation law $d  P_{j}(F)=0$, where $ P_{j}(F) $ is the $ j $th Chern character.

For a $1$-form $A$ with the field strength $F=dA+A^{2}$, the integral of $ P_{j}(F) $ over a closed $\mathcal{M}_{2j} $ gives an even-dimensional extended observable $ E(F,\mathcal{M}_{2j} ) $. $ E(F,\mathcal{M}_{2j} )=0  $ when $A$ is globally defined in $\mathbb{R}^{d}$. For the generic $A$, $ E(F,\mathcal{M}_{2j} ) \in  \mathbb{Z}$. When $ E(F,\mathcal{M}_{2j}) \neq 0$, a $2$-form $ b=F$ can be obtained with $Q_{2j+1}(F)=d P_{j}(F) $ supported on a $ (d-2j-1) $-cycle $c_{d-2j-1} $ (the Poincar\'e dual of $Q_{2j+1}(F)$). $A$ bundles in $  \mathbb{R}^{d} $ are classified by 
\begin{equation}
E(F,\mathcal{M}_{2j}) \in  \mathbb{Z}\cong \Pi_{2j-1}(G) \;.
\end{equation}
We will take $F$ in the same class as the gauge equivalent fields in a $2$-form theory. Then the globally defined $ A $ will produce an $ F $ that is gauge equivalent to $0$. Just like $ O(u^{-1}du,\mathcal{M}_{2j-1}) $ and $ P_{j}(f) $ are $0$-form limits of $O(A,\mathcal{M}_{2j-1}) $ and $ P_{j}(F) $, we may also expect $ E(F,\mathcal{M}_{2j}) $ and $ Q_{2j+1}(F) $ are $1$-form limits of some $ E(B,\mathcal{M}_{2j}) $ and $ Q_{2j+1}(B)   $ for the $2$-form $B$.

In a $2$-form theory, defect operators $ U^{(2j-1)} ( c_{d-2j} )  $ are all local gauge transformations. For a $ (d-2j)$-chain $z_{d-2j}$,  $ U^{(2j-1)} ( z_{d-2j} )  $ is nontrivial only when $ \partial  z_{d-2j}= c_{d-2j-1} \neq 0$. All $ U^{(2j-1)} ( z_{d-2j} )   $ with the same $  \partial  z_{d-2j} $ can be identified as $   \mathcal{U}^{(2j)} ( c_{d-2j-1} )  $, which are defect operators in $2$-form theories. $  \mathcal{U}^{(2j)} ( c_{d-2j-1} )   $ carries a $(d-2j-1)$-form charge measured by $  E(B,\mathcal{M}_{2j}) $. In $1$-form theories, topological sectors and vacuum structures are classified by pure gauges $ u^{-1}du   $ at $ \partial \mathbb{R}^{d} $ and $\mathbb{R}^{d}$. With $dA+A^{2}$ taken as the pure gauge of $ B $, we will also give a corresponding classification for the $2$-form theory.

The rest of the paper is organized as follows. In section \ref{5}, we classify nontrivial $0$-form bundles in nonabelian $1$-form theories and construct the related defect operators. In section \ref{8}, we discuss electric and magnetic higher-form symmetries in $1$-form gauge theories. In section \ref{3}, we compute the commutation relations between defect operators and Chern-Simons operators. In section \ref{n12}, we classify nontrivial $1$-form bundles in nonabelian $2$-form theories and construct defect operators. In section \ref{n123}, we study topological sectors in $2$-form theories. The conclusion and discussion are in section \ref{n1234}.

\section{Nontrivial $0$-form bundles}\label{5}

Consider a $( d+1 )$-dimensional $1$-form gauge theory in $ \mathbb{R}^{d,1} $ with the gauge group $ G=SU(N) $ or $U(N)$ and the Lie algebra $ g $. The $1$-form field is $ A $ with the field strength $ F=dA+A^{2} $. The $ 0 $-form $ \phi \in g $ is represented by a globally defined $ u=e^{\phi} \in G $. $ u $ and $ \phi $ are not in one-to-one correspondence. For example, when $ G=SU(2) $ and $ \tau^{a} $ are three Pauli matrices, $ e^{i \pi \tau^{a}n_{a}}=-1 $, $\forall \;\sum^{3}_{a=1}n^{2}_{a}=1 $; when $ G=U(1) $, $ e^{2m\pi i} =1$, $ \forall \; m \in  \mathbb{Z}  $. So a globally defined $ u $ may result in $ \phi $ with branch cuts. The gauge equivalence relation is $ A \sim u^{-1}(A+d)u $, where $u=e^{\phi}  $ for a globally defined $ \phi $.

The $ j $th Chern character of the $1$-form $A$ is \cite{Ge}
\begin{equation}
 P_{j}(F)= \frac{1}{j!}tr\left(  \frac{iF}{2\pi}\right)^{j} \;. 
\end{equation}
$ d P_{j}(F)=0 $. Locally, $P_{j}(F)= d  \text{CS}_{2j-1}(A)$, where $ \text{CS}_{2j-1}(A) $ is the Chern-Simons $(2j-1)$-form.
 \begin{eqnarray}
 && \text{CS}_{1}(A)=\frac{i}{2\pi} tr A \;,\\ && \text{CS}_{3}(A)=\frac{1}{2}(\frac{i}{2\pi})^{2} tr \left(  AdA+\frac{2}{3}A^{3}\right) \;,  \\  && \text{CS}_{5}(A)=\frac{1}{6}(\frac{i}{2\pi})^{3} tr \left[ A(dA)^{2}+\frac{3}{2}A^{3}dA+\frac{3}{5}A^{5}  \right] \;,  \\  \nonumber &&  \cdots
\end{eqnarray} 
The integral of $ \text{CS}_{2j-1}(A) $ over a closed $(2j-1)$-manifold $ \mathcal{M}_{2j-1}  $ gives a $(2j-1)$-dimensional extended observable 
\begin{equation}
O(A,\mathcal{M}_{2j-1} )=\int_{\mathcal{M}_{2j-1}  }\text{CS}_{2j-1}(A) \;.
\end{equation}
If $  \mathcal{M}_{2j-1}  $ is also exact with $ \partial \mathcal{M}_{2j} =\mathcal{M}_{2j-1}  $, 
\begin{equation}\label{234}
O(A,\mathcal{M}_{2j-1} )=\int_{\mathcal{M}_{2j}  }P_{j}(F)\;.
\end{equation}
In string theory, $  P_{j}(F) $ is the $D(d-2j)$-brane charge carried by $ Dd $-branes \cite{5}, so $ O(A,\mathcal{M}_{2j-1} ) $ is also a brane charge operator measuring the $D(d-2j)$ charge enclosed by $\mathcal{M}_{2j-1}$.

For the generic $ A $, $ O(A,\mathcal{M}_{2j-1} ) $ takes continuous values, but when $A= u^{-1} du$, 
\begin{eqnarray}
\nonumber  O(u^{-1}du,\mathcal{M}_{2j-1} )&=&\int_{\mathcal{M}_{2j-1}  }\text{CS}_{2j-1}(u^{-1}du)= \int_{\mathcal{M}_{2j-1}  }(-1)^{j-1}\frac{(j-1)!}{(2j-1)!}(\frac{i}{2 \pi})^{j}tr[(u^{-1}du)^{2j-1}] \\ &=&  w(u,\mathcal{M}_{2j-1} ) \in  \mathbb{Z}  \cong \Pi_{2j-1} (G)
\end{eqnarray}
is the winding number of $ G $ around $  \mathcal{M}_{2j-1}  $, characterizing the homotopy group $\Pi_{2j-1} (G)  $. (\ref{tab}) is a list of $ \Pi_{k} (G) $ for $ G=SU(N) $ or $U(N)$. $  \Pi_{k} [U(N)]=\Pi_{k} [SU(N)]\oplus \Pi_{k} [U(1)] $. 

\begin{equation}\label{tab}
\begin{tabular}{|l|c|c|c|c|c|c|c|}
     \hline
         $ k $   &  $1$    & $ 2 $ &  $3$    & $ 4 $  &  $5$    & $ 6 $  \\\hline
            $\Pi_{k}[U(1)]$              & $\mathbb{Z}$ & $0 $  & $0$ & $0 $& $0$ & $0 $  \\\hline
             $\Pi_{k}[SU(2)]$            & $0$     & $0$      & $\mathbb{Z}$     & $\mathbb{Z}_{2}$ & $\mathbb{Z}_{2}$     & $\mathbb{Z}_{12}$    \\\hline $\Pi_{k}[SU(3)]$            & $0$     & $0$      & $\mathbb{Z}$     & $0$ & $\mathbb{Z}$     & $\mathbb{Z}_{6}$       \\\hline$\Pi_{k}[SU(N)]\;\;N>3$            & $0$     & $0$      & $\mathbb{Z}$     & $0$ & $\mathbb{Z}$     & $0$      \\\hline
   \end{tabular}
\end{equation}
\

\noindent In this sense, Chern-Simons forms are characteristic classes of $u$. The globally defined $ u $ in $\mathbb{R}^{d}$ are classified by 
\begin{equation}
\left\lbrace  w(u,\mathcal{M}_{2j-1} ) \;|\; \forall \; \mathcal{M}_{2j-1} \subset  \mathbb{R}^{d},  \partial\mathcal{M}_{2j-1} =0, j=1,2,\ldots\right\rbrace   \;.
\end{equation}
For $ u $ and $u'$,
\begin{equation}
w (uu', \mathcal{M}_{n} )=w (u, \mathcal{M}_{n} )+w (u', \mathcal{M}_{n} )\;.
\end{equation}
If $ w (u, \mathcal{M}_{2j-1} )= w (u', \mathcal{M}_{2j-1} )  $ for the arbitrary closed $ \mathcal{M}_{2j-1} $, $ u $ can be continuously deformed into $ u' $. In particular, the topologically trivial $ u $ that could be continuously deformed to $ 1 $ should have $w (u, \mathcal{M}_{2j-1} )=0  $ everywhere.

For a globally defined $ u $, let $ a=u^{-1}du $ and $ f=da+a^{2} $, 
\begin{equation}
 w(u,\mathcal{M}_{2j-1} ) =\int_{\mathcal{M}_{2j}  }P_{j}(f)  \in  \mathbb{Z}  \;,
\end{equation}
so $ P_{j}(f) \in  \mathbb{Z}   $ on $ \mathcal{M}_{2j} $. When $ P_{j}(f)(x) \neq 0  $, $ x $ is a singularity. In $ \mathbb{R}^{d} $, the typical $ P_{j}(f)  $ is 
\begin{equation}\label{2.14aa}
[ P_{j}(f)]_{m_{1}\cdots m_{2j}}(x)=\frac{1}{(d-2j)!}\int_{c_{d-2j}}\epsilon_{l_{1}\cdots l_{d-2j}m_{1}\cdots m_{2j}}\delta^{(d)}(\vec{x}-\vec{y})\;dy^{l_{1}}\wedge \cdots\wedge dy^{l_{d-2j}}  \;,
\end{equation}
where $ c_{d-2j} $ is a $(d-2j)$-cycle (linear combinations of the closed $(d-2j)$-manifolds with integer coefficients). $\partial c_{d-2j}=0  $ since $ d P_{j}(f)=0 $. Therefore, a $ 0$-form with the non-zero $ w(u,\mathcal{M}_{2j-1} ) $ could produce a $1$-form $ a $ whose $ j $th Chern character $ P_{j}(f) $ is supported on a $(d-2j)$-cycle $ c_{d-2j} $ (the Poincar\'e dual of $ P_{j}(f) $). $  P_{j}(F) $ is also the $D(d-2j)$ charge in $ Dd $, so the configuration $a$ describes $D(d-2j) $-branes localized at $ c_{d-2j} $ in the $  \mathbb{R}^{d}  $ volume of the $ Dd $-branes.

We can solve for a $0$-form $ \phi $ from $ u=e^{\phi} $, but $ \phi $ must have branch cuts at some $(d-2j+1)$-chain $ z_{d-2j+1} $ with $\partial  z_{d-2j+1}=  c_{d-2j}  $. Depending on how $ \phi $ is selected, $z_{d-2j+1}  $ can be an arbitrary $(d-2j+1)$-chain with the fixed boundary $  c_{d-2j}  $. If the branch cut of $ \phi $ lives at $z_{d-2j+1}  $ with $ \partial  z_{d-2j+1}=0 $, then $ f=0 $ everywhere. If $ z_{d-2j+1}  $ is also exact with $z_{d-2j+1}=\partial   z_{d-2j+2}$, then $ z_{d-2j+1}   $ can shrink to $ 0 $ and $ u $ is just an ordinary local gauge transformation. If $  z_{d-2j+1}  $ is closed but not exact, the gauge transformation with the transformation matrix $ u $ will be a $ (2j-1) $-form global symmetry transformation that will be discussed in section \ref{8}.

In $\mathbb{R}^{d}$, for a given $ u $ with the nontrivial $w (u, \mathcal{M}_{2j-1} )  $, a related gauge transformation operator $ U^{(2j-1)} ( c_{d-2j} )$ can be constructed as \footnote{When there are matter fields, $ U^{(2j-1)} ( c_{d-2j} )$ also acts on matter. For example, for $ \Phi $ in the adjoint representation of $G$,
 \begin{eqnarray}\label{ru}
 && \label{ru1} U^{(2j-1)-1} ( c_{d-2j} )  \Phi  U^{(2j-1)} ( c_{d-2j} )=\Phi^{u}=u^{-1}\Phi u  \;, \\ \label{ru2} &&  U^{(2j-1)-1} ( c_{d-2j} )  D\Phi  U^{(2j-1)} ( c_{d-2j} )=D^{u}\Phi^{u}=u^{-1}D\Phi u \;,
\end{eqnarray} 
 where $D$ is the covariant derivative.   } 
\begin{equation}\label{2.14av}
U^{(2j-1)-1} ( c_{d-2j} )  A  U^{(2j-1)} ( c_{d-2j} )=A^{u}=u^{-1}(A+d)u\;.
\end{equation}
For $ F=dA+A^{2} $, 
\begin{equation}\label{2.14a}
U^{(2j-1)-1} ( c_{d-2j} )  F  U^{(2j-1)} ( c_{d-2j} )=d A^{u} +(A^{u})^{2}=u^{-1}Fu+f\;.
\end{equation}
Away from $c_{d-2j}  $, $ u $ is regular with $ f=0 $, so $ U^{(2j-1)} ( c_{d-2j} ) $ acts as an ordinary local gauge transformation. Under the action of $ U^{(2j-1)} ( c_{d-2j} ) $, $ \text{CS}_{2j-1}(A) $ and $O(A,\mathcal{M}_{2j-1} ) $ transform as 
\begin{eqnarray}
\nonumber &&  U^{(2j-1)-1} ( c_{d-2j} )   \text{CS}_{2j-1}(A)   U^{(2j-1)} ( c_{d-2j} )=\text{CS}_{2j-1}(A^{u})  \\\nonumber  &=&  \text{CS}_{2j-1}(A) +  \text{CS}_{2j-1}(u^{-1}du)+d\alpha_{2j-2}
\end{eqnarray}
for some $(2j-2)$-form $\alpha_{2j-2}$, and  
\begin{eqnarray}\label{619}
\nonumber &&   U^{(2j-1)-1} ( c_{d-2j} )  O(A,\mathcal{M}_{2j-1} ) U^{(2j-1)} ( c_{d-2j} )=O(A^{u},\mathcal{M}_{2j-1} )\\  &=&O(A,\mathcal{M}_{2j-1} )+ w(u,\mathcal{M}_{2j-1} )=O(A,\mathcal{M}_{2j-1} )+ L(c_{d-2j},\mathcal{M}_{2j-1} )\;,
\end{eqnarray}
where $ L(c_{d-2j},\mathcal{M}_{2j-1} )=w(u,\mathcal{M}_{2j-1} )$ is the linking number between $c_{d-2j}$ and $\mathcal{M}_{2j-1}$. $O(A^{u},\mathcal{M}_{2j-1} )=O(A,\mathcal{M}_{2j-1} ) $ if $ u=e^{\phi} $ for a globally defined $ \phi $. From (\ref{619}), we have 
\begin{equation}\label{519a}
 [O(A,\mathcal{M}_{2j-1} ) ,U^{(2j-1)} ( c_{d-2j} )]= L(c_{d-2j},\mathcal{M}_{2j-1} )U^{(2j-1)} ( c_{d-2j} )
\end{equation}
and
\begin{eqnarray}\label{519}
\nonumber \exp  \{ig O(A,\mathcal{M}_{2j-1} )\}U^{(2j-1)} ( c_{d-2j} )&=&  U^{(2j-1)} ( c_{d-2j} )\exp  \{ig O(A,\mathcal{M}_{2j-1} )\} \\  &&  \exp  \{ig  L(c_{d-2j},\mathcal{M}_{2j-1} ) \}\;,
\end{eqnarray}
where $ g $ is an arbitrary constant acting as the parameter of a $U(1)$ transformation. (\ref{519a}) shows that $U^{(2j-1)} ( c_{d-2j} )$ carries the $D(d-2j)$ charge. (\ref{519}) is an extension of the Wilson-'t Hooft commutation relation to the generic order-disorder operators. In $ \mathbb{R}^{d} $, $ U^{(2j-1)} ( c_{d-2j} ) $ is a codimension-$ 2j $ disorder operator creating a defect ($D(d-2j)$) localized at $  c_{d-2j}  $ carrying the $ j$th chern number ($D(d-2j)$ charge). For example, $ U^{(1)} ( c_{d-2} ) $ creates a codimension-$ 2$ defect in $c_{d-2} $ carrying the vortex number \cite{6,7,7a, th1}; $ U^{(3)} ( c_{d-4} ) $ creates a codimension-$ 4$ defect in $c_{d-4} $ carrying the instanton number \cite{8,9}.

More comments should be added to the situation when $j=1$, where (\ref{519}) becomes
\begin{equation}\label{ggh}
\exp  \{ig O(A,\mathcal{M}_{1} )\}U^{(1)} ( c_{d-2} )=U^{(1)} ( c_{d-2} )\exp  \{ig O(A,\mathcal{M}_{1} )\}\exp  \{ig   L(c_{d-2},\mathcal{M}_{1} ) \}\;.
\end{equation}
$ U^{(1)} ( c_{d-2} ) $ is the 't Hooft operator in $ \mathbb{R}^{d} $, but unless $G=U(1) $,
\begin{equation}
\exp  \{ig O(A,\mathcal{M}_{1} )\}= \exp  \{-\frac{g}{2\pi} \int_{\mathcal{M}_{1}  } tr A \} 
\end{equation}
is not a Wilson loop, which should be  
\begin{equation}\label{weer}
W_{R}(\mathcal{M}_{1})=tr P \exp  \{-\frac{1}{2\pi} \int_{\mathcal{M}_{1}  }  A_{R} \} 
\end{equation}
for $  A_{R}$ in the R-representation of $U(N)$. Since $u$ is globally defined, $ U^{(1)} ( c_{d-2} ) $ commutes with $W_{R}(\mathcal{M}_{1})$:
\begin{equation}
W_{R}(\mathcal{M}_{1})U^{(1)} ( c_{d-2} )=U^{(1)} ( c_{d-2} )W_{R}(\mathcal{M}_{1})
\end{equation}
just like (\ref{ggh}) when $g=2\pi $. Let 
\begin{equation}\label{322}
O(H_{R},A,\mathcal{M}_{1})=\frac{i}{2\pi} \int_{\mathcal{M}_{1}  } tr (H_{R}A) \;,
\end{equation}
where $ A $ is in the fundamental representation and $ H_{R}=diag(m_{1} ,\cdots,m_{N})$ with $m_{i}  \in \mathbb{Z} $, $ m_{1} \geq m_{2} \geq \cdots \geq m_{N} \geq 0$ is in one-to-one correspondence with the representation $R$. (\ref{weer}) can be written as \cite{Wl, Wl1, Wl2} 
\begin{equation}\label{wl1}
W_{R}(\mathcal{M}_{1})=\frac{1}{\mathcal{N}}\int DU\; U\exp \{2\pi i O(H_{R},A,\mathcal{M}_{1})\}U^{-1}\;,
\end{equation}
where $U$ is a local gauge transformation operator and the integration covers all of the gauge transformations on $\mathcal{M}_{1}$ with $\mathcal{N}  $ a normalization constant. In (\ref{wl1}), $ g $ is taken to be $2\pi$ with the possible integer multiplications absorbed in $H_{R}$. If $ g \neq 2\pi m  $ for $ m \in \mathbb{Z} $, from (\ref{ggh}), the integration over $U$ will make $ W_{R}(\mathcal{M}_{1})=0 $. The vacuum expectations of $ W_{R}(\mathcal{M}_{1}) $ and $\exp \{2\pi i O(H_{R},A,\mathcal{M}_{1})\}$ are the same:
\begin{equation}
\langle \Omega \vert W_{R}(\mathcal{M}_{1})\vert \Omega \rangle \sim \langle \Omega \vert \exp \{2\pi i O(H_{R},A,\mathcal{M}_{1})\}\vert \Omega \rangle\;,
\end{equation}
so $\exp \{2\pi i O(H_{R},A,\mathcal{M}_{1})\}$ can also act as an order operator.

In the following, we will give a classification of the globally defined $ u$ with the non-zero $  w( u,\mathcal{M}_{2j-1} )   $ for $ j=1,2,3 $ in each $ (d+1) $-dimensional nonabelian $1$-form theory and construct the corresponding defect operator $  U^{(2j-1)} ( c_{d-2j} )$ explicitly.

\subsection{$ w( u,\mathcal{M}_{1} ) \neq 0$ }

\vspace{0.2em}

\paragraph{3d gauge theory}~{}

\vspace{1em}

In the $2$-dimensional space $ \mathbb{R}^{2} $, suppose $ w( u,\mathcal{M}_{1} ) \neq 0$ for a loop $ \mathcal{M}_{1} $. $ \mathcal{M}_{1} $ can continuously deform into a point $ P \in \mathbb{R}^{2}$. In this process, $  \mathcal{M}_{1}  $ may cross some point-like singularities with $ w( u,\mathcal{M}_{1} ) $ changing by integers. When shrinking to $ P $, if $  w( u,\mathcal{M}_{1} )  $ does not reduce to $ 0 $, then $ P $ will also be a singularity. So in $\mathbb{R}^{2} $, $ u $ has singularities at points. $P_{1}(f)=\frac{i}{2\pi} tr f $ is the linear combination of
\begin{equation}
\frac{i}{2\pi} tr f_{mn}(x)=\epsilon_{mn}\delta^{(2)}(\vec{x}-\vec{y})  
\end{equation}
with integer coefficients.

The basic $ u $ has a singularity at a point $ \Sigma_{0} $. Their products give the generic $ u $ with the singularities supported at the 0-cycles $ c_{0} $. The corresponding transformation operator is denoted by $ U^{(1)} (c_{0} )$.
\begin{equation}\label{3.4}
\exp  \{ig O(A,\mathcal{M}_{1} )\} U^{(1)} (c_{0} )= U^{(1)} (c_{0} )\exp  \{ig O(A,\mathcal{M}_{1} )\}\exp  \{ig L( c_{0}, \mathcal{M}_{1} ) \}\;,
\end{equation}
where $L( c_{0}, \mathcal{M}_{1} ) =w( u, \mathcal{M}_{1} )  $ is the linking number between $ c_{0} $ and $ \mathcal{M}_{1} $.

$  U^{(1)} (c_{0} ) $ are just monopole operators ('t Hooft operators) \cite{6, 7}. If the coordinate of $ \mathbb{R}^{2} $ is $ (x^{1},x^{2}) $ and $G=U(N)$, the transformation matrix of $ U^{(1)} (a^{1},a^{2} ) $ can be selected as 
\begin{equation}
u(x^{1},x^{2})=\frac{(x^{1}-a^{1})-i(x^{2}-a^{2})}{ [ \sum^{2}_{i=1}(x^{i}-a^{i})^{2} ]^{1/2}    } \oplus 1_{N-1}\;.
\end{equation}
$ U^{(1)} (a^{1},a^{2} ) $ is a monopole operator at $ (a^{1},a^{2})$ with the charge $ 1 $. From $ u=e^{\phi} $, $ \phi $ can be solved as $  \phi =-iH \theta$ with
\begin{equation}
 H=diag(1,\underbrace{0,\cdots,0}_{N-1}) =1 \oplus 0_{N-1} \;. 
\end{equation}
Depending on how $ \theta $ is selected, there is a branch cut of $ \phi $ at a curve $ \Sigma_{1} $ with $ \partial  \Sigma_{1} =\Sigma_{0}= (a^{1},a^{2} )$. From $ u $, $ f $ is calculated as
\begin{equation}\label{uh2}
 f_{mn}(x)=-2\pi i H \epsilon_{mn}\delta^{(2)}(\vec{x}-\vec{y})  \;.
\end{equation}

\vspace{1em}

\paragraph{4d gauge theory}\label{4d}~{}

\vspace{1em}

In $ \mathbb{R}^{3} $, suppose $ w( u,\mathcal{M}_{1} )  \neq 0$ for a loop $ \mathcal{M}_{1} $ and $ \mathcal{M}_{2} $ is a surface with $\partial \mathcal{M}_{2} =\mathcal{M}_{1}  $. $ u $ must have point-like singularities in $ \mathcal{M}_{2}  $. When $ \mathcal{M}_{2}  $ deforms in $ \mathbb{R}^{3} $ with the boundary fixed, the singularities move along with it continuously. The deformed $  \mathcal{M}_{2}  $ could cover $ \mathbb{R}^{3} $, with all singularities together composing a 1-cycle. $P_{1}(f)=\frac{i}{2\pi} tr f $ is the linear combination of
\begin{equation}
\frac{i}{2\pi}tr f_{mn}(x)=\int_{\Sigma_{1}}\epsilon_{lmn}\delta^{(3)}(\vec{x}-\vec{y})dy^{l}
\end{equation}
with integer coefficients. $ \partial \Sigma_{1}=0 $.

The basic $ u $ is singular at a closed 1-manifold $ \Sigma_{1} $. Their products produce the generic $ u $ with singularities supported at 1-cycles $ c_{1} $. The corresponding transformation operator is denoted by $ U^{(1)} (c_{1} )$. 
\begin{equation}\label{3.5}
\exp  \{ig O(A,\mathcal{M}_{1} )\}U^{(1)} (c_{1} )=U^{(1)} (c_{1} )\exp  \{ig O(A,\mathcal{M}_{1} )\}\exp  \{ig L( c_{1}, \mathcal{M}_{1} ) \}\;.
\end{equation}

$  U^{(1)} (c_{1} ) $ are 't Hooft loop/line operators \cite{6, 7a, th1}. If the coordinate of $ \mathbb{R}^{3} $ is $ (x^{1},x^{2}, x^{3}) $ and $G=U(N)$, the transformation matrix of $U^{(1)} (a^{1},a^{2} )  $ can be selected as 
\begin{equation}
u(x^{1},x^{2},x^{3})=\frac{(x^{1}-a^{1})-i(x^{2}-a^{2})}{ [ \sum^{2}_{i=1}(x^{i}-a^{i})^{2} ]^{1/2}  } \oplus 1_{N-1}\;.
\end{equation}
$ U^{(1)} (a^{1},a^{2} )  $ is a 't Hooft line with charge $1$ localized at $ \Sigma_{1}=\{x^{1}=a^{1},x^{2}=a^{2}, x^{3}\in \mathbb{R}\} $. $ \phi $ solved from $ u=e^{\phi} $ must have a branch cut at some $ \Sigma_{2} $ with $\partial  \Sigma_{2}=\Sigma_{1} $.

\begin{equation}\label{uh22}
 f_{mn}(x)=-2\pi i H\int_{\Sigma_{1}}\epsilon_{lmn}\delta^{(3)}(\vec{x}-\vec{y})dy^{l}\;.
\end{equation}

\vspace{1em}

\paragraph{5d gauge theory}~{}

\vspace{1em}

In $ \mathbb{R}^{4} $, suppose $ w( u,\mathcal{M}_{1} )  \neq 0$ for a loop $ \mathcal{M}_{1} $. By the same reasoning, $ u $ has singularities at the 2-cycle of $ \mathbb{R}^{4} $. $P_{1}(f)=\frac{i}{2\pi} tr f $ is the linear combination of
\begin{equation}
\frac{i}{2\pi}tr f_{mn}(x)=\frac{1}{2!}\int_{\Sigma_{2}}\epsilon_{l_{1} l_{2}mn}\delta^{(4)}(\vec{x}-\vec{y})dy^{l_{1}}\wedge  dy^{l_{2}}
\end{equation}
with integer coefficients. $ \partial \Sigma_{2}=0 $.

The corresponding transformation operator with singularities supported at the 2-cycle $c_{2}  $ is denoted by $ U^{(1)} (c_{2} )$.
\begin{equation}\label{3.6}
\exp  \{ig O(A,\mathcal{M}_{1} )\}U^{(1)} (c_{2} )=U^{(1)} (c_{2} )\exp  \{ig O(A,\mathcal{M}_{1} )\}\exp  \{ig L( c_{2}, \mathcal{M}_{1} ) \}\;.
\end{equation}

$  U^{(1)} (c_{2} ) $ are 't Hooft surface operators. If the coordinate of $ \mathbb{R}^{4} $ is $ (x^{1},x^{2}, x^{3},x^{4}) $ and $G=U(N)$, the transformation matrix of $U^{(1)} (a^{1},a^{2} )  $ can be selected as 
\begin{equation}
u(x^{1},x^{2},x^{3},x^{4})=\frac{(x^{1}-a^{1})-i(x^{2}-a^{2})}{ [ \sum^{2}_{i=1}(x^{i}-a^{i})^{2} ]^{1/2}  } \oplus 1_{N-1}\;.
\end{equation}
$ U^{(1)} (a^{1},a^{2} )  $ is a 't Hooft plane with charge $1$ localized at $ \Sigma_{2}=\{x^{1}=a^{1},x^{2}=a^{2}, x^{3}, x^{4}\in \mathbb{R}\} $. $ \phi $ solved from $ u=e^{\phi} $ must have a branch cut at $ \Sigma_{3} $ with $\partial  \Sigma_{3}=\Sigma_{2} $. 
\begin{equation}
 f_{mn}(x)=-i \pi H\int_{\Sigma_{2}}\epsilon_{l_{1} l_{2}mn}\delta^{(4)}(\vec{x}-\vec{y})dy^{l_{1}}\wedge  dy^{l_{2}}\;.
\end{equation}

\subsection{$ w( u,\mathcal{M}_{3} ) \neq 0$ }

\vspace{0.2em}

\paragraph{5d gauge theory}~{}\label{5d}

\vspace{1em}

In $ \mathbb{R}^{4} $, suppose $ w( u,\mathcal{M}_{3} ) \neq 0$ for a compact $3$-manifold $ \mathcal{M}_{3} $. $ u $ has point-like singularities in $ \mathbb{R}^{4} $. $P_{2}(f)=\frac{1}{2}(\frac{i}{2\pi})^{2} tr f^{2} $ is the linear combination of
\begin{equation}
\frac{1}{2}(\frac{i}{2\pi})^{2} tr [ f\wedge f]_{mnpq}(x)=\epsilon_{mnpq}\delta^{(4)}(\vec{x}-\vec{y})
\end{equation}
with integer coefficients. When $P_{2}(f)  $ is supported at $\vec{y}$, $ u^{-1}du $ is the configuration of a small instanton at $ \vec{y} $ with the vanishing scale size \cite{49}.

The corresponding transformation operator with singularities supported at the 0-cycle $c_{0}  $ is denoted by $ U^{(3)} (c_{0} )$.
\begin{equation}\label{3.7}
\exp  \{ig O(A,\mathcal{M}_{3} )\} U^{(3)} (c_{0} )= U^{(3)} (c_{0} )\exp  \{ig O(A,\mathcal{M}_{3} )\}\exp  \{ig L( c_{0} , \mathcal{M}_{3} ) \}\;.
\end{equation}

$ U^{(3)} (c_{0} )$ are instanton operators \cite{8, Ber, 9}. If the coordinate of $ \mathbb{R}^{4} $ is $ (x^{1},x^{2}, x^{3},x^{4}) $ and $G=U(N) $ or $ SU(N)$ for $ N\geq 2 $, the transformation matrix of $U^{(3)} (a^{1},a^{2},a^{3},a^{4} )  $ can be selected as 
\begin{equation}\label{531}
u(x^{1},x^{2},x^{3},x^{4})=\frac{ \bar{\sigma}_{n}(x^{n}-a^{n})}{[ \sum^{4}_{i=1}(x^{i}-a^{i})^{2} ]^{1/2} } \oplus 1_{N-2} \;, \;\;\;\;\;\;n=1,2,3,4\;.
\end{equation}
$ \bar{\sigma}_{n}= (-i\vec{\tau},1_{2})$, $  \sigma_{n}=(i\vec{\tau},1_{2})$, where $ \tau_{a} $, $ a=1,2,3 $ are three Pauli matrices. $ U^{(3)} (a^{1},a^{2},a^{3},a^{4} )  $ is an instanton operator with charge $1$ localized at $ (a^{1},a^{2},a^{3},a^{4} ) $. Let $ (r,\theta,\phi,\varphi) $ with $  r \in [0,+\infty) $, $ \theta,\phi \in [0,\pi] $, $ \varphi \in [0,2\pi) $ be the spherical coordinate of $ \mathbb{R}^{4} $ centered at $ (a^{1},a^{2},a^{3},a^{4} ) $, then 
\begin{equation}
u=e^{-i \theta \tau^{a}n_{a}} \oplus 1_{N-2} =\exp \{-iH^{a}n_{a}\theta\}=e^{\phi}\;,\;\;\;\;\;\;\;\;\;\;a=1,2,3,
\end{equation}
where $ H^{a}=\tau^{a} \oplus 0_{N-2}$, $   (  n_{1},  n_{2},  n_{3})=  ( \cos \phi,  \sin \phi \cos  \varphi,  \sin \phi \sin  \varphi)   $. There is a branch cut of $ \phi $ at $ \theta=\pi $, where $ e^{-i \pi\tau^{a}n_{a}} =-1_{2} $. Depending on how $ \phi $ is solved, the branch cut can be an arbitrary curve $ \Sigma_{1} $ with $ \partial  \Sigma_{1} = \Sigma_{0} =(a^{1},a^{2},a^{3},a^{4} ) $. From $a=u^{-1}du$, $ f=da+a^{2} $ is calculated as $ f=f^{a} H^{a}$ with
\begin{equation}\label{fa2}
f^{a}_{mn}(x)=\frac{2\pi}{i}\eta^{a}_{mn}\delta^{(2)}(\vec{x}-\vec{y})\;,
\end{equation}
where $-i\sqrt{6}\eta^{a}_{mn}\tau^{a}= \sigma_{mn} =\frac{1}{2}(\sigma_{m}\bar{\sigma}_{n}-\sigma_{n}\bar{\sigma}_{m}) =\frac{1}{2} \epsilon_{mnkl} \sigma^{kl}$.
\begin{equation}
 [ f^{a}\wedge f^{b}]_{mnpq}(x)=\frac{\delta^{ab}}{3}(\frac{2\pi}{i})^{2}\epsilon_{mnpq}\delta^{(4)}(\vec{x}-\vec{y})\;.
\end{equation}

\vspace{1em}

\paragraph{6d gauge theory}~{}

\vspace{1em}

In $ \mathbb{R}^{5} $, suppose $ w( u,\mathcal{M}_{3} ) \neq 0$ for a compact $3$-manifold $ \mathcal{M}_{3} $. $ u $ has 1-cycle singularities in $ \mathbb{R}^{5} $. $P_{2}(f)=\frac{1}{2}(\frac{i}{2\pi})^{2} tr f^{2} $ is the linear combination of
\begin{equation}
\frac{1}{2}(\frac{i}{2\pi})^{2} tr [ f\wedge f]_{mnpq}(x)=\int_{\Sigma_{1}}\epsilon_{lmnpq}\delta^{(5)}(\vec{x}-\vec{y})dy^{l}
\end{equation}
with integer coefficients. $ \partial \Sigma_{1}=0 $.

The related transformation operator with singularities supported at the 1-cycle $c_{1}  $ is denoted by $ U^{(3)} (c_{1} )$.
\begin{equation}\label{3.8}
\exp  \{ig O(A,\mathcal{M}_{3} )\}U^{(3)} (c_{1} )=U^{(3)} (c_{1} )\exp  \{ig O(A,\mathcal{M}_{3} )\}\exp  \{ig L( c_{1}, \mathcal{M}_{3} ) \}\;.
\end{equation}

$ U^{(3)} (c_{1} )$ are instanton loop/line operators. If the coordinate of $ \mathbb{R}^{5} $ is $ (x^{1},x^{2}, x^{3},x^{4},x^{5}) $ and $G=U(N) $ or $ SU(N)$ for $ N\geq 2 $, the transformation matrix of $U^{(3)} (a^{1},a^{2},a^{3},a^{4} )  $ can be selected as 
\begin{equation}
u(x^{1},x^{2},x^{3},x^{4},x^{5})=\frac{  \bar{\sigma}_{n}(x^{n}-a^{n})}{[ \sum^{4}_{i=1}(x^{i}-a^{i})^{2} ]^{1/2}} \oplus 1_{N-2} \;, \;\;\;\;\;\;n=1,2,3,4\;.
\end{equation}
$ U^{(3)} (a^{1},a^{2},a^{3},a^{4} )  $ is an instanton line with charge $1$ localized at $ \Sigma_{1}=\{x^{1}=a^{1},x^{2}=a^{2}, x^{3}=a^{3}, x^{4}=a^{4}, x^{5}\in \mathbb{R}\} $. $ \phi $ solved from $ u=e^{\phi} $ must have a branch cut at $ \Sigma_{2} $ with $\partial  \Sigma_{2}=\Sigma_{1} $. 
\begin{equation}
 [ f^{a}\wedge f^{b}]_{mnpq}(x)=\frac{\delta^{ab}}{3}(\frac{2\pi}{i})^{2}\int_{\Sigma_{1}}\epsilon_{lmnpq}\delta^{(5)}(\vec{x}-\vec{y})dy^{l}\;.
\end{equation}

\subsection{$ w( u,\mathcal{M}_{5} ) \neq 0$ }

\vspace{0.2em}

\paragraph{7d gauge theory}~{}

\vspace{1em}

In $ \mathbb{R}^{6} $, suppose $ w( u,\mathcal{M}_{5} ) \neq 0$ for a compact $5$-manifold $ \mathcal{M}_{5} $. $ u $ has the 0-cycle singularities in $ \mathbb{R}^{6} $. $P_{3}(f)=\frac{1}{3!}(\frac{i}{2\pi})^{3} tr f^{3} $ is the linear combination of
\begin{equation}
\frac{1}{3!}(\frac{i}{2\pi})^{3}  tr [f \wedge f\wedge f]_{mnpqrs}(x)=\epsilon_{mnpqrs}\delta^{(6)}(\vec{x}-\vec{y})
\end{equation}
with integer coefficients.

The related transformation operator with singularities supported at the 0-cycle $c_{0}  $ is denoted by $ U^{(5)} (c_{0} )$.
\begin{equation}\label{3.10}
\exp  \{ig O(A,\mathcal{M}_{5} )\} U^{(5)} (c_{0} )= U^{(5)} (c_{0} )\exp  \{ig O(A,\mathcal{M}_{5} )\}\exp  \{ig L( c_{0}, \mathcal{M}_{5} ) \}\;.
\end{equation}
From (\ref{tab}), to have $ \Pi_{5} (G)\cong \mathbb{Z}  $, $G$ can be $U(N) $ or $ SU(N)$ with $ N \geq 3$.

\section{Electric and magnetic higher-form global symmetries in $1$-form gauge theories }\label{8}

The concept of global symmetries was generalized a lot in the past ten years, from the higher-form symmetry acting on extended objects to the non-invertible symmetry following the non-group-like fusion rules \cite{G, s1, s2, s3, s4, s5, s6, s7, s8, s9, s10, s11, ss11, Ruc, s12, s13, s14, s15, s16, s17, s18, s19, s20, s21, s22, s23, s23a, ss23, s24, ss24}. In this section, we will discuss invertible higher-form electric and magnetic global symmetries in $1$-form theories, where the higher-form is the Chern-Simons form. These symmetries are dynamics-independent.

For a globally defined $ u $ in $\mathbb{R}^{d} $, we have considered the situation when $  w(u,\mathcal{M}_{2j-1}) \neq 0  $ for a compact and thus exact $  \mathcal{M}_{2j-1}$. $ u $ must have the non-vanishing $ f $ at some closed $(d-2j)$-manifold $ \Sigma_{d-2j}  $, and is a local gauge transformation only in $\mathbb{R}^{d}\setminus\Sigma_{d-2j}$. On the other hand, $ u $ with $  w(u,\mathcal{M}_{2j-1}) \neq 0  $ for a closed but unbounded (closed but not exact) $  \mathcal{M}_{2j-1}$ will have $ f=0 $ in $\mathbb{R}^{d}$. Such $ \mathcal{M}_{2j-1}  $ ends on $ S_{\infty}^{2j-2} \times \mathbb{R}^{d-2j+1} $ at the infinity, and the operator generating the transformation $ u $ can be denoted as $U^{(2j-1)}  [S_{\infty}^{2j-2} \times \mathbb{R}^{d-2j+1}]  $. Locally, $U^{(2j-1)}  [S_{\infty}^{2j-2} \times \mathbb{R}^{d-2j+1}]   $ is an ordinary gauge transformation everywhere, commuting with gauge invariant operators like the Hamiltonian, as well as $ \exp  \{ig O(A,\mathcal{M}_{2k-1} )\}  $ for the exact $\mathcal{M}_{2k-1}   $. Its nontrivial action on $\exp  \{ig O(A,\mathcal{M}_{2j-1} )\}  $ ending at $ S_{\infty}^{2j-2} \times \mathbb{R}^{d-2j+1}  $ makes it into an electric $ (2j-1) $-form global symmetry transformation. 
\begin{eqnarray}
\nonumber && U^{(2j-1)-1}  [S_{\infty}^{2j-2} \times \mathbb{R}^{d-2j+1}] \exp  \{ig O(A,\mathcal{M}_{2j-1} )\}  U^{(2j-1)}  [S_{\infty}^{2j-2} \times \mathbb{R}^{d-2j+1}]  \\\nonumber  &=&\exp  \{ig O(A,\mathcal{M}_{2j-1} )\}  \exp \{ig w(u,\mathcal{M}_{2j-1})\}\;,
\end{eqnarray}
where $ w(u,\mathcal{M}_{2j-1})\neq 0 $ if $ \mathcal{M}_{2j-1} $ ends on $S_{\infty}^{2j-2} \times \mathbb{R}^{d-2j+1}$. In subsections \ref{1q2} and \ref{1q22}, we will study the situation when $j=1, 2$ and construct $ U^{(2j-1)}  [S_{\infty}^{2j-2} \times \mathbb{R}^{d-2j+1}]$ explicitly.

As for the magnetic global symmetry, in $(d+1)$-dimensional nonabelian $1$-form gauge theories, $ dP_{j} (F)=0$ indicates 
\begin{equation}\label{3.1q}
E(F,\mathcal{M}_{2j} )=\int_{\mathcal{M}_{2j}  }P_{j}(F)=O(A,\partial  \mathcal{M}_{2j}  )
\end{equation}
for an unbounded $  \mathcal{M}_{2j}$ with $\partial  \mathcal{M}_{2j} $ at the infinity is a conserved $(d-2j)$-form charge, the $D(d-2j)$-brane charge. On a supergravity background, $ Dd $ branes couple with the R-R $(d-2j+1)$-form field $C^{(d-2j+1)} $ via the $D(d-2j)$-brane charge density \cite{5, Li}:
\begin{equation}\label{rr}
\delta S \sim \int_{\mathbb{R}^{d,1} } C^{(d-2j+1)} \wedge P_{j}(F)\;.
\end{equation}
The global magnetic $(d-2j)$-form symmetry transformation operator is $\exp  \{ig E(F,\mathcal{M}_{2j} )\}   $ with unbounded $  \mathcal{M}_{2j}$ acting on the charged objects $ U^{(2j-1)} ( \Sigma_{d-2j} ) $:
\begin{equation}\label{3.2q}
\exp  \{ig E(F,\mathcal{M}_{2j} )\}   U^{(2j-1)} (\Sigma_{d-2j} )\exp  \{-ig E(F,\mathcal{M}_{2j} )\}=   U^{(2j-1)} (\Sigma_{d-2j}  )\exp  \{ig I(\Sigma_{d-2j}, \mathcal{M}_{2j} ) \}\;, 
\end{equation}
where $ I(\Sigma_{d-2j}, \mathcal{M}_{2j} ) $ is the intersection number between $ \Sigma_{d-2j}$ and $ \mathcal{M}_{2j} $.

Under a local magnetic $(d-2j)$-form transformation implemented by 
\begin{equation}
M[\alpha^{(d-2j)}]=\exp  \{i\int_{\mathbb{R}^{d} }  \alpha^{(d-2j)} \wedge P_{j}(F)\}\;,
\end{equation}
from (\ref{2.14a}) and (\ref{2.14aa}), 
\begin{eqnarray}
\nonumber  && U^{(2j-1)-1} ( \Sigma_{d-2j} ) M[\alpha^{(d-2j)}]U^{(2j-1)} ( \Sigma_{d-2j} )  \\\nonumber  &=& \exp  \{i\int_{\mathbb{R}^{d} }  \alpha^{(d-2j)} \wedge P_{j}(F+ufu^{-1})\}  \\\nonumber  &=&\exp  \{i\int_{\mathbb{R}^{d} }  \alpha^{(d-2j)} \wedge P_{j}(F)\} \exp  \{i\int_{\mathbb{R}^{d} }  \alpha^{(d-2j)} \wedge P_{j}(f)\} \\  &=&M[\alpha^{(d-2j)}] \exp  \{i\int_{\Sigma_{d-2j}}  \alpha^{(d-2j)} \}\;,
\end{eqnarray}
so
\begin{equation}
M[\alpha^{(d-2j)}]U^{(2j-1)} ( \Sigma_{d-2j} ) M^{-1}[\alpha^{(d-2j)}]= \exp  \{i\int_{\Sigma_{d-2j}}  \alpha^{(d-2j)} \}U^{(2j-1)} ( \Sigma_{d-2j} ) \;.
\end{equation}
$U^{(2j-1)} ( \Sigma_{d-2j} )   $ is a magnetic object localized at $\Sigma_{d-2j}$. The Chern-Simons operator
\begin{equation}
\exp\{ig O(A,\mathcal{M}_{2j-1} )\}=\exp\{ig \int_{\mathcal{M}_{2j}  }P_{j}(F) \}
\end{equation}
for a compact $\mathcal{M}_{2j-1}$ is also a local magnetic $(d-2j)$-form transformation with $  \alpha^{(d-2j)} $ taken as 
\begin{equation}
\alpha^{(d-2j)}_{m_{1}\cdots m_{d-2j}}(x)=\frac{g}{(2j)!}\int_{\mathcal{M}_{2j}}\epsilon_{l_{1}\cdots l_{2j}m_{1}\cdots m_{d-2j}}\delta^{(d)}(\vec{x}-\vec{y})\;dy^{l_{1}}\wedge \cdots\wedge dy^{l_{2j}}  \;. 
\end{equation}
On the other hand, the defect operator $ U^{(2j-1)} ( \Sigma_{d-2j} ) $ can be viewed as a local electric $(2j-1)$-form transformation for $  \text{CS}_{2j-1}(A) $ with 
\begin{equation}
O(A,\mathcal{M}_{2j-1} )\rightarrow  O(A,\mathcal{M}_{2j-1} )+ L(c_{d-2j},\mathcal{M}_{2j-1} )\;.
\end{equation}
Unless the theory is topological, these local transformations do not commute with the Hamiltonian and are not symmetries.

In modern approaches, global symmetries can be identified as the topological operators in spacetime \cite{G}. For the magnetic global symmetry, let 
\begin{equation}
\tilde{E}(F,\mathcal{M}_{2j} )=\int_{\mathcal{M}_{2j}  }P_{j}(F)\;, 
\end{equation}
where $\mathcal{M}_{2j}    $ is a closed manifold in $\mathbb{R}^{d,1}$, then $ \exp  \{ig \tilde{E}(F,\mathcal{M}_{2j} )\} $ is a topological operator invariant under the deformation of $ \mathcal{M}_{2j}    $ since $d P_{j}(F) =0$ in $\mathbb{R}^{d,1}$.

\subsection{Electric $1$-form global symmetry operators}\label{1q2}

In a $(d+1)$-dimensional $1$-form gauge theory with $ G=U(N) $, the operator in $\mathbb{R}^{d}$ generating a gauge transformation $ u $ with $  w(u,\mathcal{M}_{1}) \neq 0  $ for a closed but unbounded $ \mathcal{M}_{1} $ is an electric  $1$-form global symmetry transformation.

When $ d=1 $, $ \mathcal{M}_{1}=\mathbb{R}^{1} $, consider
\begin{equation}\label{9111}
u(x^{1})=\exp \left\lbrace - 2\pi i k(x^{1}) \right\rbrace \oplus 1_{N-1}\;,
\end{equation}
where $k$ is an arbitrary function satisfying $k(-\infty)=0 $, $ k(+\infty)=1  $. $ \forall \;x^{1} $, $u  $ is a local gauge transformation. Globally, $  O(u^{-1}du,\mathbb{R}^{1} )= w(u,\mathbb{R}^{1} )=1$. All $ u $'s in (\ref{9111}) are equivalent differing by the multiplication of a local gauge transformation, among which, a special class of $ u $ can be selected with
\begin{equation}
 k(x^{1}) = \begin{cases}
	    0\;, & \text{when $x^{1}\leq c$} \\
	    1\;, &  \text{when $x^{1}\geq c$}	
		   \end{cases}\;.
\end{equation}
$ \phi =- 2\pi i k \oplus 0_{N-1}$ has a branch point at $\Delta_{0}=\{x^{1}=c\}$. $ a=u^{-1} du$ vanishes everywhere except at $  \Delta_{0}$. The related transformation operator $U^{(1)}  [S_{\infty}^{0} ] \sim  U^{(1)}[\Delta_{0}]$ acts nontrivially only on the Wilson line $ \exp  \{ig O(A,\mathbb{R}^{1} )\}   $.

For the general $ d $, $  u$ can be taken as
\begin{equation}\label{91119}
u(x^{1},\cdots,x^{d})=\exp \left\lbrace -  2\pi i k(x^{1},\cdots,x^{d}) \right\rbrace \oplus 1_{N-1}
\end{equation}
with $ k $ satisfying $ k(-\infty, x^{2},\cdots,x^{d})=0 $, $ k(+\infty, x^{2},\cdots,x^{d})=1  $. All $ u $'s in (\ref{91119}) differ by a local gauge transformation, and a special class of $ u $ can be selected with
\begin{equation}
k(x^{1},\cdots,x^{d}) = \begin{cases}
	    0\;, & \text{when $x^{1}\leq h(x^{2},\cdots,x^{d})$} \\
	    1\;, &  \text{when $x^{1}\geq h(x^{2},\cdots,x^{d}) $}	
		   \end{cases}\;.
\end{equation}
$ \phi $ has a branch cut at a $(d-1)$-manifold $ \Delta_{d-1} = \{x^{1}= h(x^{2},\cdots,x^{d}) , x^{2},\cdots,x^{d} \in  \mathbb{R}\}$. $ a=u^{-1} du$ is non-vanishing only at $  \Delta_{d-1}$, but $ f=0 $ everywhere since $\partial  \Delta_{d-1}=0  $. The related transformation operator $U^{(1)}  [S_{\infty}^{0} \times \mathbb{R}^{d-1}] \sim  U^{(1)}[\Delta_{d-1}]$ acts nontrivially only on Wilson lines $ \exp  \{ig O(A,\mathcal{M}^{1} )\}   $ ending at $ S_{\infty}^{0} \times \mathbb{R}^{d-1} $, or equivalently, intersecting $ \Delta_{d-1} $. 
\begin{equation}
U^{(1)-1}[\Delta_{d-1}]O(A, \mathcal{M}_{1} )U^{(1)}[\Delta_{d-1}]=O(A, \mathcal{M}_{1} )+1\;,
\end{equation}
\begin{equation}\label{316}
U^{(1)-1}[\Delta_{d-1}]\exp \{igO(A, \mathcal{M}_{1} )\}U^{(1)}[\Delta_{d-1}]=\exp \{igO(A, \mathcal{M}_{1} )\}\exp \{ig\}\;,
\end{equation}
if $ \mathcal{M}_{1} $ intersects $\Delta_{d-1}  $ once.

When $ \Delta_{d-1} $ is divided into $ \Delta^{1}_{d-1} $ and $  \Delta^{2}_{d-1}$ with $  \Delta_{d-1}=\Delta^{1}_{d-1}+  \Delta^{2}_{d-1} $, $\partial \Delta^{1}_{d-1}= \Sigma_{d-2}=-\partial \Delta^{2}_{d-1}$, 
\begin{equation}
U^{(1)}[\Delta_{d-1}]=U^{(1)}(\Sigma_{d-2})U^{(1)}(-\Sigma_{d-2})\;, 
\end{equation}
where $ U^{(1)}(\Sigma_{d-2})$ and $U^{(1)}(-\Sigma_{d-2}) $ are defect operators localized at $\Sigma_{d-2}$ and $-\Sigma_{d-2}$, respectively. So two defects with the $0$ net charge can also combine into a $1$-form global symmetry operator.

$ U^{(1)}[\Delta_{d-1}] $ is the electric $1$-form symmetry operator in $(d+1)$-dimensional gauge theories \cite{G}. When $ G=U(1) $, 
\begin{equation}\label{boun}
U^{(1)}[\Delta_{d-1}]= \exp \left\lbrace - 2\pi \int_{\Delta_{d-1}} \;\ast  \Pi \right\rbrace
\end{equation}
where $ \Pi $ is the conjugate momentum of $ A $ with $ [A_{i},\Pi^{j}]=i \delta_{i}^{j}$ and $ \ast $ is the Hodge star in $\mathbb{R}^{d}$. The deformation of $\Delta_{d-1}  $ amounts to the multiplication of a local gauge transformation.

$  U^{(1)}[\Delta_{d-1}]  $ generates a gauge transformation in $\mathbb{R}^{d}$ with $ u=1_{N} $ and $ u=e^{-2 \pi i} \oplus 1_{N-1}$ on two sides of the domain wall $\Delta_{d-1}$. If we impose the boundary condition that at $\partial \mathbb{R}^{d}$, $ u $ must approach a constant, then the transformation with $ u=1_{N} $ and $ u=e^{-2 \pi g i} \oplus 1_{N-1}$ on two sides, which is obtained via replacing $2\pi$ by $2\pi g$ in (\ref{boun}), is not allowed. So the boundary condition makes the electric $1$-form symmetry discrete.

\subsection{Electric $3$-form global symmetry operators}\label{1q22}

The electric $3$-form global symmetry operators exist in $ 3 $-form gauge theories. Nevertheless, in $(d+1)$-dimensional $1$-form gauge theories with $ G=U(N) $ or $SU(N)$ for $ N\geq 2 $, the operator in $\mathbb{R}^{d}$ generating a gauge transformation $ u $ with $  w(u,\mathcal{M}_{3}) \neq 0  $ for a closed but unbounded $ \mathcal{M}_{3} $ acts nontrivially only on the $ 3 $-dimensional extended operators $ \exp \{igO(A, \mathcal{M}_{3} )\} $ and thus can also be regarded as an electric $3$-form global symmetry transformation. Here, the $3$-form is the composite field $\text{CS}_{3}(A)$.

When $ d=3 $, $\mathcal{M}_{3}   =\mathbb{R}^{3} $, consider 
\begin{equation}\label{912}
u(x^{1},x^{2},x^{3})=\exp \left\lbrace - \pi i k^{a}(x^{1},x^{2},x^{3}) \tau^{a} \right\rbrace \oplus 1_{N-2}\;,\;\;\;\;\;\;\;\;\;a=1,2,3
\end{equation}
with the boundary condition $ k^{a}\rightarrow \hat{x}^{a}=x^{a} / [ \sum^{3}_{i=1}(x^{i})^{2} ]^{1/2} $ when $  \sum^{3}_{i=1}(x^{i})^{2}\rightarrow \infty$. So at $  S_{\infty}^{2} $, $ u=  e^{- \pi i\hat{x} ^{a}\tau^{a}} \oplus 1_{N-2}=-1_{2} \oplus 1_{N-2}$. $ O(u^{-1}du,\mathbb{R}^{3} )=w(u,\mathbb{R}^{3})=1$. All $ u $'s are equivalent differing by a local gauge transformation, among which, a particular set of $ u $ can be selected with 
\begin{equation}
k^{a}=\frac{x^{a}-c^{a}}{ [ \sum^{3}_{i=1}(x^{i}-c^{i})^{2} ]^{1/2}  }\;.
\end{equation}
There is a branch point of $ \phi $ at $\Delta_{0}=\{x^{a}=c^{a}\;|\;a=1,2,3\}$. Since $k^{a}k^{a}=1$, $ u^{-1} du$ is non-vanishing only at $  \Delta_{0}$, but $ f=0 $ everywhere since $\partial  \Delta_{0}=0  $. $ \text{CS}_{3} (u^{-1} du)=\delta^{(3)}(x^{a}-c^{a})dx^{1}\wedge dx^{2}\wedge dx^{3} $. The related transformation operator $U^{(3)}  [S_{\infty}^{2} ] \sim  U^{(3)}[\Delta_{0}]$ acts nontrivially only on $ \exp  \{ig O(A,\mathbb{R}^{3} )\}   $.

For the general $ d $, $ u $ can be selected as 
\begin{equation}
u(x^{1},\cdots,x^{d})=\exp \left\lbrace - \pi i k^{a}(x^{1},\cdots,x^{d}) \tau^{a} \right\rbrace \oplus 1_{N-2}\;,\;\;\;\;\;\;\;\;\;a=1,2,3
\end{equation}
with $ k^{a}$ satisfying $ k^{a}\rightarrow \hat{x}^{a}=x^{a} /[ \sum^{3}_{i=1}(x^{i})^{2} ]^{1/2}$ when $ \sum^{3}_{i=1}(x^{i})^{2} \rightarrow \infty$. At $  S_{\infty}^{2} \times \mathbb{R}^{d-3}$, $ u=  e^{- \pi i\hat{x} ^{a}\tau^{a}} \oplus 1_{N-2}=-1_{2} \oplus 1_{N-2}$. When
\begin{equation}
k^{a}=\frac{x^{a}-h^{a}(x^{4},\cdots,x^{d})}{\{ \sum^{3}_{i=1}[x^{i}-h^{i}(x^{4},\cdots,x^{d})]^{2} \}^{1/2} }\;,
\end{equation}
there is a branch cut of $ \phi $ at a $(d-3)$-manifold $ \Delta_{d-3} = \{x^{a}= h^{a}(x^{4},\cdots,x^{d}) , x^{4},\cdots,x^{d} \in \mathbb{R}\;|\;a=1,2,3 \}$. $ u^{-1} du$ is non-vanishing only at $  \Delta_{d-3}$, but $ f=0 $ everywhere since $\partial  \Delta_{d-3}=0  $. The related transformation operator $U^{(3)}  [S_{\infty}^{2} \times \mathbb{R}^{d-3}] \sim  U^{(3)}[\Delta_{d-3}]$ acts nontrivially only on $ \exp \{igO(A, \mathcal{M}_{3} )\} $ ending at $ S_{\infty}^{2} \times \mathbb{R}^{d-3} $, or equivalently, intersecting $ \Delta_{d-3} $. 
\begin{equation}
U^{(3)-1}[\Delta_{d-3}]O(A, \mathcal{M}_{3} )U^{(3)}[\Delta_{d-3}]=O(A, \mathcal{M}_{3} )+1\;,
\end{equation}
\begin{equation}\label{3166}
U^{(3)-1}[\Delta_{d-3}]\exp \{igO(A, \mathcal{M}_{3} )\}U^{(3)}[\Delta_{d-3}]=\exp \{igO(A, \mathcal{M}_{3} )\}\exp \{ig\}\;,
\end{equation}
if $ \mathcal{M}_{3} $ intersects $\Delta_{d-3}  $ once. With $  \Delta_{d-3}=\Delta^{1}_{d-3}+  \Delta^{2}_{d-3} $ and $\partial \Delta^{1}_{d-3}= \Sigma_{d-4}=-\partial \Delta^{2}_{d-3}$, 
\begin{equation}
U^{(3)}[\Delta_{d-3}]=U^{(3)}(\Sigma_{d-4})U^{(3)}(-\Sigma_{d-4})\;.
\end{equation}

In a $4d$ Yang-Mills theory, the semi-classical vacua in $\mathbb{R}^{3}$ are pure gauges $\vert u^{-1}du \rangle$ classified by the winding number $ w(u,\mathbb{R}^{3}) $, and $ U^{(3)}[S_{\infty}^{2} ]$ is the operator increasing the winding by $ 1 $.  
For $ \vert u^{-1}du \rangle $ with
\begin{equation}\label{wn1}
O(A,\mathbb{R}^{3})\vert u^{-1}du \rangle=w(u,\mathbb{R}^{3}) \vert u^{-1}du \rangle\;,
\end{equation}
$U^{(3)}[S_{\infty}^{2} ]\vert u^{-1}du \rangle= \vert u'^{-1}du' \rangle $ has the winding $w(u',\mathbb{R}^{3})= w(u,\mathbb{R}^{3}) +1 $:
\begin{equation}
O(A,\mathbb{R}^{3})U^{(3)}[S_{\infty}^{2} ]\vert u^{-1}du \rangle=[w(u,\mathbb{R}^{3}) +1]U^{(3)}[S_{\infty}^{2} ]\vert u^{-1}du \rangle\;.
\end{equation}
$ U^{(3)}[S_{\infty}^{2} ] $ commutes with the Hamiltonian, so the physical vacuum should be an eigenstate with 
\begin{equation}
 U^{(3)}[S_{\infty}^{2} ] \vert \theta \rangle = e^{i\theta} \vert \theta \rangle\;,
\end{equation}
which is the $\theta$-vacuum \cite{Jac, qcd}.

Similar to the $1$-form case, $  U^{(3)}[\Delta_{d-3}]  $ generates a gauge transformation in $\mathbb{R}^{d}$ with $ u=e^{- \pi i\hat{x}^{a}\pi^{a}} \oplus 1_{N-2}$ in the $ 3d $ space transverse to $ \Delta_{d-3} $. Since $  u=e^{- \pi g i\hat{x}^{a}\pi^{a}} \oplus 1_{N-2} $ is angle-dependent, the boundary condition of a constant $ u $ at $\partial \mathbb{R}^{d}$ makes the electric $3$-form symmetry discrete.

\subsection{$  U^{(2j-1)}[\Delta_{d-2j+1}]  $ as a superselection operator and the spontaneously symmetry breaking}

The electric $ (2j-1) $-form global symmetry transformation $  U^{(2j-1)}[\Delta_{d-2j+1}]  $ is a ``large" gauge transformation, in contrast to ``small" gauge transformations with $ O(u^{-1}du, \mathcal{M}_{2j-1} )=0 $ for all closed $ \mathcal{M}_{2j-1}$. The Hilbert space (invariant under the ``small" gauge transformations) can be decomposed into the direct sum of common eigenspaces of $ U^{(2j-1)}[\Delta_{d-2j+1}]    $ for all $\Delta_{d-2j+1}$. $ U^{(2j-1)}[\Delta_{d-2j+1}]    $ is unitary, so the eigenvalues are complex phases and the different eigenspaces are orthogonal. Gauge invariant operators $\mathcal{O}$ are also diagonal with respect to this decomposition. $  \langle \Psi_{1}\vert \mathcal{O}\vert \Psi_{2}\rangle =0$ if $\vert \Psi_{1}\rangle$ and $\vert \Psi_{2}\rangle$ belong to different eigenspaces. So $U^{(2j-1)}[\Delta_{d-2j+1}] $ acts as a superselection operator separating the large Hilbert space into distinct superselection sectors, between which no superpositions are possible.

For a state $  \vert \Psi \rangle $ in the eigenspace with the eigenvalue $1$, i.e. 
\begin{equation}
U^{(2j-1)}[\Delta_{d-2j+1}]     \vert \Psi \rangle = \vert \Psi \rangle\;,\;\;\;\;\;\;\;\;\;\;\forall\; \Delta_{d-2j+1}\;,
\end{equation}
let  
\begin{equation}\label{330}
\vert \Psi \rangle_{[\theta,\mathcal{M}_{2j-1}]}=\exp \{-i \theta O(A, \mathcal{M}_{2j-1} )\} \vert \Psi\rangle\;,\;\;\;\;\;\;\;\;\;\;\partial \mathcal{M}_{2j-1}=0\;.
\end{equation}
Then for $\mathcal{M}_{2j-1}$ and $\mathcal{M}'_{2j-1}$ differing by an exact $(2j-1)$-manifold, $\vert \Psi \rangle_{[\theta,\mathcal{M}_{2j-1}]}$ and $\vert \Psi \rangle_{[\theta,\mathcal{M}'_{2j-1}]}$ differ by a phase. From (\ref{316}) and (\ref{3166}), 
\begin{eqnarray}\label{331}
\nonumber && U^{(2j-1)}[\Delta_{d-2j+1}]    \vert \Psi \rangle_{[\theta,\mathcal{M}_{2j-1}]} \\\nonumber&=& U^{(2j-1)}[\Delta_{d-2j+1}]    \exp \{-i \theta O(A, \mathcal{M}_{2j-1} )\}  U^{(2j-1)-1}[\Delta_{d-2j+1}]    \vert \Psi \rangle\\ &=& \exp \{i \theta I(\Delta_{d-2j+1},\mathcal{M}_{2j-1})\} \vert \Psi \rangle_{[\theta,\mathcal{M}_{2j-1}]}\;,
\end{eqnarray}
where $ I(\Delta_{d-2j+1},\mathcal{M}_{2j-1})$ is the intersection number of $ \Delta_{d-2j+1}$ and $\mathcal{M}_{2j-1} $. $ \vert \Psi \rangle_{[\theta,\mathcal{M}_{2j-1}]} $ is in an eigenspace with the eigenvalue $\exp \{i \theta I(\Delta_{d-2j+1},\mathcal{M}_{2j-1})\}$. Successive actions of $\exp \{-i \theta O(A, \mathcal{M}_{2j-1} )\}  $ with the generic $\theta$ and $\mathcal{M}_{2j-1}$ produce eigenstates $\vert \Psi \rangle_{[\theta,\mathcal{M}_{2j-1};\theta',\mathcal{M}'_{2j-1};\cdots]}  $ for all possible eigenvalues.

When $j=2$, $d=3$, with $\vert u^{-1}du \rangle$ in (\ref{wn1}) denoted by $\vert n\rangle$ for $w(u,\mathbb{R}^{3})=n$, the $ \theta $-vacuum for $ \theta=0 $ is $  \vert 0\rangle = \sum_{n}\;\vert n\rangle $, and the action of $ \exp \{-i \theta O(A, \mathbb{R}^{3} )\} $ gives
\begin{equation}
\vert \theta \rangle=\exp \{-i \theta O(A, \mathbb{R}^{3} )\} \vert 0\rangle= \sum_{n}\;\exp \{-i \theta O(A, \mathbb{R}^{3} )\}\vert n\rangle= \sum_{n}\;\exp \{-i n\theta \}\vert n\rangle
\end{equation}
with $  U^{(3)}[\Delta_{0} ] \vert \theta \rangle = e^{i\theta} \vert \theta \rangle$.

$  U^{(2j-1)}[\Delta_{d-2j+1}]   $ commutes with the Hamiltonian. The electric global symmetry is unbroken provided that the vacuum $ \vert  \Omega \rangle $ is also an eigenstate of $ U^{(2j-1)}[\Delta_{d-2j+1}] $. Then for the arbitrary charged operator $\exp \{-i \theta O(A, \mathcal{M}_{2j-1} )\}$, $ \langle \Omega \vert \exp \{-i \theta O(A, \mathcal{M}_{2j-1} )\}  \vert  \Omega \rangle = 0 $.\footnote{Similarly, for the magnetic $(d-2j)$-form global symmetry, if $  \vert  \Omega \rangle$ is an eigenstate of $\exp  \{ig E(F,\mathcal{M}_{2j} )\}$ with $E(F,\mathcal{M}_{2j})$ given by (\ref{3.1q}), then the symmetry will be unbroken. For a charged operator $U^{(2j-1)} (\Sigma_{d-2j} ) $, $ \langle \Omega \vert U^{(2j-1)} (\Sigma_{d-2j} )  \vert  \Omega \rangle = 0$ according to (\ref{3.2q}).} Otherwise, the symmetry is spontaneously broken with $\langle \Omega \vert \exp \{-i \theta O(A, \mathcal{M}_{2j-1} )\} \vert  \Omega \rangle \neq 0$ for some closed but not exact $\mathcal{M}_{2j-1}$. In the following, we will first study the situation when the symmetry is unbroken.

In the symmetric phase, physical states transform by a phase under $  U^{(2j-1)}[\Delta_{d-2j+1}]    $, and the physical Hilbert space is one eigenspace. $\exp \{-i \theta O(A, \mathcal{M}_{2j-1} )\}$ is an operator implementing a transformation between different eigenspaces, and has the vanishing matrix elements in each space. $\exp \{-i \theta O(A, \mathcal{M}_{2j-1} )\}  $ also makes the Hamiltonian $H$ transform as
\begin{equation}\label{3311}
H\rightarrow H'=\exp \{-i \theta O(A, \mathcal{M}_{2j-1} )\}  H \exp \{i \theta O(A, \mathcal{M}_{2j-1} )\}   \;,
 \end{equation}
where $H'$ is the Hamiltonian obtained from the action $ S'=S+\delta S $ with 
\begin{equation}
\delta S \sim \theta \int_{\mathcal{M}_{2j-1} \times \mathbb{R}}P_{j}(F)=\int_{ \mathbb{R}^{d,1}} \bar{C}^{(d-2j+1)}\wedge P_{j}(F)=\int_{ \mathbb{R}^{d,1}} (\bar{C}^{(d-2j+1)}+d \alpha^{(d-2j)} )\wedge P_{j}(F)\;.
\end{equation}
$\ast P_{j}(F)  $ is the Noether current of the dual magnetic $(d-2j)$-form symmetry. $ \bar{C}^{(d-2j+1)}/\theta $ is the Poincar\'e dual of the closed $2j$-manifold $ \mathcal{M}_{2j-1} \times \mathbb{R} \subset \mathbb{R}^{d,1}$. $ d \bar{C}^{(d-2j+1)}=0 $. With a suitable $  \alpha^{(d-2j)}$ selected, $ \bar{C}^{(d-2j+1)}+d \alpha^{(d-2j)} $ can be a smooth closed $(d-2j+1)$-form in $\mathbb{R}^{d,1}$. So physical Hilbert spaces are in one-to-one correspondence with the closed $(d-2j+1)$-forms $ \bar{C}^{(d-2j+1)} \sim \bar{C}^{(d-2j+1)}+d \alpha^{(d-2j)}  $ in $ \mathbb{R}^{d,1}$; selecting a physical space amounts to specifying a closed background field for the magnetic $(d-2j)$-form global symmetry. When $d=3$, $j=2$, the familiar conclusion on $ \theta $-sector and $ \theta $-angle in $4d$ Yang-Mills theory is recovered: $\exp \{-i \theta O(A, \mathbb{R}^{3} )\}   $ is the operator implementing a T-transformation which changes the $\theta$-angle and makes one $ \theta $-sector transform into another. Appendix \ref{AAx} contains a brief review on the $ \theta $-vacuum and $ \theta $-angle in the path integral formalism.

When $d=2j-1$, $  \bar{C}^{(0)} =\theta$, the electric $d$-form global symmetry cannot be spontaneously broken because the physical vacuum must be the $\theta$-vacuum due to the requirement of the cluster decomposition principle \cite{Callan}. The statement is well known in $4d$ Yang-Mills theories and can also be extended to $2d$ and $6d$ theories \cite{qed2}. The only charged operator is $\exp \{-i \alpha O(A, \mathbb{R}^{d} )\}  $, and $\langle \theta \vert \exp \{-i \alpha O(A, \mathbb{R}^{d} )\}  \vert  \theta \rangle = 0$.

When $ d >2j-1 $, the symmetry can be broken or not. For example, in gauge theories, the area law of the Wilson loop indicates a vanishing expectation value of the Wilson line and the preserved electric $1$-form symmetry, while a perimeter or Coulomb law implies a spontaneously symmetry breaking \cite{G}. As a simple demonstration, consider a $ 4d $ $U(1)$ free gauge theory with $ H=\frac{1}{2}(g^{2} \Pi_{i}\Pi^{i}+g^{-2}B_{i} B^{i}    ) $. $ \exp \{-i \theta O(A, \mathcal{M}_{1} )\} $ and $U^{(1)} (\Sigma_{1} )$ are Wilson and 't Hooft lines. When $ g= \infty $, $\Pi_{i} \vert \Omega \rangle =0  $, so $  U^{(1)}[\Delta_{2}] \vert \Omega \rangle = \vert \Omega \rangle $, $\langle \Omega \vert \exp \{-i \theta O(A, \mathcal{M}_{1} )\} \vert \Omega \rangle=0$, $ \langle \Omega \vert    U^{(1)} (\Sigma_{1} )  \vert \Omega \rangle =1 $, the electric symmetry is preserved and the magnetic symmetry is broken. When $ g = 0 $, $B_{i} \vert \Omega \rangle =0 $. Let $\vert \Omega \rangle = \sum_{\alpha} \vert A=d\alpha \rangle$ for $ \alpha $ vanishing at the infinity. Then $\langle \Omega \vert \exp \{-i \theta O(A, \mathcal{M}_{1} )\} \vert \Omega \rangle=1$, $ \langle \Omega \vert    U^{(1)} (\Sigma_{1} )  \vert \Omega \rangle = 0 $, the breaking pattern is reversed. For the finite $g$, both symmetries are broken. In Appendix \ref{AAx}, we study the electric $3$-form symmetry of $5d$ Yang-Mills theories in the path integral formalism and show that there are no physical reasons, like the cluster decomposition principle, ensuring that the vacuum must be an eigenstate of $ U^{(3)}[\Delta_{1}] $, so the symmetry can be broken.

We should also mention the difference between electric global symmetries considered in this paper and those in other literature. In an abelian $1$-form pure gauge theory with the action $ S(F) $, the equation of motion is $d \ast G=0$, where $G(F)$ is a $2$-form constructed from the field strength $F$. There is an electric $1$-form global symmetry acting on Wilson lines with $G$ the Noether current. $ G_{0i} =\Pi_{i}$, so the symmetry operator is just (\ref{boun}). Up to this point, two kinds of symmetries are the same. With the charged matter introduced, the electric symmetry is explicitly broken, since Wilson lines are endable and $d \ast G=\ast J$. The conclusion holds for $n$-form abelian theories and can also induce the completeness-of-spectrum conjecture as a result of the no-global-symmetries conjecture \cite{Ruc, s12}. However, such symmetry-breaking mechanism does not apply in our situation. As a ``large" gauge transformation, $ U^{(1)}[\Delta_{d-1}] $ acts on both matter and the gauge field, and still commutes with the Hamiltonian. Of course, in the presence of the charged matter or when $ j>1 $, Noether current does not exist, and $ U^{(2j-1)}[\Delta_{d-2j+1}] $ cannot be explicitly written as an integral of a closed form over $ \Delta_{d-2j+1} $. It is a symmetry merely because the Hamiltonian is gauge invariant.

\subsection{Fate of symmetries when coupled to gravity}

The electric and magnetic global symmetries are ubiquitous in $1$-form gauge theories. However, it is conjectured that in a consistent theory of quantum gravity, global symmetries, discrete or continuous, including the higher-form and spontaneously broken ones, are not allowed \cite{g1, g2, g3, g4, g5, g6, g7, g8}. In this section, we will consider whether these global symmetries can be gauged or explicitly broken for the theory to couple with gravity.

In a quantum gravity theory, the existence of global symmetries may result in contradictions. For magnetic global symmetries, take a $5d$ gauge theory coupling with gravity as an example. Consider a gauge field configuration with the instanton charge $Q$ localized in a small region. $E\sim Q$, so when $ Q $ is big enough, a black hole can be produced. The black hole may evaporate, loosing mass with the fixed $Q$, ending up as a problematic remnant of Planckian size. So such symmetry is not allowed. The magnetic global symmetry is one kind of ``Chern-Weil" symmetries whose Noether current conservation follows from Bianchi identities. The gauging and explicit breaking mechanisms of ``Chern-Weil" symmetries have been systematically presented in \cite{cw} (see also \cite{cw1}). When embedded in string theory, which is a consistent quantum gravity, $U(N) $ gauge theory becomes the worldvolume theory on $D$-branes, where the magnetic global symmetries are gauged through the coupling (\ref{rr}) with the dynamical R-R supergravity fields.\footnote{The equation of motion is 
\begin{equation}
d \ast H^{(d-2j+2)}= P_{j}(F) \wedge \delta^{Dd}_{9-d}\;,
\end{equation}
where $ H^{(d-2j+2)} $ is the field strength of $ C^{(d-2j+1)} $, $ \ast $ is the Hodge star in $10d$ spacetime, and $\delta^{Dd}_{9-d}$ is the Poincar\'e dual of the $ Dd $ brane worldvolume $\Sigma_{d+1}  $. $\partial \Sigma_{d+1}  =0$. For an arbitrary closed manifold $\mathcal{M}_{2j} \subset \Sigma_{d+1} $, one can always find a closed manifold $\mathcal{M}_{2j+9-d}  $ in $10d$ so that $\mathcal{M}_{2j}=\mathcal{M}_{2j+9-d} \cap \Sigma_{d+1} $. Therefore,   
\begin{equation}
\int_{\mathcal{M}_{2j}  }P_{j}(F)=\int_{\mathcal{M}_{2j+9-d} \cap \Sigma_{d+1}  }P_{j}(F)=\int_{\mathcal{M}_{2j+9-d}   }P_{j}(F) \wedge \delta^{Dd}_{9-d}=0\;.
\end{equation}
The symmetry is gauged.}

In theories with global symmetries, states like $ q \cdots  \vert \Omega \rangle $, where $q$ is a charged operator and $ \vert \Omega \rangle$ is the vacuum, can be produced, which lead to inconsistencies in quantum gravity as already demonstrated. When the global symmetry is gauged, physical states are $ \mathcal{O}  \cdots  \vert \Omega \rangle $ with $\mathcal{O}$ a gauge singlet, so the above black hole argument is fine. Now consider a gauge theory with electric global symmetries. In the symmetric phase, physical states are $ \mathcal{O} \cdots  \vert   \bar{C}^{(d-1)},  \bar{C}^{(d-3)},\cdots    \rangle $, where $  \vert   \bar{C}^{(d-1)},  \bar{C}^{(d-3)},\cdots    \rangle $ is a generalized $ \theta $-vacuum labeled by the background closed forms. $\mathcal{O} $ is gauge invariant, while a charged operator will generate states like $ \exp \{-i \alpha O(A, \mathcal{M}_{2j-1} )\} \vert   \bar{C}^{(d-1)},  \bar{C}^{(d-3)},\cdots    \rangle $ in another Hilbert space. When the symmetry is spontaneously broken, physical states are $ \mathcal{O}  \cdots  \vert \Omega \rangle $ for some broken vacuum $ \vert \Omega \rangle $. Both situations are not quite different from the gauged case, so we may expect that no contradictions will arise neither.

In string theory, the low energy effective theory on $ N $ coincident $D$-branes is a $U(N)$ gauge theory, where all global symmetries with Noether currents are gauged via the coupling with supergravity, but $U^{(2j-1)}[\Delta_{d-2j+1}]$ remains as a global symmetry of the Hamiltonian. As long as the $U(N)$ gauge symmetry is not excluded in quantum gravity, the electric $(2j-1)$-form symmetries persist. Nevertheless, the construction of $U^{(2j-1)}[\Delta_{d-2j+1}]$ depends on the space topology, i.e. the existence of the closed but not exact $\mathcal{M}_{2j-1}$ (a nontrivial homology group $H_{2j-1}(M)$). For example, when the space $ M $ is compact, $ U^{(1)}[\Delta_{d-1}] $ exists in $T^{d}$ but not $S^{d}$. So, in quantum gravity with the topology change, the electric global symmetry is broken.

In the following, we will briefly discuss what happens to the electric global symmetry when the dual magnetic symmetry is gauged. Appendix \ref{AAx} contains a path integral analysis when $d=3$, $j=2$. With a dynamical $ (d-2j+1) $-form $ C^{(d-2j+1)}  $ introduced, the total action is
\begin{equation}
S[ C^{(d-2j+1)},A,\Phi]=S[d  C^{(d-2j+1)}]+S[A,\Phi]-\int_{ \mathbb{R}^{d,1}} (C^{(d-2j+1)}+ \bar{C}^{(d-2j+1)})\wedge P_{j}(F)\;,
\end{equation}
where $S[d  C^{(d-2j+1)}]  $ is the action for $ C^{(d-2j+1)}  $, $S[A,\Phi]$ is the action of the gauge field $A$ coupling with the matter $\Phi$, and $\bar{C}^{(d-2j+1)}$ is a background closed form. $ S[ C^{(d-2j+1)},A,\Phi] $ is invariant under the shift
\begin{equation}\label{shif}
C^{(d-2j+1)} \rightarrow C^{(d-2j+1)}+\bar{c}^{(d-2j+1)}\;,\;\;\;\;\;\;\;\bar{C}^{(d-2j+1)}\rightarrow \bar{C}^{(d-2j+1)}-\bar{c}^{(d-2j+1)}
\end{equation}
with $d\bar{c}^{(d-2j+1)}=0  $. As a result, the background $ \bar{C}^{(d-2j+1)} $, which labels the selected superselection sector, is not specified.

The electric $(2j-1)$-form symmetry operator is still $U^{(2j-1)}[\Delta_{d-2j+1}]   $, under which, the Hilbert space is decomposed into eigenspaces. In (\ref{330}) and (\ref{331}), $\exp \{-i \theta O(A, \mathcal{M}_{2j-1} )\}  $ can be replaced by 
\begin{equation}
\exp \{-i \theta O(A, \Pi^{(d-2j+1)},\mathcal{M}_{2j-1} )\} \equiv \exp \{-i \theta O(A, \mathcal{M}_{2j-1} )\}\exp \{-i \theta \int_{\mathcal{M}_{2j-1} } \ast\Pi^{(d-2j+1)}    \}\;,
\end{equation}
where $ \Pi^{(d-2j+1)} $ is the conjugate momentum of $ C^{(d-2j+1)}$. Instead of (\ref{3311}), 
\begin{equation}
 H'=\exp \{-i \theta O(A, \Pi^{(d-2j+1)},\mathcal{M}_{2j-1} )\}  H \exp \{i \theta O(A, \Pi^{(d-2j+1)},\mathcal{M}_{2j-1} )\} =H  
 \end{equation}
due to the symmetry (\ref{shif}). The action of $\exp \{-i \theta O(A, \Pi^{(d-2j+1)},\mathcal{M}_{2j-1} )\} $ makes one eigenspace transform into another with $ H $ invariant. So when the magnetic $(d-2j)$-form symmetry is gauged, there is no privileged superselection sector.

\section{Commutation relations of extended operators}\label{3}

Now, we are going to compute the commutation relations between the extended operators $\exp  \{ig O(A,\mathcal{M}_{2k-1} )\}  $ and $ U^{(2j-1)}(\Sigma_{d-2j}) $ in $ \mathbb{R}^{d}$ with $\partial \mathcal{M}_{2k-1}= \partial\Sigma_{d-2j} =0  $. When $ \mathcal{M}_{2k-1} $ is compact, $\exp  \{ig O(A,\mathcal{M}_{2k-1} )\} $ and $U^{(2j-1)}(\Sigma_{d-2j}) $ are also magnetic $(d-2k)$-form and electric $(2j-1)$-form local transformation operators, so the commutation relations present their actions on each other. We will list the main result with the proof given in Appendix \ref{A}.

Chern-Simons terms can be calculated from \cite{Ge}
\begin{equation}
\text{CS}_{2j-1} (A,F)=\frac{1}{(j-1)!}(\frac{i}{2\pi})^{j}\int^{1}_{0}\delta t \; str(A,F_{t}^{j-1})\;,
\end{equation}
where $ F_{t} = tF+(t^{2}-t)A^{2} $ and ``$ str $" is the symmetric trace
\begin{equation}
 str (M_{1},\cdots ,M_{n})= \frac{1}{n!}\sum_{(i_{1},\cdots,i_{n})}tr(M_{i_{1}}\cdots M_{1_{n}})
\end{equation}
with the sum over all permutations.
\begin{equation}
O (A,\mathcal{M}_{2j-1})=\int_{\mathcal{M}_{2j-1}}\text{CS}_{2j-1} (A,F)\;.
\end{equation}
For $ H  $ in the Lie algebra of $G$, we may define
\begin{equation}
\text{CS}_{2j-1} (H,\cdots,H,A,F)= \frac{1}{(j-1)!}(\frac{i}{2\pi})^{j}\int^{1}_{0}\delta t \; str(H,\cdots,H,A,F_{t}^{j-1})
\end{equation}
as well as
\begin{equation}
O (H,\cdots,H,A,\mathcal{M}_{2j-1})=\int_{\mathcal{M}_{2j-1}}\text{CS}_{2j-1} (H,\cdots,H,A,F)\;.
\end{equation}
(\ref{322}) is an example when $j=1$. 

\begin{itemize}

\item If $u $ is the transformation matrix of $ U^{(1)}(\Sigma_{d-2}) $ with 
\begin{equation}
 f(x)=\frac{\pi H}{i(d-2)!}\int_{\Sigma_{d-2}}\epsilon_{l_{1}\cdots l_{d-2}m_{1}m_{2}}\delta^{(d)}(\vec{x}-\vec{y})\;dy^{l_{1}}\wedge \cdots\wedge dy^{l_{d-2}}\wedge dx^{m_{1}} \wedge  dx^{m_{2}}
\end{equation}
and $tr H=1 $, from (\ref{a1}), we will have
\begin{eqnarray}\label{o1}
\nonumber && U^{(1)-1}(\Sigma_{d-2})O (A,\mathcal{M}_{2j-1})U^{(1)}(\Sigma_{d-2}) \\\nonumber  &=&O (u^{-1}du,\mathcal{M}_{2j-1})+O (A,\mathcal{M}_{2j-1})
+ \int_{\mathcal{M}_{2j-1}\cap \Sigma_{d-2}} \frac{1}{(j-2)!}(\frac{i}{2 \pi})^{j-1}\int_{0}^{1}  \delta t \;str( H ,A,F^{j-2}_{t})\\  &=&L (\Sigma_{d-2},\mathcal{M}_{2j-1})+O (A,\mathcal{M}_{2j-1})
+O(H,A,\mathcal{M}_{2j-1}\cap \Sigma_{d-2}) \;,
\end{eqnarray}
and
\begin{eqnarray}\label{o11}
\nonumber  \exp \{igO (A,\mathcal{M}_{2j-1})\}U^{(1)}(\Sigma_{d-2})&=& U^{(1)}(\Sigma_{d-2})\exp \{igO (A,\mathcal{M}_{2j-1})\}\\\nonumber &&\exp \{igL (\Sigma_{d-2},\mathcal{M}_{2j-1})\}\exp \{ig
O(H,A,\mathcal{M}_{2j-1}\cap \Sigma_{d-2}) \}\;.\\
\end{eqnarray}
In (\ref{o1}) and (\ref{o11}), $L (\Sigma_{d-2},\mathcal{M}_{2j-1})=0  $ if $j \neq 1$; $ O(H,A,\mathcal{M}_{2j-1}\cap \Sigma_{d-2})=0 $ if $j=1$.

\begin{itemize}

\item[1)]

When $ j=1 $, 
\begin{equation}
\exp \{igO (A,\mathcal{M}_{1})\}U^{(1)}(\Sigma_{d-2})=U^{(1)}(\Sigma_{d-2})\exp \{igO (A,\mathcal{M}_{1})\}\exp \{igL(\Sigma_{d-2},\mathcal{M}_{1})\}\;.
\end{equation}
Actions of $\exp \{igO (A,\mathcal{M}_{1})\}$ and $U^{(1)}(\Sigma_{d-2})$ on each other produce a phase. 
\item[2)]

When $ j\geq 2 $, 
\begin{eqnarray}\label{6.100}
\nonumber \exp \{igO (A,\mathcal{M}_{2j-1})\}U^{(1)}(\Sigma_{d-2}) &=& U^{(1)}(\Sigma_{d-2})\exp \{igO (A,\mathcal{M}_{2j-1})\} \\ &&\exp \{igO(H,A,\mathcal{M}_{2j-1}\cap \Sigma_{d-2})\}.
\end{eqnarray}
Actions of $\exp \{igO (A,\mathcal{M}_{2j-1})\}$ and $U^{(1)}(\Sigma_{d-2})$ on each other produce a lower-dimensional Chern-Simons operator at the intersection. 
In particular, when $j=2$, 
\begin{eqnarray}\label{711}
\nonumber  \exp \{igO (A,\mathcal{M}_{3})\}U^{(1)}(\Sigma_{d-2})&=& U^{(1)}(\Sigma_{d-2})\exp \{igO (A,\mathcal{M}_{3})\}\\  &&\exp \{- \frac{g}{2 \pi}\int_{\mathcal{M}_{3}\cap \Sigma_{d-2}} tr( HA)\}\;.
\end{eqnarray}

In a $4$-dimensional gauge theory, $ d=3 $, $ \exp \{igO (A,\mathbb{R}^{3})\} $ is the operator generating a T-transformation with
\begin{equation}
 \exp \{igO (A,\mathbb{R}^{3})\}\Pi_{i} \exp \{-igO (A,\mathbb{R}^{3})\}= \Pi_{i}+\frac{g}{8\pi^{2}}\epsilon_{ijk}F^{jk}\;,
\end{equation}
and $ U^{(1)}(\Sigma_{1}) $ is the 't Hooft loop/line at $ \Sigma_{1} $. $ \Pi_{i} $ is the conjugate momentum of $ A_{i} $ with $[A_{i},\Pi^{j} ]=i \delta_{i}^{j} $. (\ref{711}) becomes 
\begin{equation}\label{712}
\exp \{igO (A,\mathbb{R}^{3})\}U^{(1)}(\Sigma_{1})\exp \{-igO (A,\mathbb{R}^{3})\}=U^{(1)}(\Sigma_{1})\exp \{- \frac{g}{2 \pi}\int_{ \Sigma_{1}} tr( HA)\}\;.
\end{equation}
This is the Witten effect, namely, under the T-transformation which changes the $ \theta $-angle, a 't Hooft loop/line is multiplied by a Wilson loop/line at the same locus \cite{wi}.

In a $ 3d $ gauge theory, we have monopole operators $U^{(1)}(\Sigma_{0})  $ in the $2d$ space whose time evolution yields $ U^{(1)}(\Sigma_{1})  $ in the $3d$ spacetime. In this case, the interpretation of (\ref{712}) is that when the action contains a Chern-Simons term, monopole operators get a $ 0 $-form electric charge to couple with $ A $ \cite{7}.

\item[3)]
When $ j\geq 2 $ and $ H=\frac{1}{N} 1_{N}$, $ O(H,A,\mathcal{M}_{2j-1}\cap \Sigma_{d-2})=\frac{1}{N}O (A,\mathcal{M}_{2j-1}\cap \Sigma_{d-2})  $, (\ref{6.100}) becomes 
\begin{eqnarray}
\nonumber  \exp \{igO (A,\mathcal{M}_{2j-1})\}U^{(1)}(\Sigma_{d-2})&=& U^{(1)}(\Sigma_{d-2})\exp \{igO (A,\mathcal{M}_{2j-1})\}\\  &&\exp \{
 \frac{ig}{N}O (A,\mathcal{M}_{2j-1}\cap \Sigma_{d-2}) \}\;.
\end{eqnarray}

In particular, when $ d=2j-1 $,
\begin{equation}\label{615}
\exp \{igO (A,\mathbb{R}^{d})\}U^{(1)}(\Sigma_{d-2})\exp \{-igO (A,\mathbb{R}^{d})\}= U^{(1)}(\Sigma_{d-2})\exp \{
 \frac{ig}{N}O (A, \Sigma_{d-2}) \}\;.
\end{equation}

\item[4)]
In a $(d+1)$-dimensional gauge theory, when the space dimension $d$ is odd, the operator $  \exp \{igO (A,\mathbb{R}^{d})\}$ implements a T-transformation with 
\begin{eqnarray}\label{ka}
\nonumber &&  \exp \{igO (A,\mathbb{R}^{d})\}\Pi_{i}  \exp \{-igO (A,\mathbb{R}^{d})\}\\  &=&  \Pi_{i}-\frac{2g}{(\frac{d-1}{2})!}(\frac{i}{4\pi})^{\frac{d+1}{2}}\epsilon_{i l_{1} \cdots l_{d-1}}F^{l_{1}l_{2}}\cdots F^{l_{d-2}l_{d-1}}\;,
\end{eqnarray}
which changes the coefficient of the $ \theta $-term $tr F^{\frac{d+1}{2}} $. In type IIB string theory, the effective action on $ Dd $-branes for the odd $d$ may contain $tr F^{\frac{d+1}{2}} $ coupling with the background R-R $0$-form $ C^{(0)} \sim \theta$, so (\ref{ka}) is the T-transformation on $ Dd $-branes. Under (\ref{ka}), according to (\ref{6.100}), $U^{(1)}(\Sigma_{d-2})  $ transforms as
\begin{equation}\label{6100}
\exp \{igO (A,\mathbb{R}^{d})\}U^{(1)}(\Sigma_{d-2})\exp \{-igO (A,\mathbb{R}^{d})\}= U^{(1)}(\Sigma_{d-2})\exp \{igO(H,A, \Sigma_{d-2})\}\;.
\end{equation}
This is a higher dimensional Witten effect. In an even-dimensional gauge theory, when the $ \theta $-angle changes, the 't Hooft operator $ U^{(1)}(\Sigma_{d-2})  $ is multiplied by $ \exp \{igO(H,A, \Sigma_{d-2})\} $ at the same locus. In an odd-dimensional gauge theory, when the action contains a Chern-Simons term, the 't Hooft operator $ U^{(1)}(\Sigma_{d-2})  $ gets a $ (d-2 )$-form electric charge to couple with $\text{CS}_{d-1} (A)$.

When $ H=\frac{1}{N} 1_{N}  $, (\ref{6100}) reduces to (\ref{615}). When $  H=\frac{1}{N} 1_{N} $, up to the multiplication of a local gauge transformation, $  U^{(1)}(\Sigma_{d-2})  $ can be written as 
\begin{eqnarray}\label{kaa}
\nonumber   U^{(1)}(\Sigma_{d-2})&=& \exp \left\lbrace - \frac{2\pi}{(d-1)!N}\int_{\Sigma_{d-1}} \epsilon_{l_{1}\cdots l_{d-1}m}\;tr \Pi^{m}\;dy^{l_{1}}\wedge \cdots\wedge dy^{l_{d-1}} \right\rbrace\\  &=& \exp \left\lbrace - \frac{2\pi}{N}\int_{\Sigma_{d-1}} \;\ast tr \Pi \right\rbrace\;,
\end{eqnarray}
where $ \Sigma_{d-1} $ is an arbitrary $ (d-1) $-manifold with $ \partial \Sigma_{d-1}=\Sigma_{d-2} $. With (\ref{ka}) plugged in (\ref{kaa}), we arrive at (\ref{615}):
\begin{eqnarray}
\nonumber && \exp \{igO (A,\mathbb{R}^{d})\} U^{(1)}(\Sigma_{d-2})  \exp \{-igO (A,\mathbb{R}^{d})\}\\\nonumber &=& U^{(1)}(\Sigma_{d-2}) \exp \left\lbrace  \frac{ig}{N}\int_{\Sigma_{d-1}} \;P_{\frac{d-1}{2}}(F)\right\rbrace=U^{(1)}(\Sigma_{d-2}) \exp \left\lbrace  \frac{ig}{N}\int_{\Sigma_{d-2}} \;\text{CS}_{d-2}(A)\right\rbrace \\&=&  U^{(1)}(\Sigma_{d-2})\exp \{
 \frac{ig}{N}O (A, \Sigma_{d-2}) \}\;.
\end{eqnarray}

When the gauge group is $ U(1) $, one can also define an S-transformation with
\begin{equation}
\Pi_{i} \rightarrow -\frac{2}{(\frac{d-1}{2})!}(\frac{i}{4\pi})^{\frac{d+1}{2}}\epsilon_{i l_{1} \cdots l_{d-1}}F^{l_{1}l_{2}}\cdots F^{l_{d-2}l_{d-1}}
\end{equation}
so that
\begin{equation}
 U^{(1)}(\Sigma_{d-2})\rightarrow \exp \{
iO (A, \Sigma_{d-2}) \}\;.
\end{equation}

\end{itemize}

\item

If $ u $ is the transformation matrix of $U^{(3)}(\Sigma_{d-4}) $ with $ f=f^{a}H_{a} $, $ H_{a} =\tau_{a} \oplus 0_{N-2} $ and
\begin{equation}
f^{a}\wedge f^{b}=-\frac{4\pi^{2}\delta^{ab}}{3\cdot 4!(d-4)!}\int_{\Sigma_{d-4}}\epsilon_{l_{1}\cdots l_{d-4}m_{1}\cdots m_{4}}\delta^{(d)}(\vec{x}-\vec{y})\;dy^{l_{1}}\wedge \cdots\wedge dy^{l_{d-4}}\wedge dx^{m_{1}} \wedge\cdots \wedge dx^{m_{4}} \;,
\end{equation}
from (\ref{a2}), we will have
\begin{eqnarray}\label{oo1}
\nonumber && U^{(3)-1}(\Sigma_{d-4})O (A,\mathcal{M}_{2j-1})U^{(3)}(\Sigma_{d-4}) \\\nonumber  &=&O (u^{-1}du,\mathcal{M}_{2j-1})+O (A,\mathcal{M}_{2j-1})\\\nonumber  &+&\frac{1}{6(j-3)!}(\frac{i}{2 \pi})^{j-2} \int_{\mathcal{M}_{2j-1}\cap \Sigma_{d-4}}\int_{0}^{1}  \delta t\;str(gH_{a}g^{-1},gH_{a}g^{-1}, A,F^{j-3}_{t} )\\\nonumber&=& L (\Sigma_{d-4},\mathcal{M}_{2j-1})+O (A,\mathcal{M}_{2j-1})+ \frac{1}{6} O(gH_{a}g^{-1},gH_{a}g^{-1},A,\mathcal{M}_{2j-1}\cap \Sigma_{d-4})   \;,\\
\end{eqnarray}
and
\begin{eqnarray}\label{oo11}
\nonumber && \exp \{ig O (A,\mathcal{M}_{2j-1})\}U^{(3)}(\Sigma_{d-4}) \\ \nonumber  &=&U^{(3)}(\Sigma_{d-4}) \exp \{igO (A,\mathcal{M}_{2j-1})\}\\ \nonumber  &&\exp \{igL (\Sigma_{d-4},\mathcal{M}_{2j-1})\}\exp \{ \frac{ig}{6} O(gH_{a}g^{-1},gH_{a}g^{-1},A,\mathcal{M}_{2j-1}\cap \Sigma_{d-4}) \}  \;.\\
\end{eqnarray}
In (\ref{oo1}) and (\ref{oo11}), $L (\Sigma_{d-4},\mathcal{M}_{2j-1})=0  $ if $j \neq 2$; $ O(gH_{a}g^{-1},gH_{a}g^{-1},A,\mathcal{M}_{2j-1}\cap \Sigma_{d-4}) =0 $ if $ j <3$. 

\begin{itemize}

\item[1)]

When $j=1$,
\begin{equation}
\exp \{igO (A,\mathcal{M}_{1})\}U^{(3)}(\Sigma_{d-4})  =U^{(3)}(\Sigma_{d-4})\exp \{igO (A,\mathcal{M}_{1}) \}\;.
\end{equation}

\item[2)]

When $j=2$, 
\begin{equation}
\exp \{igO (A,\mathcal{M}_{3})\}U^{(3)}(\Sigma_{d-4})  =U^{(3)}(\Sigma_{d-4})\exp \{igO (A,\mathcal{M}_{3}) \}\exp \{igL(\Sigma_{d-4},\mathcal{M}_{3})\}\;.
\end{equation}

\item[3)]

When $j\geq 3$, 
\begin{eqnarray}\label{j}
\nonumber \exp \{igO (A,\mathcal{M}_{2j-1})\}U^{(3)}(\Sigma_{d-4})&=& U^{(3)}(\Sigma_{d-4})\exp \{igO (A,\mathcal{M}_{2j-1})\} \\  \nonumber&&\exp \{ \frac{ig}{6} O(gH_{a}g^{-1},gH_{a}g^{-1},A,\mathcal{M}_{2j-1}\cap \Sigma_{d-4}) \}  \;.\\
\end{eqnarray}
In particular, when $ j=3 $,
\begin{eqnarray}\label{7111}
\nonumber \exp \{igO (A,\mathcal{M}_{5})\}U^{(3)}(\Sigma_{d-4})  &=& U^{(3)}(\Sigma_{d-4}) \exp \{igO (A,\mathcal{M}_{5})\}\\  &&\exp \{-\frac{g}{2 \pi} \int_{\mathcal{M}_{5}\cap \Sigma_{d-4}}tr(\frac{1}{6}H^{a}H^{a}A )\}  \;,
\end{eqnarray}
where $ \frac{1}{6}H^{a}H^{a}=(\frac{1}{2})_{2}\oplus 0_{N-2} $.

In a $6$-dimensional gauge theory, $ d=5 $, $ \exp \{igO (A,\mathbb{R}^{5})\} $ is the operator generating the T-transformation
\begin{equation}\label{t}
\Pi_{i}\rightarrow \Pi_{i}+\frac{ig}{(4\pi)^{3}}\epsilon_{ijklm}F^{jk}F^{lm}\;,
\end{equation}
and $ U^{(3)}(\Sigma_{1}) $ is the instanton loop/line at $ \Sigma_{1} $, (\ref{7111}) becomes 
\begin{equation}\label{7122}
 \exp \{igO (A,\mathbb{R}^{5})\}U^{(3)}(\Sigma_{1})\exp \{-igO (A,\mathbb{R}^{5})\}  = U^{(3)}(\Sigma_{1}) \exp \{-\frac{g}{2 \pi} \int_{ \Sigma_{1}}tr( \frac{1}{6}H^{a}H^{a}A )\}  \;.
\end{equation}
Compared with (\ref{712}), this is the Witten effect for instantons. Under (\ref{t}) which changes the $ \theta $-term $ \theta F^{3} $, the instanton loop/line is multiplied by a Wilson loop/line at the same locus.

In a $ 5d $ gauge theory, we have instanton operators $U^{(3)}(\Sigma_{0})  $ in the $4d$ space whose time evolution gives $ U^{(3)}(\Sigma_{1})  $ in the $5d$ spacetime. The interpretation of (\ref{7122}) is that when the action contains a Chern-Simons term, instanton operators get the $ 0 $-form electric charge to couple with $A$ \cite{8}.

\item[4)]

In a $(d+1)$-dimensional gauge theory with the odd $ d $, from (\ref{j}), 
\begin{eqnarray}\label{61000}
\nonumber &&\exp \{igO (A,\mathbb{R}^{d})\}U^{(3)}(\Sigma_{d-4})\exp \{-igO (A,\mathbb{R}^{d})\}  \\  &=&U^{(3)}(\Sigma_{d-4})\exp \{ \frac{ig}{6} O(gH_{a}g^{-1},gH_{a}g^{-1},A, \Sigma_{d-4}) \}  \;.
\end{eqnarray}
So under the T-transformation (\ref{ka}), the instanton operator $U^{(3)}(\Sigma_{d-4})  $ is multiplied by $ \exp \{ \frac{ig}{6} O(gH_{a}g^{-1},gH_{a}g^{-1},A, \Sigma_{d-4}) \} $ at the same locus. This is the higher dimensional Witten effect for instantons. In a $d$-dimensional gauge theory with the odd $ d $, when the action contains a Chern-Simons term, the instanton operator $ U^{(3)}(\Sigma_{d-5}) $ gets a $ (d-5 )$-form electric charge to couple with $  \text{CS}_{d-4} (A)$.

\end{itemize}

\end{itemize}

Finally, we may consider a concrete example. In a $6d$ $U(N)$ gauge theory for $N \geq 2$, extended operators in $\mathbb{R}^{5}$ include $\exp  \{ig O(A,\mathcal{M}_{1} )\}  $, $\exp  \{ig O(A,\mathcal{M}_{3} )\}  $, $\exp  \{ig O(A,\mathbb{R}^{5} )\}  $, $ U^{(1)}(\Sigma_{3}) $ and $ U^{(3)}(\Sigma_{1}) $. Nontrivial commutation relations are
\begin{eqnarray}
\nonumber && \exp  \{ig O(A,\mathcal{M}_{1} )\}U^{(1)} (\Sigma_{3} )=U^{(1)} (\Sigma_{3} )\exp  \{ig O(A,\mathcal{M}_{1} )\}\exp  \{ig L( \Sigma_{3}, \mathcal{M}_{1} ) \}\;,\\\nonumber &&\exp  \{ig O(A,\mathcal{M}_{3} )\}  U^{(1)}(\Sigma_{3})=U^{(1)}(\Sigma_{3})\exp  \{ig O(A,\mathcal{M}_{3} )\} \exp \{- \frac{g}{2 \pi}\int_{\mathcal{M}_{3}\cap \Sigma_{3}} tr( HA)\}\;,\\ &&\exp  \{ig O(A,\mathbb{R}^{5} )\}  U^{(1)}(\Sigma_{3})=U^{(1)}(\Sigma_{3})\exp  \{ig O(A,\mathbb{R}^{5} )\}  \exp  \{ig O(H,A,\Sigma_{3}  )\} \;,
\end{eqnarray}
where $ H=1\oplus 0_{N-1} $ and 
\begin{eqnarray}
\nonumber &&\exp  \{ig O(A,\mathcal{M}_{3} )\}U^{(3)} (\Sigma_{1} )=U^{(3)} (\Sigma_{1} )\exp  \{ig O(A,\mathcal{M}_{3} )\}\exp  \{ig L( \Sigma_{1}, \mathcal{M}_{3} ) \} \;,\\ \nonumber &&\exp  \{ig O(A,\mathbb{R}^{5} )\}  U^{(3)}(\Sigma_{1})=U^{(3)}(\Sigma_{1})\exp  \{ig O(A,\mathbb{R}^{5} )\}   \exp \{-\frac{g}{2 \pi} \int_{ \Sigma_{1}}tr(\frac{1}{6}H^{a}H^{a}A )\}\;,\\
\end{eqnarray}
where $ \frac{1}{6}H^{a}H^{a}=(\frac{1}{2})_{2}\oplus 0_{N-2}$.

\section{Nontrivial $ 1$-form bundles}\label{n12}

In previous sections, $ A $ and $ u $ are both globally defined. Now, we will consider the generic $ A $ and $ u $ in $\mathbb{R}^{d}$ that may not have a global definition.

For the given $ A $ in $\mathbb{R}^{d}$ with $ F=dA+A^{2} $, the integral of $ P_{j}(F)$ over an exact $2j$-manifold $ \mathcal{M}_{2j}  $ gives an even-dimensional observable 
\begin{equation}
E(F,\mathcal{M}_{2j} )=\int_{\mathcal{M}_{2j}  }P_{j}(F)\;.
\end{equation}
Locally, $ P_{j}(F)= d  \text{CS}_{2j-1}(A) $, so for a globally defined $ A $, $ E(F,\mathcal{M}_{2j} )=0 $ everywhere. The non-zero $ E(F,\mathcal{M}_{2j} )  $ can be obtained when the $ A $ bundle is nontrivial on $ \mathcal{M}_{2j} $. Suppose $ P_{N} $ and $ P_{S} $ are two patches covering $ \mathcal{M}_{2j}  $. The $ 1 $-forms in $ P_{N} $ and $ P_{S} $ are $ A_{N} $ and $ A_{S} $, respectively. At $ P_{N} \cap P_{S} \sim \mathcal{M}_{2j-1} \times I$, 
\begin{equation}\label{62}
A_{N} =u^{-1}(A_{S}+d)u
\end{equation}
and $P_{j}(F_{N})=P_{j}(F_{S})$ \footnote{$F_{N} =u^{-1}F_{S}u$, so $F$ may not be globally defined on $\mathcal{M}_{2j}$. However, we can always find the suitable $A$ bundle so that $F$ is globally defined on $\mathcal{M}_{2j}$.}, where the transition function $ u  $ is globally defined in $ \mathcal{M}_{2j-1} \times I $. Let $  \mathcal{M}_{2j} =\mathcal{M}^{N}_{2j} +\mathcal{M}^{S}_{2j} $ with $\mathcal{M}^{N}_{2j}  \subset P_{N}  $, $\mathcal{M}^{S}_{2j}  \subset P_{S}  $ and $\partial\mathcal{M}^{N}_{2j} =\mathcal{M}_{2j-1}=-\partial\mathcal{M}^{S}_{2j} $, then 
\begin{eqnarray}\label{4.26}
\nonumber  E(F,\mathcal{M}_{2j} )  &=&\int_{\mathcal{M}^{N}_{2j}  }P_{j}(F_{N}) +\int_{\mathcal{M}^{S}_{2j}  }P_{j}(F_{S})\\ \nonumber &=& \int_{\mathcal{M}_{2j-1}  }\text{CS}_{2j-1}(A_{N})-\int_{\mathcal{M}_{2j-1}  }\text{CS}_{2j-1}(A_{S})  \\ &=& \int_{\mathcal{M}_{2j-1}  }\text{CS}_{2j-1}(u^{-1}du)=w(u,\mathcal{M}_{2j-1}  ) \;.
\end{eqnarray}
So $E(F,\mathcal{M}_{2j} ) \in \mathbb{Z}\cong \Pi_{2j-1}(G)  $.

The $1$-form $ A $ in $\mathbb{R}^{d}$ are classified by 
\begin{equation}\label{c}
\left\lbrace  E(F,\mathcal{M}_{2j} )\;|\; \forall \; \mathcal{M}_{2j} \subset \mathbb{R}^{d},  \partial\mathcal{M}_{2j} =0, j=1,2,\ldots\right\rbrace   \;.
\end{equation}
If $ E(F,\mathcal{M}_{2j} )= E(F',\mathcal{M}_{2j} )  $ for the arbitrary $ \mathcal{M}_{2j} $, $ A $ and $ A' $ are in the same class and could be continuously deformed into each other. Trivial $ A $ bundle that could be continuously deformed to $ 0 $ should have $E(F,\mathcal{M}_{2j} )=0  $ everywhere. From a nontrivial $ A $ bundle with the non-zero $ E(F,\mathcal{M}_{2j} ) $, a $ (2j+1) $-form $ Q_{2j+1}(F) $ can be extracted via 
\begin{equation}
E(F,\mathcal{M}_{2j} )=\int_{\mathcal{M}_{2j}  }P_{j}(F)=\int_{\mathcal{M}_{2j+1}  }Q_{2j+1}(F)\;,
\end{equation}
where $\partial \mathcal{M}_{2j+1}  =\mathcal{M}_{2j}    $. $ Q_{2j+1}(F)=d P_{j}(F)$. $ E(F,\mathcal{M}_{2j} ) $ are integers, so $Q_{2j+1}(F) \in  \mathbb{Z}  $ in $ \mathcal{M}_{2j+1} $. When $ Q_{2j+1}(F)(x) \neq 0  $, $ x $ is a singularity. In $ \mathbb{R}^{d} $, the typical $ Q_{2j+1}(F) $ is 
\begin{equation}
Q_{m_{1}\cdots m_{2j+1}}(x)=\frac{1}{(d-2j-1)!}\int_{c_{d-2j-1}}\epsilon_{l_{1}\cdots l_{d-2j-1}m_{1}\cdots m_{2j+1}}\delta^{(d)}(\vec{x}-\vec{y})dy^{l_{1}}\wedge \cdots\wedge dy^{l_{d-2j-1}}\;.
\end{equation}
$ dQ_{2j+1}(F)=0 $, so $\partial c_{d-2j-1}=0  $. The nontrivial $A$ bundle could produce a $2$-form $b=F $ whose $ Q_{2j+1}(b) $ is supported on a $(d-2j-1)$-cycle $ c_{d-2j-1} $ (the Poincar\'e dual of $ Q_{2j+1}(b) $). In particular, $ Q_{3}(b) =\frac{i}{2\pi} tr h $, where $h$ is the field strength of the $2$-form $b$.

In a $2$-form theory, if the characteristic class (\ref{c}) of $A$ is regarded as the gauge equivalence class of $F$, for $A$ and $A'$ in the same class, $F$ and $F'$ will be gauge equivalent. In particular, the trivial $A$ bundle will produce an $ F $ that is a pure gauge. For a nonabelian $2$-form $B$ with the $3$-form field strength $H$, if $  B=F=dA+A^{2} $ for some $A$, then $ H=0 $ \footnote{In higher gauge theories, a 2-connection consists of a $1$-form $A$ and a $2$-form $B$ valued in Lie algebras $g$ and $h$. If $g=h$, the fake curvature condition will make $B=F$, in which case, the 2-curvature $H=dB+[A,B]=0$, see section 4.4 of \cite{00}.}. $ Q_{2j+1}(F) $ are gauge invariants.

Just as $ \text{CS}_{2j-1}(u^{-1}du) $ is the $0$-form limit of $ \text{CS}_{2j-1}(A) $, we may expect $ P_{j}(F) $ is also the $1$-form limit of some $ P_{j}(B)  $, and the even-dimensional observables of the $ 2 $-form $B$ are 
\begin{equation}
E(B,\mathcal{M}_{2j} )=\int_{\mathcal{M}_{2j}  }P_{j}(B)=\int_{\mathcal{M}_{2j+1}  }Q_{2j+1}(B)  
\end{equation}
with $\partial  \mathcal{M}_{2j+1}=\mathcal{M}_{2j} $, which may take continuous values. The gauge invariant closed form $ Q_{2j+1}(B)=d  P_{j}(B)     $ could act as the characteristic class of $B$, whose integration over a closed $\mathcal{M}_{2j+1}  $ is non-zero only when the $ B $ bundle is nontrivial on $\mathcal{M}_{2j+1} $. When $ j=1 $,
\begin{equation}
P_{1}(B) =\frac{i}{2\pi}tr B\;,\;\;\;\;\;\;\;Q_{3}(B)=dP_{1}(B)=\frac{i}{2\pi}tr H
\end{equation}
are $ U(1) $ components of $ B $ and $H$. We do not know the explicit form of $  P_{j}(B)   $ for $ j >1$ which may involve the field strength $ H $. In a $ U(1) $ $ 2$-form gauge theory, the only gauge invariant extended observable is 
\begin{equation}\label{e}
\int_{\partial \mathcal{M}_{3} }B=\int_{\mathcal{M}_{3}  } H\;,
\end{equation}
since $ H^{k} =0$ when $ k >1  $.

$ U^{(2j-1)} ( c_{d-2j} )$ in section \ref{5} with $ \partial c_{d-2j} =0 $ are transformations with the globally defined $ u $ that cannot change the characteristic class of $A$, so 
\begin{equation}
E(F,\mathcal{M}_{2k} )U^{(2j-1)} ( c_{d-2j} )=U^{(2j-1)} ( c_{d-2j} )E(F,\mathcal{M}_{2k} ) \;.
\end{equation}
In a $2$-form theory, $  U^{(2j-1)} ( c_{d-2j} ) $ is not a defect operator but only a local gauge transformation. We can also construct operators generating nontrivial principal bundles in $ \mathbb{R}^{d} $, which could change the class of $ A $. With $  \mathcal{M}_{2j}  $ divided into $\mathcal{M}^{N}_{2j}  $ and $\mathcal{M}^{S}_{2j} $ with $ \partial\mathcal{M}^{N}_{2j} =\mathcal{M}_{2j-1}=-\partial\mathcal{M}^{S}_{2j}  $, let $ u=v $ and $ u=1_{N}$ in $\mathcal{M}^{N}_{2j} $ and $\mathcal{M}^{S}_{2j}   $, respectively. The $ u $ bundle in $\mathcal{M}_{2j}   $ is characterized by the transition function $ v $ at $ \mathcal{M}_{2j-1} $, which is classified by $w(v, \mathcal{M}_{2j-1})\sim \Pi_{2j-1} (G)$. With $\mathcal{M}_{2j}    $ continuously deforming and finally covering the whole $ \mathbb{R}^{d} $, we get a nontrivial principal bundle in $\mathbb{R}^{d}$. Let $ a=u^{-1} du$ and $ f=da+a^{2} $. $ P_{j}(f) $ is supported on $ \Sigma_{d-2j} $ which is closed only when $w(v, \mathcal{M}_{2j-1})=0  $ so that the bundle is trivial. The corresponding transformation operator is $  U^{(2j-1)} ( \Sigma_{d-2j} ) $ with $\partial  \Sigma_{d-2j} = \Sigma_{d-2j-1}   $. Product of $  U^{(2j-1)} ( \Sigma_{d-2j} ) $ gives $   U^{(2j-1)} ( z_{d-2j} )$, where $ z_{d-2j} $ is a $(d-2j)$-chain with $ \partial  z_{d-2j} = c_{d-2j-1}    $. All $   U^{(2j-1)} ( z_{d-2j} )$ sharing the same $\partial  z_{d-2j}   $ are gauge equivalent and could be denoted by $  \mathcal{U}^{(2j)} ( c_{d-2j-1} )  $. $  \mathcal{U}^{(2j)} ( c_{d-2j-1} )  $ is a defect operator in the $2$-form gauge theory localized at $ c_{d-2j-1} $. 
\begin{equation}
\exp  \{ig E(F,\mathcal{M}_{2j} )\} \mathcal{U}^{(2j)} (c_{d-2j-1} )= \mathcal{U}^{(2j)} (c_{d-2j-1} )\exp  \{ig E(F,\mathcal{M}_{2j} )\}\exp  \{ig L( c_{d-2j-1}, \mathcal{M}_{2j} ) \}\;,
\end{equation}
where $L( c_{d-2j-1}, \mathcal{M}_{2j} )  $ is the Linking number between $  c_{d-2j-1}$ and $ \mathcal{M}_{2j} $.

In the following, for $ (d+1) $-dimensional nonabelian gauge theories, we will give a classification of the $A$ bundles with the non-zero $  E( F,\mathcal{M}_{2j} )   $ in $\mathbb{R}^{d}$, which may produce $ 2 $-forms with non-vanishing field strengths. We will also construct defect operators $ \mathcal{U}^{(2j)} (c_{d-2j-1} ) $.

\subsection{$ E( F,\mathcal{M}_{2} ) \neq 0$ }

\vspace{0.2em}

\paragraph{4d gauge theory}~{}

\vspace{1em}

In $ \mathbb{R}^{3} $, suppose $ E( F,\mathcal{M}_{2} ) \neq 0$ for a compact $2$-manifold $ \mathcal{M}_{2} $. $ \mathcal{M}_{2} $ could continuously deform into a point $ P \in \mathbb{R}^{3}$. In this process, $  \mathcal{M}_{2}  $ may cross some point-like singularities with $ E( F,\mathcal{M}_{2} )  $ changing by integers. When shrinking to $P$, if $  E( F,\mathcal{M}_{2} ) $ does not reduce to $ 0 $, then $P$ will also be a singularity. $ A $ has point-like singularities in $ \mathbb{R}^{3} $. The nontrivial $ A $ bundle produces a $ 2 $-form $ b=F $. $Q_{mnp}(b)=\frac{i}{2\pi} tr h$ is the linear combination of
\begin{equation}
Q_{mnp}(x)=\frac{i}{2\pi}(d \;tr F)_{mnp}(x)=\epsilon_{mnp}\delta^{(3)}(\vec{x}-\vec{y})
\end{equation}
with integer coefficients. The Dirac monopole is an example of such configurations \cite{1,2,3}.

One can also construct a non-globally defined $ u $ with $ w(v, \mathcal{M}_{1}) \neq 0$ for the transition function $ v $ at the overlap $\mathcal{M}_{1} \times I$. The related transformation operator is $U^{(1)} (z_{1} )$ with $ \partial z_{1}=c_{0} $. $ P_{1}(f) $ is supported on $z_{1}$. The action of $ U^{(1)} (z_{1} ) $ could make $ A $ transform into $A'$ in another class.
\begin{equation}
\exp  \{ig E(F,\mathcal{M}_{2} )\}   U^{(1)} (z_{1} )=   U^{(1)} (z_{1} )\exp  \{ig E(F,\mathcal{M}_{2} )\}\exp  \{ig I( z_{1}, \mathcal{M}_{2} ) \}\;, 
\end{equation}
where $ I( z_{1}, \mathcal{M}_{2} )  $ is the intersection number between $ z_{1} $ and $\mathcal{M}_{2}  $, the linking number $ L( c_{0}, \mathcal{M}_{2} ) $ between $ c_{0} $ and $ \mathcal{M}_{2} $ and also the winding number $ w( v, \mathcal{M}_{1} )  $ around $\mathcal{M}_{1}  $. 
\begin{equation}
 I( z_{1}, \mathcal{M}_{2} ) =L( c_{0}, \mathcal{M}_{2} )= w( v, \mathcal{M}_{1} )  \;.
\end{equation}
All $  U^{(1)} (z_{1} ) $ with the same $ \partial z_{1} $ can be identified as a single operator $ \mathcal{U}^{(2)} (c_{0} )$ with
\begin{equation}
\exp  \{ig E(F,\mathcal{M}_{2} )\} \mathcal{U}^{(2)} (c_{0} )= \mathcal{U}^{(2)} (c_{0} )\exp  \{ig E(F,\mathcal{M}_{2} )\}\exp  \{ig L( c_{0}, \mathcal{M}_{2} ) \}\;.
\end{equation}

If the coordinate of $ \mathbb{R}^{3} $ is $ (x^{1},x^{2}, x^{3}) $ and $G=U(N)$, the transformation matrix of $U^{(1)} (\Sigma_{1} )  $ with $ \Sigma_{1}=\{x^{1}=a^{1},x^{2}=a^{2}, x^{3} \geq a^{3} \} $ can be selected as 
\begin{equation}
	u(x^{1},x^{2},x^{3}) = \begin{cases}
	    \frac{[(x^{1}-a^{1})-i(x^{2}-a^{2})]}{[ \sum^{2}_{i=1}(x^{i}-a^{i})^{2} ]^{1/2}}   \oplus 1_{N-1}\;, & \text{when $ x^{3} \geq a^{3} $} \\
	      1_{N}, &  \text{when $ x^{3} < a^{3} $}	
		   \end{cases}\;.
\end{equation}
$ U^{(1)} (\Sigma_{1} )  \sim \mathcal{U}^{(2)} (\Sigma_{0} )  $, where $ \Sigma_{0}=\partial \Sigma_{1} $ is the point $(a^{1},a^{2},a^{3} )   $.

\vspace{1em}

\paragraph{5d gauge theory}~{}

\vspace{1em}

In $ \mathbb{R}^{4} $, suppose $ E( F,\mathcal{M}_{2} )  \neq 0$ for a compact $2$-manifold $ \mathcal{M}_{2} $ with $\partial \mathcal{M}_{3} =\mathcal{M}_{2}   $. $ A $ has point-like singularities in $ \mathcal{M}_{3}  $. When $ \mathcal{M}_{3}  $ deforms in $ \mathbb{R}^{4} $ with the boundary fixed, singularities move along with it continuously. The deformed $  \mathcal{M}_{3}  $ could cover $ \mathbb{R}^{4} $, with all singularities together composing a 1-cycle in $ \mathbb{R}^{4} $. The nontrivial $ A $ bundle produces a $ 2 $-form $ b=F $. $Q_{mnp}(b)=\frac{i}{2\pi} tr h$ is the linear combination of
\begin{equation}
Q_{mnp}(x)=\frac{i}{2\pi} (d \;tr F)_{mnp}(x)=\int_{\Sigma_{1}}\epsilon_{lmnp}\delta^{(4)}(\vec{x}-\vec{y})dy^{l}
\end{equation}
with integer coefficients. $ \partial \Sigma_{1}=0 $.

$  U^{(1)} (z_{2} ) $ is nontrivial only when $  \partial z_{2}= c_{1} \neq 0$. All $  U^{(1)} (z_{2} ) $ with the same $ \partial z_{2} $ can be identified as $ \mathcal{U}^{(2)} (c_{1} )$. 
\begin{equation}
\exp  \{ig E(F,\mathcal{M}_{2} )\} \mathcal{U}^{(2)} (c_{1} )= \mathcal{U}^{(2)} (c_{1} )\exp  \{ig E(F,\mathcal{M}_{2} )\}\exp  \{ig L( c_{1}, \mathcal{M}_{2} ) \}\;.
\end{equation}
$ \mathcal{U}^{(2)} (c_{1} ) $ is a defect operator supported at $ c_{1} $ which could change the characteristic class of $A$.

If the coordinate of $ \mathbb{R}^{4} $ is $ (x^{1},x^{2}, x^{3},x^{4}) $ and $G=U(N)$, the transformation matrix of $U^{(1)} (\Sigma_{2} )  $ with $ \Sigma_{2}=\{x^{1}=a^{1},x^{2}=a^{2}, x^{3} \geq a^{3} ,x^{4} \in \mathbb{R}\} $ can be selected as 
\begin{equation}
	u(x^{1},x^{2},x^{3},x^{4}) = \begin{cases}
	   \frac{[(x^{1}-a^{1})-i(x^{2}-a^{2})]}{[ \sum^{2}_{i=1}(x^{i}-a^{i})^{2} ]^{1/2}}  \oplus 1_{N-1}\;, & \text{when $ x^{3} \geq a^{3} $} \\
	      1_{N}, &  \text{when $ x^{3} < a^{3} $}	
		   \end{cases}\;.
\end{equation}
$ U^{(1)} (\Sigma_{2} )  \sim \mathcal{U}^{(2)} (\Sigma_{1} )  $, where $ \Sigma_{1}=\partial \Sigma_{2} =\{x^{1}=a^{1},x^{2}=a^{2}, x^{3} = a^{3} ,x^{4} \in \mathbb{R}\}$. 

\vspace{1em}

\paragraph{6d gauge theory}~{}

\vspace{1em}

In $ \mathbb{R}^{5} $, suppose $ E( F,\mathcal{M}_{2} )   \neq 0$ for a compact $2$-manifold $ \mathcal{M}_{2} $. By the same reasoning, $A$ has singularities supported at a 2-cycle $ c_{2} $. A $ 2 $-form $ b=F $ can be obtained. $Q_{mnp}(b)=\frac{i}{2\pi} tr h$ is the linear combination of
\begin{equation}
Q_{mnp}(x)=\frac{i}{2\pi}(d \;tr F)_{mnp}(x)=\frac{1}{2!}\int_{\Sigma_{2}}\epsilon_{l_{1}l_{2}mnp}\delta^{(5)}(\vec{x}-\vec{y})dy^{l_{1}}\wedge dy^{l_{2}}
\end{equation}
with integer coefficients. $ \partial \Sigma_{2}=0 $.

$  U^{(1)} (z_{3} ) $ is nontrivial only when $  \partial z_{3}= c_{2} \neq 0$. All $  U^{(1)} (z_{3} ) $ with the same $ \partial z_{3} $ can be identified as $ \mathcal{U}^{(2)} (c_{2} )$. 
\begin{equation}\label{623}
\exp  \{ig E(F,\mathcal{M}_{2} )\} \mathcal{U}^{(2)} (c_{2} )= \mathcal{U}^{(2)} (c_{2} )\exp  \{ig E(F,\mathcal{M}_{2} )\}\exp  \{ig L( c_{2}, \mathcal{M}_{2} ) \}\;,
\end{equation}
where $ L( c_{2}, \mathcal{M}_{2} ) $ is the linking number between $ c_{2} $ and $ \mathcal{M}_{2} $. (\ref{623}) looks like the Wilson-'t Hooft commutation relation in a $6d$ $2$-form theory \cite{6d}, with $\exp  \{ig E(F,\mathcal{M}_{2} )\}   $ and $\mathcal{U}^{(2)} (c_{2} )  $ playing roles of Wilson and 't Hooft surface operators, respectively.

If the coordinate of $ \mathbb{R}^{5} $ is $ (x^{1},x^{2}, x^{3},x^{4},x^{5}) $ and $G=U(N)$, the transformation matrix of $U^{(1)} (\Sigma_{3} )  $ with $ \Sigma_{3}=\{x^{1}=a^{1},x^{2}=a^{2}, x^{3} \geq a^{3} ,x^{4},x^{5} \in \mathbb{R}\} $ can be selected as 
\begin{equation}
	u(x^{1},x^{2},x^{3},x^{4},x^{5}) = \begin{cases}
	     \frac{[(x^{1}-a^{1})-i(x^{2}-a^{2})]}{[ \sum^{2}_{i=1}(x^{i}-a^{i})^{2} ]^{1/2}} \oplus 1_{N-1}\;, & \text{when $ x^{3} \geq a^{3} $} \\
	      1_{N}, &  \text{when $ x^{3} < a^{3} $}	
		   \end{cases}\;.
\end{equation}
$ U^{(1)} (\Sigma_{3} )  \sim \mathcal{U}^{(2)} (\Sigma_{2} )  $, where $ \Sigma_{2}=\partial \Sigma_{3}=\{x^{1}=a^{1},x^{2}=a^{2}, x^{3} = a^{3} ,x^{4},x^{5} \in \mathbb{R}\} $.

\subsection{$ E( F,\mathcal{M}_{4} ) \neq 0$ }

\vspace{0.2em}

\paragraph{6d gauge theory}~{}

\vspace{1em}

In $ \mathbb{R}^{5} $, suppose $ E( F,\mathcal{M}_{4} ) \neq 0$ for a compact $ \mathcal{M}_{4} $. $ A $ has point-like singularities in $ \mathbb{R}^{5} $. A $ 2 $-form $ b=F $ can be obtained with $Q_{mnpqr}(b)$ the linear combination of
\begin{equation}
Q_{mnpqr}(x)=\frac{1}{2}(\frac{i}{2\pi})^{2}(d \;tr F\wedge F)_{mnpqr}(x)=\epsilon_{mnpqr}\delta^{(5)}(\vec{x}-\vec{y})
\end{equation}
with integer coefficients. The instanton configuration on the Euclidean $ \mathbb{R}^{5} $ constructed in \cite{8} is one example.

$  U^{(3)} (z_{1} ) $ is nontrivial only when $  \partial z_{1}= c_{0} \neq 0$. All $  U^{(3)} (z_{1} ) $ with the same $ \partial z_{1} $ can be identified as $ \mathcal{U}^{(4)} (c_{0} )$. 
\begin{equation}
\exp  \{ig E(F,\mathcal{M}_{4} )\} \mathcal{U}^{(4)} (c_{0} )= \mathcal{U}^{(4)} (c_{0} )\exp  \{ig E(F,\mathcal{M}_{4} )\}\exp  \{ig L( c_{0}, \mathcal{M}_{4} ) \}\;.
\end{equation}

If the coordinate of $ \mathbb{R}^{5} $ is $ (x^{1},x^{2}, x^{3},x^{4},x^{5}) $ and $G=U(N) $ or $ SU(N)$ with $ N\geq 2 $, the transformation matrix of $U^{(3)} (\Sigma_{1} )  $ with $ \Sigma_{1}  =\{x^{1}=a^{1},x^{2}=a^{2}, x^{3} = a^{3} ,x^{4}=a^{4},x^{5}\geq a^{5}\} $ can be selected as 
\begin{equation}
	u(x^{1},x^{2},x^{3},x^{4},x^{5}) = \begin{cases}
	 \frac{\bar{\sigma}_{n}(x^{n}-a^{n})}{[ \sum^{4}_{i=1}(x^{i}-a^{i})^{2} ]^{1/2}}     \oplus 1_{N-2} \;, \;\;n=1,2,3,4\;, & \text{when $ x^{5} \geq a^{5} $} \\
	      1_{N}, &  \text{when $ x^{5} < a^{5} $}	
		   \end{cases}\;.
\end{equation}
$ U^{(3)} (\Sigma_{1} )  \sim \mathcal{U}^{(4)} (\Sigma_{0} )  $, where $ \Sigma_{0}=\partial \Sigma_{1} $ is the point $(a^{1},a^{2},a^{3},a^{4},a^{5} )   $.

\section{Topological sectors of $2$-form gauge theories}\label{n123}

In $1$-form gauge theories, pure gauges play an important role in the classification of the topological sectors. A gauge field configuration in $\mathbb{R}^{d} $ with the finite energy must have $ A \rightarrow u^{-1}du $ at $ \partial \mathbb{R}^{d} $. As a result, when $ d $ is even, depending on the boundary $ u^{-1}du  $, finite energy configurations are eigenstates of the conserved charge
\begin{equation}\label{3.2a}
E(F,\mathbb{R}^{d})=\int_{\mathbb{R}^{d}} P_{d/2}(F)=\int_{\partial \mathbb{R}^{d}} \text{CS}_{d-1} (u^{-1}du )=w(u,\partial \mathbb{R}^{d} )\in \mathbb{Z} \cong \Pi_{d-1}(G)
\end{equation}
with integer eigenvalues.

For example, in $ 3d $ $ U(N) $ $1$-form theories, the finite energy states fall into the homotopy sectors labeled by $ \Pi_{1} [U(N)]  $, the vortex charge. In $ 5d $ $1$-form gauge theories with $ G=U(N) $ or $SU(N)$ for $ N\geq 2 $, the finite energy states are in the homotopy class $ \Pi_{3} (G)$ labeled by the instanton number.

Semi-classical vacua are pure gauges. When $ d $ is odd, with $\partial \mathbb{R}^{d}  $ identified as one point so that $ \mathbb{R}^{d} \sim S^{d}$, pure gauge configurations in $ \mathbb{R}^{d} $ fall into the sectors labeled by the Chern-Simons charge $ O(A, \mathbb{R}^{d})$, which are integers for pure gauges:
\begin{equation}
O( u^{-1}du  , \mathbb{R}^{d})=\int_{\mathbb{R}^{d}}  \text{CS}_{d} ( u^{-1}du  )=w(u,S^{d} )\in \mathbb{Z} \cong \Pi_{d}(G)\;.
\end{equation}
The physical vacuum, the so-called $ \theta $-vacuum, is the superposition of them. For example, in $ 4d $ $1$-form gauge theories, pure gauges are classified by homotopy sectors of $ \Pi_{3} (G)$, and the solutions interpolating different sectors are instantons.

In this section, with the assumption that the pure gauge of the $ 2 $-form is $ dA+A^{2} $, we will study the topological sectors in the $B$ configuration space. The discussion does not apply to the chiral $2$-form, for which the self-duality condition will impose the constraint like $\epsilon^{ijklm} [B_{pq}(x),H_{klm}(y)]\sim  \delta^{(5)}(x-y)(\delta_{p}^{i}\delta_{q}^{j}-\delta_{q}^{i}\delta_{p}^{j})$, reducing the configuration space.

\subsection{$4d$ $U(N)$ $2$-form gauge theory}

First, consider a $U(1)$ $2$-form theory, which in $4d$ is dual to a $0$-form theory. The gauge field is $ B$ with $B \sim B + dA $ for a globally defined $ A $, and the field strength is $ H=dB $. The theory contains a conserved topological charge 
\begin{equation}
E(B,\partial \mathbb{R}^{3})=\frac{i}{2\pi}\int_{\mathbb{R}^{3}} H=\frac{i}{2\pi}\int_{\partial \mathbb{R}^{3}} B\;.
\end{equation}
The finite energy configurations in $ 3d $ must have $ B \rightarrow F $ at $\partial \mathbb{R}^{3}  $ for some closed $ F $, in which case,
\begin{equation}
E(B,\partial \mathbb{R}^{3})=\frac{i}{2\pi}\int_{\partial \mathbb{R}^{3}} F \in \mathbb{Z} \cong \Pi_{1} [U(1)]   \;.
\end{equation}
Therefore, depending on the boundary field $F$, $B$ configurations in $\mathbb{R}^{3}$ are divided into the topological sectors carrying the integer $ E(B,\partial \mathbb{R}^{3}) $ charges.

When $ N>1 $, 
\begin{equation}
E(B,\partial \mathbb{R}^{3})=\frac{i}{2\pi}\int_{\mathbb{R}^{3}} trH=\frac{i}{2\pi}\int_{\partial \mathbb{R}^{3}} tr B
\end{equation}
is a charge carried by the $U(1)$ factor. So based on the configurations of $ tr B$, $ U(N) $ $ B$ fields also fall into sectors labeled by $ \Pi_{1} [U(N)]\cong \Pi_{1} [U(1)] $.

\subsection{$5d$ nonabelian $2$-form gauge theory}

In a $ 5d $ $2$-form gauge theory with $ G=U(N) $ or $SU(N)$ for $ N\geq 2 $, the vacuum configurations with the vanishing field strength are pure gauges, which are expected to take the form of $B= dA+A^{2}$. So all configurations in a $ 5d $ $1$-form gauge theory are vacuum states of a $2$-form theory.

In a $ 5d $ $1$-form theory, the finite energy configurations in $\mathbb{R}^{4} $ are divided into sectors labeled by instanton numbers 
\begin{equation}\label{3.2}
E( F  , \mathbb{R}^{4})=\int_{\mathbb{R}^{4}}  P_{2}(F)=\int_{\partial \mathbb{R}^{4}}  \text{CS}_{3} ( u^{-1}du  )=w(u,\partial \mathbb{R}^{4} )\in \mathbb{Z} \cong \Pi_{3}(G)\;.
\end{equation}
Alternatively, with $\partial \mathbb{R}^{4}  $ identified as one point so that $ \mathbb{R}^{4} \sim S^{4}$, $ A $ bundles in $\mathbb{R}^{4}   $ fall into the sectors labeled by 
\begin{equation}
E( F  , S^{4})=\int_{S^{4}}  P_{2}(F)=\int_{S^{3}}  \text{CS}_{3} ( u^{-1}du  )=w(u,S^{3} )\in \mathbb{Z} \cong \Pi_{3}(G)\;.
\end{equation}
Such $A$ bundle can be constructed following (\ref{62})-(\ref{4.26}), with $ u $ the transition function at the overlap $ S^{3} \times I $. The $A$ bundle in (\ref{3.2}), although is defined in $ \mathbb{R}^{4} $, has $ F=0 $ at $\partial \mathbb{R}^{4}   $ and thus can also be mapped into $ S^{4} $, with $ S^{3} $ an infinitesimally small sphere surrounding the north pole.

The semiclassical vacua of the $2$-form theory fall into disjoint homotopy classes labeled by the instanton number. In a $5d$ $1$-form gauge theory, the instanton number is conserved topologically. $A$ and $A'$ with the different instanton numbers cannot be continuously deformed into each other so that the instanton transferring process is forbidden. On the other hand, in a $5d$ $2$-form theory, pure gauges $ F $ and $F'$ of different instanton numbers can be interpolated by the $2$-form $B$, so the transition between different vacua is possible. The physical vacuum is also a superposition labeled by an angle $ \theta $. The effect of the $ \theta $ vacuum can be encoded by a $5$-form $ Q_{5} (B)$ in the action, while the direct generalization of the $ 4d $ $ \theta $-term $tr( F\wedge F )$ gives $ tr(H \wedge H) $ in $6d$, which is $0$.

\subsection{$6d$ nonabelian non-chiral $2$-form gauge theory}

Consider the non-chiral $ 6d $ $2$-form gauge theory with $ G=U(N) $ or $SU(N)$ for $ N\geq 2 $. Again, we may assume the finite energy configurations in $\mathbb{R}^{5}$ have $ B \rightarrow dA+A^{2}  $ at $ \partial \mathbb{R}^{5} $. Depending on the boundary $F$, $ B $ fields fall into the sectors labeled by the charge $ E(F,\partial \mathbb{R}^{5} ) \cong  \Pi_{3}[G] $. For example, the $4d$ $ F $ configurations in a $ 5d $ $1$-form gauge theory are the proper boundary limit that could be mapped into $   \partial \mathbb{R}^{5} \sim  S^{4} \sim \mathbb{R}^{4}$.

$ P_{2} (F)$ may be the $1$-form limit of some $ P_{2}(B) $ with $Q_{5} (B)=d P_{2}(B)  $ a $5$-form gauge invariant. $d Q_{5} (B)=0  $, which yields a conserved charge 
\begin{equation}
E(B,\partial \mathbb{R}^{5} )=\int_{\mathbb{R}^{5}} Q_{5} (B)=\int_{\partial \mathbb{R}^{5}} P_{2}(B) = \int_{\partial \mathbb{R}^{5}} P_{2}(F) =  E(F,\partial \mathbb{R}^{5} )\;.
\end{equation}
The finite energy $B$ fields are divided into the sectors carrying the distinct integer charges, just like in $5d$ $1$-form theories, $A$ fields fall into the sectors labeled by instanton numbers. However, $ E(B,\partial \mathbb{R}^{5} ) $ does not impose a minimum energy bound on each sector, because $Q_{5} (B)  $ and $H \wedge \ast H$ do not have the same dimension. Even in $1$-form gauge theories, (\ref{3.2a}) sets an energy bound only when $d=4 $.

\section{Conclusion and discussion}\label{n1234}

In $(d+1)$-dimensional nonabelian $1$-form and $2$-form theories, the nontrivial $0$-form and $1$-form bundles are characterized by the homotopy group $\Pi_{k}(G)$ for all closed manifolds in $\mathbb{R}^{d}$. The discussion can be extended to $n$-form theories, where the nontrivial $(n-1)$-form bundles are also characterized by $\Pi_{k}(G)$, although the gauge transformation rule and the field strength definition are unknown when $n >1$. When $n=1$, a nontrivial $0$-form bundle produces a $1$-form $ a $ in $\mathbb{R}^{d}$ with $ P_{j} (f)$ supported on a $(d-2j)$-cycle $ c_{d-2j} $. In string theory, $  P_{j} (F) $ is the $D(d-2j)$-brane charge carried by $Dd$ branes, so $ a $ is the configuration of $D(d-2j)$-branes localized at $ c_{d-2j} $ (see Appendix \ref{ABx} for a summary of the stability and supersymmetry of these branes). Generically, a nontrivial $(n-1)$-form bundle could produce an $n$-form with some closed $(2j-1+n)$-form $ Q_{2j-1+n} $ supported on a $(d-2j+1-n)$-cycle $ c_{d-2j+1-n} $. These configurations also carry $(d-2j+1-n)$-form charges which do not have a string theory realization. Just as $   P_{j} (F) $ is the characteristic class of the $1$-form, we may expect $Q_{2j-1+n}  $ will be the characteristic class of the $n$-form.

In $\mathbb{R}^{d}$, $D(d-2j)$-brane configurations are in one-to-one correspondence with defect operators $U^{(2j-1)} ( c_{d-2j} )  $, which gives a classification of the disorder operators in $1$-form gauge theories. $ U^{(2j-1)} ( c_{d-2j} )  $ can also act as a local electric $(2j-1)$-form transformation, where the $(2j-1)$-form is the composite field $\text{CS}_{2j-1} (A)$. On the other hand, from the current conservation $dP_{j}(F)=0  $ in $\mathbb{R}^{d,1}$, $\exp  \{ig E(F,\mathcal{M}_{2j} )\}=\exp  \{ig O(A,\partial\mathcal{M}_{2j} )\}$ with $\mathcal{M}_{2j} \subset \mathbb{R}^{d}$ is a local magnetic $(d-2j)$-form transformation in space. (\ref{519}) shows that $U^{(2j-1)} ( c_{d-2j} )  $ and $\exp  \{ig O(A,\partial\mathcal{M}_{2j} )\}$ satisfy the required commutation relation. In $1$-form theories, these local transformations are not symmetries, but $ U^{(2j-1)} ( c_{d-2j} )   $ will become a local symmetry operator in $2$-form theories, where the defect operators are $\mathcal{U}^{(2j)}(c_{d-2j-1} )  $. Generically, in $n$-form theories, defect operators constructed from the nontrivial $(n-1)$-form bundles are $ \mathcal{U}^{(2j+n-2)}(c_{d-2j+1-n} ) $.

The classification of the topological sectors in $2$-form gauge theories is entirely based on the assumption that if $B= dA+A^{2} $, then the field strength $H=0$. We do not know the exact definition of $H$ for the nonabelian $2$-form. The assumption is made by taking the characteristic class equivalence relation (\ref{c}) of the $1$-form $A$ as a gauge equivalence relation of the $2$-form $F$.

\section*{Acknowledgements}
The author would like to thank Andreas Gustavsson for helpful comments and discussions. The work is supported in part by NSFC under the Grant No. 11605049.

\vspace{6mm}

\begin{appendix}

\section{$ \theta $-vacuum and the generalizations}\label{AAx}

In a $4d$ Yang-Mills theory, with $\vert u^{-1}du \rangle$ in (\ref{wn1}) for $w(u,\mathbb{R}^{3})=n$ denoted by $\vert n\rangle$, then 
\begin{equation}\label{hig}
\vert \theta \rangle =  \sum_{n}\;e^{-in\theta}\vert n\rangle\;. 
\end{equation}
For an arbitrary gauge invariant operator $\mathcal{O}$, $ \langle \theta' |\mathcal{O}|\theta\rangle =0 $, if $\theta' \neq\theta$, $ \langle m |\mathcal{O}|n\rangle =\langle m+1 |\mathcal{O}|n+1\rangle$. The cluster decomposition principle requires that the physical vacuum must be a $\theta$-vacuum \cite{Callan}. The vacuum expectation value of $\mathcal{O}(A)$ is \cite{qcd}
\begin{eqnarray}
\nonumber  \langle \theta |\mathcal{O}(A)|\theta\rangle &=&\frac{\sum_{m,n}e^{im\theta}e^{-in\theta}\langle m |\mathcal{O}(A) |n\rangle}{\sum_{m,n}e^{im\theta}e^{-in\theta}\langle m |n\rangle}=\frac{\sum_{Q}e^{iQ\theta}[\sum_{n}\langle n+Q |\mathcal{O}(A)  |n\rangle]}{\sum_{Q}e^{iQ\theta}[\sum_{n}\langle n+Q |n\rangle]}\\ &=& \frac{\sum_{Q}e^{iQ\theta}\int\mathcal{D}A_{Q}\;\mathcal{O}(A_{Q}) e^{-S[A_{Q}]}
}{\sum_{Q}e^{iQ\theta}\int\mathcal{D}A_{Q} \;e^{-S[A_{Q}]}}=\frac{\int\mathcal{D}A\;\mathcal{O} (A)e^{-S_{\theta}[A]}
}{ \int  \mathcal{D}A\;e^{-S_{\theta}[A]}}\;,
\end{eqnarray}
where $ A_{Q} $ is the $4d$ gauge field configuration with the instanton number $Q=m-n$, and 
\begin{equation}\label{Saz8}
S_{\theta}[A]=S[A]-i\theta \int_{\mathbb{R}^{3,1}} \;P_{2}(F)\;.
\end{equation}
So selecting a $ \theta $-vacuum amounts to specifying the $ \theta $ value in (\ref{Saz8}).

Now consider a $5d$ Yang-Mills theory coupling with a closed background $1$-form $ C^{(1)} $ which does not affect the equations of motion. The action is
\begin{equation}
S_{C^{(1)}}[A]=S[A]-i \int_{\mathbb{R}^{4,1}} \;C^{(1)}\wedge P_{2}(F)=S[A]-i \theta\int_{\mathcal{M}_{3} \times \mathbb{R}}  P_{2}(F)\;.
\end{equation}
$ C^{(1)}/\theta $ is gauge equivalent to the Poincar\'e dual of $ \mathcal{M}_{3} \times \mathbb{R} $, where $ \mathcal{M}_{3}  $ is a closed but not exact $3$-manifold in $\mathbb{R}^{4}$ and $\mathbb{R}$ is the time dimension. The vacuum expectation value of $\mathcal{O}(A)$ is 
\begin{eqnarray}\label{a5a}
\nonumber  \frac{\int\mathcal{D}A\;\mathcal{O}(A) e^{-S_{C^{(1)}}[A]}
}{ \int  \mathcal{D}A\;e^{-S_{C^{(1)}}[A]}} &=&\frac{\int\mathcal{D}A\;\mathcal{O}(A) e^{-S[A]+i \theta\int_{\mathcal{M}_{3} \times \mathbb{R}}  P_{2}(F)}
}{\int\mathcal{D}A \;e^{-S[A]+i \theta\int_{\mathcal{M}_{3} \times \mathbb{R}}  P_{2}(F)}}=\frac{\sum_{Q}e^{iQ\theta}\int\mathcal{D}A_{Q}\;\mathcal{O}(A_{Q}) e^{-S[A_{Q}]}
}{\sum_{Q}e^{iQ\theta}\int\mathcal{D}A_{Q} \;e^{-S[A_{Q}]}} \;.\\
\end{eqnarray}
A charge $Q$ instanton configuration in $\mathcal{M}_{3} \times \mathbb{R}  $ extending along the transverse direction gives an instanton string in $\mathbb{R}^{4,1}$, which is the sector $A_{Q}$ in (\ref{a5a}). (\ref{a5a}) can be formally interpreted as $\langle  C^{(1)} |\mathcal{O}(A)| C^{(1)}\rangle$, where $ | C^{(1)}\rangle $ is the higher dimensional analogue of (\ref{hig}). However, unless the spacetime has the topology of $ \mathbb{R}^{3,1} \times S^{1} $, $S[A_{Q}]$ is infinite with $e^{-S[A_{Q}]}=0$ when $Q \neq 0$. As a result, the $n$-changing process, which involves the infinitely heavy instanton strings, is forbidden. $ \langle m |\mathcal{O}(A) |n\rangle=0 $ when $m \neq n$. (\ref{a5a}) can also be taken as $   \langle n |\mathcal{O}(A) |n\rangle$. Furthermore, in the $5d$ path integral, the finite action configurations do not include the instanton string sector $ A_{Q} $, so the cluster decomposition principle argument, which yields the $ \theta $-vacuum in $4d$, cannot ensure the $C^{(1)}$-vacuum here. The vacuum does not need to preserve the $3$-form symmetry.

Finally, with a massless axion $a$ added to (\ref{Saz8}), the action becomes 
\begin{equation}
S_{\theta}[A,a]=S[A]+\int_{\mathbb{R}^{3,1}} \; da \wedge \ast da-i \int_{\mathbb{R}^{3,1}} (\theta+a) P_{2}(F)\;,
\end{equation}
which has a shift symmetry $S_{\theta}[A,a]=S_{\theta-c}[A,a+c]  $ for a constant $c$. The vacuum expectation value of $ \mathcal{O}(A) $ is
\begin{eqnarray}
\nonumber  \langle \theta   |\mathcal{O}(A)|\theta  \rangle &=&\frac{\sum_{m,n}e^{im\theta}e^{-in\theta}\langle m   |\mathcal{O}(A)  |n  \rangle}{\sum_{m,n}e^{im\theta}e^{-in\theta}\langle m   |n  \rangle}=\frac{\int \mathcal{D}A\mathcal{D}a\;\mathcal{O}(A)   e^{-S_{\theta}[A,a]}
}{ \int  \mathcal{D}A\mathcal{D}a\;e^{-S_{\theta}[A,a]}}\\ &=&\frac{\int \mathcal{D}A\mathcal{D}a\;\mathcal{O}(A) e^{-S_{\theta-c}[A,a]}
}{ \int  \mathcal{D}A\mathcal{D}a\;e^{-S_{\theta-c}[A,a]}}=\langle \theta -c  |\mathcal{O} (A)  |\theta-c\rangle \;.
\end{eqnarray}
There is no specified $\theta$.

\section{The action of $ U^{(2j-1)}(\Sigma_{d-2j}) $ on $ O(A,\mathcal{M}_{2k-1} ) $}\label{A}

Let
\begin{equation}
A_{t}=A_{0}+t(A_{1}-A_{0})\;\;\;\;\;\;\;\;\;\;F_{t}=dA_{t}+A_{t}^{2}\;.
\end{equation}
Define an operator $ l_{t} $ by
\begin{equation}
l_{t}A_{t}=0\;\;\;\;\;\;\;\;\;\; l_{t} F_{t}=\delta t(A_{1}-A_{0})\;.
\end{equation}
For any polynomial $ S(A,F) $ of $A$ and $F$, we have Cartan’s homotopy formula \cite{Ge}
\begin{equation}\label{4.33}
S(A_{1},F_{1})-S(A_{0},F_{0})=(dk_{01}+k_{01}d)S(A_{t},F_{t})
\end{equation}
where the homotopy operator $k_{01}  $ is defined by
\begin{equation}
k_{01}S(A_{t},F_{t})\equiv \int^{1}_{0}  l_{t}S(A_{t},F_{t})\;.
\end{equation}

Consider a singular gauge transformation with the transformation matrix $ u $:
\begin{equation}
A\rightarrow A^{u}=u^{-1}(A+d)u\;,\;\;\;\;\;\;\;\;\;F\rightarrow F^{u}=d A^{u} +(A^{u})^{2}=u^{-1}Fu+f\;,
\end{equation}
where $f=d(u^{-1}du)+(u^{-1}du)^{2}$. Suppose 
\begin{equation}
A^{u}_{t}=tu^{-1}Au+u^{-1}du\;\;\;\;\;\;\;\;\;\;F^{u}_{t}=d A^{u}_{t} +(A^{u}_{t})^{2} = u^{-1}F_{t}u+f\;,
\end{equation}
where $ F_{t} = tF+(t^{2}-t)A^{2} $, $ A_{t} =tA$. 
\begin{equation}
 A^{u}_{0}=u^{-1}du\;,\;\;\;\;\; A^{u}_{1}=u^{-1}Au+u^{-1}du =A^{u}\;,\;\;\;\;\; F^{u}_{0}=f\;,\;\;\;\;\; F^{u}_{1}=u^{-1}Fu +f=F^{u}\;.
\end{equation}
Let $ S(A,F)=\text{CS}_{2j-1} (A,F)$, where 
\begin{equation}
\text{CS}_{2j-1} (A,F)=\frac{1}{j!}(\frac{i}{2\pi})^{j}\int^{1}_{0} l_{t}\; tr(F_{t}^{j})=\frac{1}{(j-1)!}(\frac{i}{2\pi})^{j}\int^{1}_{0}\delta t \; str(A,F_{t}^{j-1})\;.
\end{equation}
From (\ref{4.33}), we have
\begin{eqnarray}
\nonumber  \text{CS}_{2j-1} (A^{u},F^{u})-\text{CS}_{2j-1}(u^{-1}du,f) &=&d[k_{01}\text{CS}_{2j-1}(A^{u}_{t},F^{u}_{t})] + k_{01}d\text{CS}_{2j-1}(A^{u}_{t},F^{u}_{t})\\ &=& d[k_{01}\text{CS}_{2j-1}(A^{u}_{t},F^{u}_{t})] + k_{01}P_{j}(F^{u}_{t})
\end{eqnarray}
with
\begin{eqnarray}\label{ki}
\nonumber && k_{01}P_{j}(F^{u}_{t}) =k_{01}  P_{j}[F_{t}+ufu^{-1}] =\frac{1}{j!}(\frac{i}{2 \pi})^{j}\int_{0}^{1}  l_{t} \;tr(F_{t}+ufu^{-1})^{j}\\\nonumber  &=&\text{CS}_{2j-1} (A,F)+\frac{1}{(j-1)!}(\frac{i}{2 \pi})^{j}\int_{0}^{1}  l_{t} \;str( ufu^{-1},F^{j-1}_{t})
\\\nonumber  &&+\frac{1}{(j-2)!2!}(\frac{i}{2 \pi})^{j}\int_{0}^{1}  l_{t} \;str(ufu^{-1},ufu^{-1}, F^{j-2}_{t} )+\cdots\\\nonumber  &=&\text{CS}_{2j-1} (A,F)+\frac{1}{(j-2)!}(\frac{i}{2 \pi})^{j}\int_{0}^{1} \delta t \; str( ufu^{-1} ,A,F^{j-2}_{t})
\\  &&+\frac{1}{(j-3)!2!}(\frac{i}{2 \pi})^{j}\int_{0}^{1} \delta t \;str(ufu^{-1},ufu^{-1}, A,F^{j-3}_{t} )+\cdots
\end{eqnarray}
For 
\begin{equation}
O (A,\mathcal{M}_{2j-1})= \int_{\mathcal{M}_{2j-1}}\text{CS}_{2j-1} (A,F)\;,
\end{equation}
there is
\begin{equation}
O (A^{u},\mathcal{M}_{2j-1})=O (u^{-1}du,\mathcal{M}_{2j-1})
+ \int_{\mathcal{M}_{2j-1}}k_{01}P_{j}(F^{u}_{t}) \;.
\end{equation}

If $ u $ is the transformation matrix of $ U^{(1)}(\Sigma_{d-2}) $, then
\begin{equation}
 f(x)=\frac{\pi H}{i(d-2)!}\int_{\Sigma_{d-2}}\epsilon_{l_{1}\cdots l_{d-2}m_{1}m_{2}}\delta^{(d)}(\vec{x}-\vec{y})\;dy^{l_{1}}\wedge \cdots\wedge dy^{l_{d-2}}\wedge dx^{m_{1}} \wedge  dx^{m_{2}} \;,
\end{equation}
where $ H$ can be selected as $1\oplus 0_{N-1} $. $ f^{k} =0$, when $ k\geq 2 $. $ufu^{-1}=f  $.
\begin{eqnarray}\label{a1}
\nonumber && U^{(1)-1}(\Sigma_{d-2})O (A,\mathcal{M}_{2j-1})U^{(1)}(\Sigma_{d-2})=O (A^{u},\mathcal{M}_{2j-1})\\\nonumber  &=&O (u^{-1}du,\mathcal{M}_{2j-1})
+ \int_{\mathcal{M}_{2j-1}}\left[  \text{CS}_{2j-1} (A,F)+\frac{1}{(j-2)!}(\frac{i}{2 \pi})^{j}\int_{0}^{1}  \delta t  \;str( ufu^{-1} ,A,F^{j-2}_{t}) \right] \\\nonumber  &=&O (u^{-1}du,\mathcal{M}_{2j-1})+O (A,\mathcal{M}_{2j-1})
+ \frac{1}{(j-2)!}(\frac{i}{2 \pi})^{j}\int_{\mathcal{M}_{2j-1}}\int_{0}^{1}  \delta t \;str( ufu^{-1} ,A,F^{j-2}_{t}) \\\nonumber  &=&O (u^{-1}du,\mathcal{M}_{2j-1})+O (A,\mathcal{M}_{2j-1})
+ \int_{\mathcal{M}_{2j-1}\cap \Sigma_{d-2}} \frac{1}{(j-2)!}(\frac{i}{2 \pi})^{j-1}\int_{0}^{1}  \delta t \;str( H ,A,F^{j-2}_{t})\;.\\
\end{eqnarray}
In (\ref{ki}), terms with more than one $ f $ do not have the contribution to (\ref{a1}). The last term in (\ref{a1}) will vanish if $ \mathcal{M}_{2j-1}\cap \Sigma_{d-2} $ is not a $(2j-3)$-manifold.

If $ u $ is the transformation matrix of $U^{(3)}(\Sigma_{d-4}) $, then $ f=f^{a}H_{a} $ with $ H_{a} =\tau_{a} \oplus 0_{N-2} $,
and 
\begin{equation}
f^{a}\wedge f^{b}=-\frac{4\pi^{2}\delta^{ab}}{3\cdot 4!(d-4)!}\int_{\Sigma_{d-4}}\epsilon_{l_{1}\cdots l_{d-4}m_{1}\cdots m_{4}}\delta^{(d)}(\vec{x}-\vec{y})\;dy^{l_{1}}\wedge \cdots\wedge dy^{l_{d-4}}\wedge dx^{m_{1}} \wedge\cdots \wedge dx^{m_{4}} \;.
\end{equation}
$ f^{k} =0$, when $ k\geq 3 $. Away from $ \Sigma_{d-4} $, $ f=0 $.
\begin{eqnarray}\label{a2}
\nonumber && U^{(3)-1}(\Sigma_{d-4})O (A,\mathcal{M}_{2j-1})U^{(3)}(\Sigma_{d-4})=O (A^{u},\mathcal{M}_{2j-1})\\\nonumber  &=&O (u^{-1}du,\mathcal{M}_{2j-1})
+ \int_{\mathcal{M}_{2j-1}}\left[ \text{CS}_{2j-1} (A,F)+\frac{1}{(j-3)!2!}(\frac{i}{2 \pi})^{j}\int_{0}^{1} \delta t  \;str(ufu^{-1},ufu^{-1}, A,F^{j-3}_{t} )\right]  \\\nonumber  &=&O (u^{-1}du,\mathcal{M}_{2j-1})+O (A,\mathcal{M}_{2j-1})+\frac{1}{(j-3)!2!}(\frac{i}{2 \pi})^{j} \int_{\mathcal{M}_{2j-1}}\int_{0}^{1}  \delta t\;str(ufu^{-1},ufu^{-1}, A,F^{j-3}_{t} )  \\\nonumber  &=&O (u^{-1}du,\mathcal{M}_{2j-1})+O (A,\mathcal{M}_{2j-1})\\  &+&\frac{1}{6(j-3)!}(\frac{i}{2 \pi})^{j-2} \int_{\mathcal{M}_{2j-1}\cap \Sigma_{d-4}}\int_{0}^{1}  \delta t\;str(uH_{a}u^{-1},uH_{a}u^{-1}, A,F^{j-3}_{t} ) \;.
\end{eqnarray}
In (\ref{ki}), terms with more than two $ f $ do not have the contribution to (\ref{a2}). For the term involving a single $ f $, 
\begin{equation}
\int_{\mathcal{M}_{2j-1}}\int_{0}^{1}  \delta t \;str( ufu^{-1} ,A,F^{j-2}_{t})\sim \int_{\mathcal{M}_{2j-1}\cap \Sigma_{d-4}} AF^{j-2}_{t} =0\;,
\end{equation}
because it is an integration of a $ (2j-3) $-form over a $(2j-5)$-manifold. The last term in (\ref{a2}) will vanish if $ \mathcal{M}_{2j-1}\cap \Sigma_{d-4}$ is not a $(2j-5)$-manifold.

\section{Stability and supersymmetry of branes within branes}\label{ABx}

The $D(d-2j)$-brane charge carried by $Dd$ branes is measured by the total flux $ \int_{\mathcal{M}_{2j}  }P_{j}(F) $. Dynamically, depending on the value of $j$, the brane charge density $P_{j}(F)$ may tend to spread out in $Dd$ or contract to a $ \delta $-function localized at $ \Sigma_{d-2j} $.

Consider $ D(d-2j) $ and $Dd$ branes parallel to each other and separated in the transverse space. Each preserve $16$ supersymmetries. As a system, the configuration preserves $8$ supersymmetries with the zero binding energy only when $j=2, 4$ \cite{z1, z2, z3, z4, z5, z6, z7}. The $ D(d-2) $-$Dd$ interaction is attractive, so $ D(d-2) $ will move toward and dissolve in $Dd$, leaving the uniformly distributed flux. The system then becomes BPS. In fact, the energy scales as $ F^{2} $ while the $D(d-2) $ charge scales as $ F$, so the minimum energy is obtained for the uniformly distributed flux. $ D(d-4) $ and $Dd$ exert no force to each other and can move toward (apart) freely. The $ D(d-4)$ charge scales as $F^{2}$, so classically, the charge density in $Dd$ can spread out (shrink to $\Sigma_{d-4}$) without costing the energy. The $ D(d-6) $-$Dd$ interaction is repulsive and the $D(d-6)  $ charge scales as $F^{3}$, so the $D(d-6) $ charge inside $Dd$ will converge at $\Sigma_{d-6} $ to produce a localized $D(d-6) $ which then moves away from $Dd$ to reduce the energy. The system can never be BPS.

The defect operator $U^{(2j-1)}(\Sigma_{d-2j})$ creates a $D(d-2j)$ brane localized at $ \Sigma_{d-2j} $ of $Dd$, which is stable and supersymmetric only when $ j=2 $ or $4$ and $\Sigma_{d-2j}$ is a $ (d-2j) $-plane in $\mathbb{R}^{d}$. In the following, we will calculate the supersymmetry transformation of $ U^{(2j-1)}(\Sigma_{d-2j}) $ in super Yang-Mills theories for $j=1, 2$. The results are consistent with the corresponding branes.

In $(d+1)$-dimensional super Yang-Mills theories, the supercharge operator in canonical formalism is  
\begin{equation}
\mathcal{Q}=\int d^{d}y \; tr(\frac{1}{2}F_{mn}\Gamma^{mn}\Gamma^{0}\Psi+\cdots)\;,
\end{equation}
where $m, n=1,\cdots,d$, $  \Gamma^{m}$, $\Gamma^{0}$, $ \Psi $ and $\mathcal{Q}$ are $10d$ Gamma matrices and spinors. $ \cdots $ are terms without involving $F_{mn}$. Under the action of a defect operator $U^{(2j-1)}(\Sigma_{d-2j})$, from (\ref{ru1})-(\ref{2.14a}), 
\begin{equation}
U^{(2j-1)-1}(\Sigma_{d-2j})\mathcal{Q}U^{(2j-1)}(\Sigma_{d-2j})=\mathcal{Q}+\int d^{d}y \; tr(\frac{1}{2}f_{mn}\Gamma^{mn}\Gamma^{0}u^{-1}\Psi u)\;, 
\end{equation}
\begin{equation}\label{uh1}
[\mathcal{Q},U^{(2j-1)}(\Sigma_{d-2j})]=\left( \int d^{d}y \; tr(\frac{1}{2}f_{mn}\Gamma^{mn}\Gamma^{0}\Psi )\right)  U^{(2j-1)}(\Sigma_{d-2j})     \;, 
\end{equation}
where we have used the fact that gauge invariant operators without involving $F_{mn}$ do not change under the action of $U^{(2j-1)}(\Sigma_{d-2j})$.

When $ j=1 $, $ d=2 $ and $\Sigma_{0}=\{x\}$, with (\ref{uh2}) plugged in (\ref{uh1}), 
\begin{equation}
[\mathcal{Q},U^{(1)}(x)]=\left( -\pi i \;tr[  H \epsilon_{mn}\Gamma^{mn}\Gamma^{0}\Psi(x)]\right) U^{(1)}(x)\;.
\end{equation}
When $ j=1 $, $ d=3 $, with (\ref{uh22}) plugged in (\ref{uh1}), 
\begin{equation}
[\mathcal{Q},U^{(1)}(\Sigma_{1})]=\left( -\pi i \int_{\Sigma_{1}}  tr[ H     \epsilon_{lmn}  \Gamma^{mn}\Gamma^{0}\Psi(x)]\;dx^{l}\right) U^{(1)}(\Sigma_{1})\;.
\end{equation}
For the generic $d$, let $\Sigma_{d-2} =\{x^{1}=a^{1},x^{2}=a^{2},x^{3},\cdots,x^{d} \in \mathbb{R}\}$ for definiteness. Then 
\begin{equation}
[\mathcal{Q},U^{(1)}(\Sigma_{d-2})]=\left( -2\pi i \int_{\Sigma_{d-2}}  tr[ H     \Gamma^{1}\Gamma^{2}\Gamma^{0}\Psi(x)]\;d \sigma^{(d-2)} \right) U^{(1)}(\Sigma_{d-2})\;.
\end{equation}
Just like $D(d-2)$ branes localized in $Dd$, $U^{(1)}(\Sigma_{d-2}) $ breaks all supersymmetries.\footnote{Of course, in SYM theories, monopole ('t Hooft) operators and 't Hooft loops/lines are BPS when the suitable matter coupling is included \cite{th1, ca1}. In our case, let 
\begin{equation}
V^{I}(\Sigma_{d-2})=\exp \left\lbrace  -2\pi i \int_{\Sigma_{d-2}}  tr[ H \Phi^{I}(x)   ]\; d\sigma^{(d-2)}\right\rbrace \;,\;\;\;\;\;\;\;\;\;I=1,\cdots,9
\end{equation}
where $ \Phi^{I} $ with $I=d+1,\cdots,9$ are scalars and the situation of $ I=1,\cdots,d $ with $\Phi^{I}=A^{I}  $ is also taken into account for completeness.
\begin{equation}
[\mathcal{Q},V^{I}(\Sigma_{d-2})]=\left( -2\pi i \int_{\Sigma_{d-2}}  tr[ H \Gamma^{I} \Psi (x) ]\;d\sigma^{(d-2)}\right)  V^{I}(\Sigma_{d-2})\;,
\end{equation}
so for the dressed operator $ V^{I}(\Sigma_{d-2})U^{(1)}(\Sigma_{d-2}) $,
\begin{eqnarray}
\nonumber  [\mathcal{Q},V^{I}(\Sigma_{d-2})U^{(1)}(\Sigma_{d-2})] &=&\left( -2\pi i \int_{\Sigma_{d-2}}  tr[ H \Gamma^{I} \Psi(x) + H     \Gamma^{1}\Gamma^{2}\Gamma^{0}\Psi (x)]\;d\sigma^{(d-2)}\right) V^{I}(\Sigma_{d-2}) U^{(1)}(\Sigma_{d-2})\\ &=&\frac{1}{2} (1 +     \Gamma)\left( -4\pi i \int_{\Sigma_{d-2}}  tr[  \Gamma^{I} H\Psi(x)]\;d\sigma^{(d-2)}\right)  V^{I}(\Sigma_{d-2}) U^{(1)}(\Sigma_{d-2})\;,
\end{eqnarray}
where $  \Gamma=\Gamma^{0}\Gamma^{1}\Gamma^{2}\Gamma^{I} $. $ (\Gamma)^{2} =1$. $ \mathcal{Q} $ can be decomposed as $\mathcal{Q}=\mathcal{Q}^{012I+}+\mathcal{Q}^{012I-}$ with $\mathcal{Q}^{012I\pm}=\frac{1}{2}(1\pm \Gamma)\mathcal{Q}$. 
\begin{equation}
 [\mathcal{Q}^{012I-},V^{I}(\Sigma_{d-2})U^{(1)}(\Sigma_{d-2})]=0\;.
\end{equation}
$ V^{I}(\Sigma_{d-2})U^{(1)}(\Sigma_{d-2}) $ is $1/2$ BPS. This is not in conflict with the statement for branes, since $ \mathcal{Q}^{012I-} $ cannot be the supersymmetry preserved by the $D(d-2)$ within $Dd$ configurations.}

When $ j=2 $, $ d=4 $, let $G=SU(2)$ for simplicity. With (\ref{fa2}) plugged in (\ref{uh1}), 
\begin{eqnarray}
\nonumber  [\mathcal{Q},U^{(3)}(\Sigma_{0})] &=& \left(  \int d^{4}y \; tr(\frac{\pi}{\sqrt{6}}\sigma_{mn}\delta^{(2)}(\vec{x}-\vec{y})\Gamma^{mn}\Gamma^{0} \Psi (y))\right) U^{(3)}(\Sigma_{0})\\\nonumber &=&\frac{1}{2}(1+\Gamma^{5}) \left(    \int d^{4}y \; \frac{\pi}{\sqrt{6}} tr(\sigma_{mn}\Gamma^{mn}\delta^{(2)}(\vec{x}-\vec{y})\Gamma^{0} \Psi (y))\right) U^{(3)}(\Sigma_{0})\;,\\
\end{eqnarray}
where we have used $ \sigma_{mn}=\frac{1}{2}\epsilon_{mnkl}\sigma^{kl}$ and $\Gamma^{5}\Gamma^{mn}=\frac{1}{2}\epsilon^{mnkl}\Gamma_{kl}$ with $\Gamma^{5}=-\Gamma^{1}\Gamma^{2}\Gamma^{3}\Gamma^{4}$. So for $\mathcal{Q}=\mathcal{Q}^{+}+\mathcal{Q}^{-}$ with $\mathcal{Q}^{\pm}=\frac{1}{2}(1\pm \Gamma^{5})\mathcal{Q}$, 
\begin{equation}
 [\mathcal{Q}^{-},U^{(3)}(\Sigma_{0})]=0\;.
\end{equation}
The instanton operator $  U^{(3)}(\Sigma_{0})$ is $1/2$ BPS \cite{Ber, 9}, and $\mathcal{Q}^{-}$ is indeed the supercharge preserved by $D0$ embedded in $D4$.

\end{appendix}





\vspace{3mm}


\begin{thebibliography}{10}











\bibitem{0}
L.~Breen and W.~Messing, ``Differential geometry of GERBES,'' Adv. Math. \textbf{198} (2005), 732, math/0106083.




\bibitem{00}
J.~C.~Baez and J.~Huerta, ``An Invitation to Higher Gauge Theory,'' Gen. Rel. Grav. \textbf{43} (2011), 2335-2392, arXiv:1003.4485.





\bibitem{000}
L.~Borsten, M.~J.~Farahani, B.~Jurco, H.~Kim, J.~Narozny, D.~Rist, C.~Saemann and M.~Wolf, ``Higher Gauge Theory,'' arXiv:2401.05275.






\bibitem{01}
C.~Teitelboim, ``Gauge Invariance for Extended Objects,'' Phys. Lett. B \textbf{167} (1986), 63-68.



\bibitem{01a}
M.~Henneaux, ``Uniqueness of the Freedman-Townsend interaction vertex for two form gauge fields,'' Phys. Lett. B \textbf{368} (1996), 83-88, hep-th/9511145.


\bibitem{001}
M.~Henneaux and B.~Knaepen, ``All consistent interactions for exterior form gauge fields,'' Phys. Rev. D \textbf{56} (1997), R6076-R6080, hep-th/9706119.







\bibitem{G}
D.~Gaiotto, A.~Kapustin, N.~Seiberg and B.~Willett, ``Generalized Global Symmetries,'' JHEP \textbf{02} (2015), 172, arXiv:1412.5148.







\bibitem{1}
P.~A.~M.~Dirac, ``Quantised singularities in the electromagnetic field,'' Proc. Roy. Soc. Lond. A \textbf{133} (1931) no.821, 60-72.




\bibitem{2}
P.~A.~M.~Dirac, ``The Theory of magnetic poles,''
Phys. Rev. \textbf{74} (1948), 817-830.



\bibitem{3}
T.~T.~Wu and C.~N.~Yang, ``Concept of Nonintegrable Phase Factors and Global Formulation of Gauge Fields,'' Phys. Rev. D \textbf{12} (1975), 3845-3857.








\bibitem{5}
M.~R.~Douglas, ``Branes within branes,'' in \textsl{Strings, Branes and Dualities}, Nato Science Series C \textbf{520}, Springer, Dordrecht, The Netherlands (1999), pp. 267–275, hep-th/9512077.




\bibitem{6}
G.~'t Hooft, ``On the Phase Transition Towards Permanent Quark Confinement,'' Nucl. Phys. B \textbf{138} (1978), 1-25.



\bibitem{7}
V.~Borokhov, A.~Kapustin and X.~k.~Wu, ``Topological disorder operators in three-dimensional conformal field theory,'' JHEP \textbf{11} (2002), 049, hep-th/0206054.





\bibitem{7a}
H.~Reinhardt, ``On 't Hooft's loop operator,'' Phys. Lett. B \textbf{557} (2003), 317-323, hep-th/0212264.



\bibitem{th1}
A.~Kapustin, ``Wilson-'t Hooft operators in four-dimensional gauge theories and S-duality,'' Phys. Rev. D \textbf{74} (2006), 025005, hep-th/0501015.




\bibitem{8}
N.~Lambert, C.~Papageorgakis and M.~Schmidt-Sommerfeld, ``Instanton Operators in Five-Dimensional Gauge Theories,'' JHEP \textbf{03} (2015), 019, arXiv:1412.2789.

\bibitem{Ber}
O.~Bergman and D.~Rodriguez-Gomez, ``A Note on Instanton Operators, Instanton Particles, and Supersymmetry,'' JHEP \textbf{05} (2016), 068, arXiv:1601.00752.




\bibitem{9}
S.~Hu, ``Quantum Spectrum of BPS Instanton States in $ 5d $ Gauge Theories,'' arXiv:2309.16394.




\bibitem{wi}
E.~Witten, ``Dyons of Charge e theta/2 pi,'' Phys. Lett. B \textbf{86} (1979), 283-287.





\bibitem{Li}
M.~Li, ``Boundary states of D-branes and Dy strings,'' Nucl. Phys. B \textbf{460} (1996), 351-361, hep-th/9510161.




\bibitem{gw1}
S.~Gukov and E.~Witten, ``Gauge Theory, Ramification, And The Geometric Langlands Program,'' hep-th/0612073.




\bibitem{gw2}
S.~Gukov and E.~Witten, ``Rigid Surface Operators,'' Adv. Theor. Math. Phys. \textbf{14} (2010) no.1, 87-178, arXiv:0804.1561.



\bibitem{Wl}
D.~Diakonov and V.~Y.~Petrov, ``A Formula for the Wilson Loop,'' Phys. Lett. B \textbf{224} (1989), 131-135.




\bibitem{Wl1} 
K.~I.~Kondo and Y.~Taira, ``NonAbelian Stokes theorem and quark confinement in SU(N) Yang-Mills gauge theory,'' Prog. Theor. Phys. \textbf{104} (2000), 1189-1265, hep-th/9911242.



\bibitem{Wl2} 
R.~Matsudo and K.~I.~Kondo, ``Non-Abelian Stokes theorem for the Wilson loop operator in an arbitrary representation and its implication to quark confinement,'' Phys. Rev. D \textbf{92} (2015) no.12, 125038, arXiv:1509.04891. 




\bibitem{Ge}
M.~Nakahara, \textit{Geometry, Topology and Physics, Second Edition.} Graduate student series in physics. Taylor \& Francis, 2003.






\bibitem{49} 
A.~A.~Belavin, A.~M.~Polyakov, A.~S.~Schwartz and Y.~S.~Tyupkin, ``Pseudoparticle Solutions of the Yang-Mills Equations,'' Phys. Lett. B \textbf{59} (1975), 85-87.





\bibitem{s1}
S.~Grozdanov, D.~M.~Hofman and N.~Iqbal, ``Generalized global symmetries and dissipative magnetohydrodynamics,'' Phys. Rev. D \textbf{95} (2017) no.9, 096003, arXiv:1610.07392.


\bibitem{s2}
F.~Benini, P.~S.~Hsin and N.~Seiberg, ``Comments on global symmetries, anomalies, and duality in (2 + 1)d,'' JHEP \textbf{04} (2017), 135, arXiv:1702.07035.



\bibitem{s3}
L.~Bhardwaj and Y.~Tachikawa, ``On finite symmetries and their gauging in two dimensions,'' JHEP \textbf{03} (2018), 189, arXiv:1704.02330.


\bibitem{s4}
D.~Gaiotto and T.~Johnson-Freyd, ``Symmetry Protected Topological phases and Generalized Cohomology,'' JHEP \textbf{05} (2019), 007, arXiv:1712.07950.



\bibitem{s5}
C.~M.~Chang, Y.~H.~Lin, S.~H.~Shao, Y.~Wang and X.~Yin, ``Topological Defect Lines and Renormalization Group Flows in Two Dimensions,'' JHEP \textbf{01} (2019), 026, arXiv:1802.04445.


\bibitem{s6}
C.~C\'ordova, T.~T.~Dumitrescu and K.~Intriligator, ``Exploring 2-Group Global Symmetries,'' JHEP \textbf{02} (2019), 184, arXiv:1802.04790. 




\bibitem{s7}
F.~Benini, C.~C\'ordova and P.~S.~Hsin, ``On 2-Group Global Symmetries and their Anomalies,'' JHEP \textbf{03} (2019), 118, arXiv:1803.09336.



\bibitem{s8}
P.~S.~Hsin, H.~T.~Lam and N.~Seiberg, ``Comments on One-Form Global Symmetries and Their Gauging in 3d and 4d,'' SciPost Phys. \textbf{6} (2019) no.3, 039, arXiv:1812.04716.





\bibitem{s9}
N.~Seiberg, ``Field Theories With a Vector Global Symmetry,'' SciPost Phys. \textbf{8} (2020) no.4, 050, arXiv:1909.10544.


\bibitem{s10}
D.~R.~Morrison, S.~Schafer-Nameki and B.~Willett, ``Higher-Form Symmetries in 5d,'' JHEP \textbf{09} (2020), 024, arXiv:2005.12296.




\bibitem{s11}
F.~Albertini, M.~Del Zotto, I.~Garc\'\i{}a Etxebarria and S.~S.~Hosseini, ``Higher Form Symmetries and M-theory,'' JHEP \textbf{12} (2020), 203, arXiv:2005.12831.


\bibitem{Ruc}
T.~Rudelius and S.~H.~Shao, ``Topological Operators and Completeness of Spectrum in Discrete Gauge Theories,'' JHEP \textbf{12} (2020), 172, arXiv:2006.10052.



\bibitem{ss11}
L.~Bhardwaj and S.~Sch\"afer-Nameki, ``Higher-form symmetries of 6d and 5d theories,'' JHEP \textbf{02} (2021), 159, arXiv:2008.09600.





\bibitem{s12}
B.~Heidenreich, J.~McNamara, M.~Montero, M.~Reece, T.~Rudelius and I.~Valenzuela, ``Non-invertible global symmetries and completeness of the spectrum,'' JHEP \textbf{09} (2021), 203, arXiv:2104.07036.








\bibitem{s13}
Y.~Choi, C.~Cordova, P.~S.~Hsin, H.~T.~Lam and S.~H.~Shao, ``Noninvertible duality defects in 3+1 dimensions,'' Phys. Rev. D \textbf{105} (2022) no.12, 125016, arXiv:2111.01139.




\bibitem{s14}
J.~Kaidi, K.~Ohmori and Y.~Zheng, ``Kramers-Wannier-like Duality Defects in (3+1)D Gauge Theories,'' Phys. Rev. Lett. \textbf{128} (2022) no.11, 111601, arXiv:2111.01141.



\bibitem{s15}
F.~Apruzzi, F.~Bonetti, I.~Garc\'\i{}a Etxebarria, S.~S.~Hosseini and S.~Schafer-Nameki, ``Symmetry TFTs from String Theory,'' Commun. Math. Phys. \textbf{402} (2023) no.1, 895-949, arXiv:2112.02092.








\bibitem{s16}
K.~Roumpedakis, S.~Seifnashri and S.~H.~Shao, ``Higher Gauging and Non-invertible Condensation Defects,'' Commun. Math. Phys. \textbf{401} (2023) no.3, 3043-3107, arXiv:2204.02407.



\bibitem{s17}
L.~Bhardwaj, L.~E.~Bottini, S.~Schafer-Nameki and A.~Tiwari, ``Non-invertible higher-categorical symmetries,'' SciPost Phys. \textbf{14} (2023) no.1, 007, arXiv:2204.06564.





\bibitem{s18}
Y.~Choi, C.~Cordova, P.~S.~Hsin, H.~T.~Lam and S.~H.~Shao, ``Non-invertible Condensation, Duality, and Triality Defects in 3+1 Dimensions,'' Commun. Math. Phys. \textbf{402} (2023) no.1, 489-542, arXiv:2204.09025.

\bibitem{s19}
J.~Kaidi, G.~Zafrir and Y.~Zheng, ``Non-invertible symmetries of $ \mathcal{N} $ = 4 SYM and twisted compactification,'' JHEP \textbf{08} (2022), 053, arXiv:2205.01104.


\bibitem{s20}
Y.~Choi, H.~T.~Lam and S.~H.~Shao, ``Noninvertible Global Symmetries in the Standard Model,'' Phys. Rev. Lett. \textbf{129} (2022) no.16, 161601, arXiv:2205.05086.



\bibitem{s21}
V.~Bashmakov, M.~Del Zotto and A.~Hasan, ``On the 6d origin of non-invertible symmetries in 4d,'' JHEP \textbf{09} (2023), 161, arXiv:2206.07073.






\bibitem{s22}
L.~Bhardwaj, S.~Schafer-Nameki and J.~Wu, ``Universal Non-Invertible Symmetries,'' Fortsch. Phys. \textbf{70} (2022) no.11, 2200143, arXiv:2208.05973.



\bibitem{s23}
F.~Apruzzi, I.~Bah, F.~Bonetti and S.~Schafer-Nameki, ``Noninvertible Symmetries from Holography and Branes,'' Phys. Rev. Lett. \textbf{130} (2023) no.12, 121601, arXiv:2208.07373.




\bibitem{s23a}
I.~Bah, E.~Leung and T.~Waddleton, ``Non-invertible symmetries, brane dynamics, and tachyon condensation,'' JHEP \textbf{01} (2024), 117, arXiv:2306.15783.



\bibitem{ss23}
J.~J.~Heckman, M.~H\"ubner, E.~Torres and H.~Y.~Zhang, ``The Branes Behind Generalized Symmetry Operators,'' Fortsch. Phys. \textbf{71} (2023) no.1, 2200180, arXiv:2209.03343.



\bibitem{s24}
J.~Kaidi, K.~Ohmori and Y.~Zheng, ``Symmetry TFTs for Non-invertible Defects,''
Commun. Math. Phys. \textbf{404} (2023) no.2, 1021-1124, arXiv:2209.11062.



\bibitem{ss24}
J.~J.~Heckman, M.~Hubner, E.~Torres, X.~Yu and H.~Y.~Zhang, ``Top down approach to topological duality defects,'' Phys. Rev. D \textbf{108} (2023) no.4, 046015, arXiv:2212.09743. 


\bibitem{Jac}
R.~Jackiw, ``Introduction to the Yang-Mills Quantum Theory,'' Rev. Mod. Phys. \textbf{52} (1980), 661-673.

\bibitem{qcd}
H.~Forkel, ``A Primer on instantons in QCD,'' hep-ph/0009136.

\bibitem{Callan}
C.~G.~Callan, Jr., R.~F.~Dashen and D.~J.~Gross, ``The Structure of the Gauge Theory Vacuum,'' Phys. Lett. B \textbf{63} (1976), 334-340.




\bibitem{qed2}
C.~Gattringer, ``$QED_{2}$ and U(1) problem,'' hep-th/9503137.









\bibitem{g1}
T.~Banks and L.~J.~Dixon, ``Constraints on String Vacua with Space-Time Supersymmetry,'' Nucl. Phys. B \textbf{307} (1988), 93-108.




\bibitem{g2}
S.~B.~Giddings and A.~Strominger, ``Loss of incoherence and determination of coupling constants in quantum gravity,'' Nucl. Phys. B \textbf{307} (1988), 854-866.



\bibitem{g3}
R.~Kallosh, A.~D.~Linde, D.~A.~Linde and L.~Susskind, ``Gravity and global symmetries,'' Phys. Rev. D \textbf{52} (1995), 912-935, hep-th/9502069.





\bibitem{g4}
N.~Arkani-Hamed, L.~Motl, A.~Nicolis and C.~Vafa, ``The String landscape, black holes and gravity as the weakest force,'' JHEP \textbf{06} (2007), 060, hep-th/0601001.





\bibitem{g5}
T.~Banks and N.~Seiberg, ``Symmetries and Strings in Field Theory and Gravity,''
Phys. Rev. D \textbf{83} (2011), 084019, arXiv:1011.5120.




\bibitem{g6}
D.~Harlow and H.~Ooguri, ``Constraints on Symmetries from Holography,''
Phys. Rev. Lett. \textbf{122} (2019) no.19, 191601, arXiv:1810.05337.



\bibitem{g7}
D.~Harlow and H.~Ooguri, ``Symmetries in quantum field theory and quantum gravity,'' Commun. Math. Phys. \textbf{383} (2021) no.3, 1669-1804, arXiv:1810.05338.





\bibitem{g8}
D.~Harlow and E.~Shaghoulian, ``Global symmetry, Euclidean gravity, and the black hole information problem,'' JHEP \textbf{04} (2021), 175, arXiv:2010.10539.









\bibitem{cw}
B.~Heidenreich, J.~McNamara, M.~Montero, M.~Reece, T.~Rudelius and I.~Valenzuela, ``Chern-Weil global symmetries and how quantum gravity avoids them,'' JHEP \textbf{11} (2021), 053, arXiv:2012.00009. 




\bibitem{cw1}
E.~Garc\'\i{}a-Valdecasas, M.~Reece and M.~Suzuki, ``Monopole Breaking of Chern-Weil Symmetries,'' arXiv:2408.00067.











\bibitem{6d}
M.~Henningson, ``Commutation relations for surface operators in six-dimensional (2, 0) theory,'' JHEP \textbf{03} (2001), 011, hep-th/0012070.










\bibitem{z1}
M.~B.~Green and M.~Gutperle, ``Light cone supersymmetry and d-branes,'' Nucl. Phys. B \textbf{476} (1996), 484-514, hep-th/9604091.







\bibitem{z2}
G.~Lifschytz, ``Comparing d-branes to black-branes,'' Phys. Lett. B \textbf{388} (1996), 720-726, hep-th/9604156.







\bibitem{z3}
A.~A.~Tseytlin, ```No force' condition and BPS combinations of p-branes in eleven-dimensions and ten-dimensions,'' Nucl. Phys. B \textbf{487} (1997), 141-154, hep-th/9609212.



\bibitem{z4}
J.~Polchinski, ``Tasi lectures on D-branes,'' hep-th/9611050.

\bibitem{z5}
U.~Danielsson, G.~Ferretti and I.~R.~Klebanov, ``Creation of fundamental strings by crossing D-branes,'' Phys. Rev. Lett. \textbf{79} (1997), 1984-1987, hep-th/9705084.



\bibitem{z6}
W.~Taylor, ``Adhering zero-branes to six-branes and eight-branes,'' Nucl. Phys. B \textbf{508} (1997), 122-132, hep-th/9705116.


\bibitem{z7}
W.~Taylor, ``Lectures on D-branes, gauge theory and M(atrices),'' hep-th/9801182.














\bibitem{ca1}
V.~Borokhov, A.~Kapustin and X.~k.~Wu, ``Monopole operators and mirror symmetry in three-dimensions,'' JHEP \textbf{12} (2002), 044, hep-th/0207074.

























\end{thebibliography}
\end{document}